\documentclass[a4paper,11pt]{article}
\pdfoutput=1

\usepackage{jcappub}
\usepackage[T1]{fontenc} 
\usepackage[utf8]{inputenc}
\usepackage[T1]{fontenc} 
\usepackage{natbib, hyperref}
\usepackage[varg]{txfonts}
\usepackage{caption}
\usepackage{comment}

\DeclareMathOperator{\csch}{csch}

\title{Cosmography of the Local Universe by Multipole Analysis of the Expansion Rate Fluctuation Field}

\author[]{Basheer Kalbouneh$^{1}$, Christian Marinoni$^{1}$, Roy Maartens$^{2,3,4}$}

\affiliation[1]{Aix Marseille Univ, Universit\'e de Toulon, CNRS, CPT, Marseille, France}
\affiliation[2]{Department of Physics \& Astronomy, University of the Western Cape, Cape Town 7535, South Africa}
\affiliation[3]{Institute of Cosmology \& Gravitation, University of Portsmouth, Portsmouth PO1 3FX, United Kingdom}
\affiliation[4]{National Institute for Theoretical \& Computational Sciences, Cape Town 7535, South Africa}

\abstract{
We establish a relationship between the multipoles of the expansion rate fluctuation field $\eta,$ which capture in an accurate way deviations from isotropy in the redshift-distance relation, and the multipoles of the {\it covariant cosmographic parameters}—Hubble $\mathbb{H}_o$, deceleration $\mathbb{Q}_o$, jerk $\mathbb{J}_o$, and curvature $\mathbb{R}_o$. These parameters, derived from the third-order expansion of the luminosity distance with respect to redshift in a generic spacetime, provide model-independent insight into the geometry and symmetries of the cosmic line element without requiring the Cosmological Principle. Moreover, we demonstrate that although this approach is fully nonperturbative and does not rely on concepts such as peculiar velocities, it has the potential to constrain the motion of the matter frame at the observer's position relative to the cosmic microwave background.

We use two analytical axisymmetric models, motivated by observational evidence, to test the formalism and the effectiveness of the third-order expansion of the luminosity distance. These models also help to predict the precision with which future surveys,  like the Zwicky Transient Facility,  will constrain the covariant cosmographic parameters in the local ($z<0.1$) Universe.
}

\begin{document}
\maketitle
\flushbottom

\section{Introduction}

Recent observational programs have become increasingly precise, yet they still struggle to converge on the fundamental parameters of the Standard Model of cosmology. Notably, tensions exist in the estimates of the Hubble parameter ($H_0$), the root mean square fluctuations of the matter density field ($\sigma_8$), and the spatial curvature of the Universe ($\Omega_k$) \cite{DiValentino:2021izs, Abdalla:2022yfr, Perivolaropoulos:2021jda, Schoneberg:2021qvd}.

These tensions suggest a need to reassess the basic assumptions of cosmology, particularly those concerning the structure of cosmic spacetime. One example is the discrepancy between local determinations of $H_0$ and estimates from the CMB, which may indicate a breakdown of the Cosmological Principle (CP) i.e. a non-trivial less symmetric nature of the cosmic line element  \cite{Schwarz:2007wf, Kashlinsky:2008ut, Antoniou:2010gw, Cai:2011xs, Kalus:2012zu, Wang:2014vqa, Yoon:2014daa, Tiwari:2015tba, Javanmardi2015, Bengaly:2015nwa, Colin:2017juj, Rameez:2017euv, Migkas:2020fza, Migkas:2021zdo, Secrest:2020has, Siewert:2020krp, Luongo:2021nqh, Krishnan:2021jmh, Sorrenti:2022zat, Aluri:2022hzs, Cowell_Dhawan_Macpherson_2023, Hu:2023eyf, Dainotti:2021pqg}.

By definition the CP does not  apply on small scales ($r \lesssim 150$ $h^{-1}$Mpc), which affects determinations of the Hubble parameter \cite{Marinoni:2012ba, Hoffman:2017ako}. Traditionally these  deviations from spatial uniformity  are addressed using perturbative techniques \cite{Strauss:1995fz}.   However, this approach has drawbacks: it is fundamentally model-dependent, since it demands the a priori assumption of the FLRW metric, i.e   the existence of well-defined smooth background variables that depend only on time. Furthermore, it assumes matter behaves as a fluid, density fluctuations are small, and velocity fluctuations are irrotational.

We explore whether it is possible to use measurements to represent the cosmic spacetime nonperturbatively, without relying  on peculiar velocities, and in a model-independent way, without any prior reference to principles or symmetry assumptions. Our ultimate goal is  to  characterize the expansion rate on local cosmic scales ($z \lesssim 0.1$) more meaningfully and comprehensively than using the $H_0$ parameter of the Standard Model alone. This program involves two critical steps. First, we need to identify and classify deviations from isotropy in the observed redshift-distance relation in an unbiased and model-independent way.  Then, we need to relate these angular distortions to geometric quantities, defined covariantly in arbitrary spacetimes, to gain insight into the effective structure of the cosmic line element in the observer's surroundings.

The first challenge was tackled by introducing the {\it expansion rate fluctuation field} $\eta$ \cite{kalbouneh_marinoni_bel_2023}.  As a scalar, Gaussian indicator, $\eta$ presents several statistical advantages and it lends itself quite readily to a decomposition on a spherical harmonic basis, facilitating straightforward signal interpretation.  Indeed, analysis of observational data within the $z<0.05$ range unveiled unexpected symmetries. Notably, we discovered an alignment of the maxima of the lowest order multipoles   (dipole, quadrupole, and octupole), along with an axisymmetric configuration of their 3D structure, shedding new light on the anisotropic nature of the local cosmos.

The second challenge is to generalize the cosmic expansion rate around an arbitrary event in a generic spacetime. This idea dates back to the seminal work by Kristian and Sachs, `Observations in Cosmology' \cite{kristian_sachs_1966}, where they represented deviations from the Hubble-Lemaître law due to anisotropic expansion through the tensorial expansion of the luminosity distance with respect to redshift. This approach yields covariant cosmographic parameters, a set of line-of-sight dependent functions that characterize the metric structure in the neighbourhood of  the observer. 
Their idea was refined and generalized by Ellis and collaborators \cite{MacCallum_Ellis_1970,ellis_2009,ellis_1983,ellis85}.
Peebles \cite{Peebles1971} highlighted the significance of this method for constraining perturbations and shearing effects in the Hubble-Lemaître law. Although there have been notable contributions building on this paradigm (e.g.  \cite{Hasse:1999,Clarkson_theses_2000,clarkson_maartens_2010,Umeh:2013,heinesen_2021,Maartens:2023tib}), progress in analyzing the covariant structure of the expansion field has been slow. This is due to technical challenges and the lack of extensive datasets necessary for rigorous application. However, with current (CF4, Pantheon+) and upcoming (ZTF, LSST, DESI) catalogs of redshift-independent distances, we are on the verge of a pivotal advancement.  

The covariant parameters, although initially unknown, possess a constrained number of degrees of freedom determined by a finite number of spherical harmonic components \cite{Hasse:1999,Clarkson_theses_2000,clarkson_maartens_2010,Umeh:2013,heinesen_2021,Maartens:2023tib}. These covariant cosmographic parameters  can thus be fully  reconstructed by fitting a limited subset of their multipoles to the data. In \cite{Maartens:2023tib}, which is the precursor to this paper, it was shown in detail for the first time how these multipoles, while being independent of coordinate choice, assume different values for observers at the same spatial position but with distinct relative motions. This  explains why several authors have found discrepancies in the amplitude of the multipoles of the covariant cosmographic parameters when estimated in different reference systems (heliocentric or CMB frames) \cite{Colin:2019ulu, Rubin:2019ywt, Dhawan_Borderies_Macpherson_Heinesen_2022, Cowell_Dhawan_Macpherson_2023}. These differences are not due to systematic errors in the measurements or biases in the statistical inferences but have a physical basis, as the measurements contain information about both the metric and the observer's kinematics.

In this paper, we build upon these previous analyses to demonstrate how the observed multipoles of expansion rate fluctuations, estimated by a generic cosmological observer in a  generic state  of motion, can be related to the multipoles of cosmographic parameters determined by a matter co-moving observer, which is more physically relevant. The rationale for adopting this approach lies in recognizing that, despite the orthogonality of the multipoles of the covariant cosmographic parameters—which results in minimal correlations among them—directly fitting them to distance modulus data up to a specified order can lead to substantial biases. This risk arises particularly when the spatial distribution of cosmic sources in the sky is non-uniform, and higher-order multipoles, which might have been disregarded in the analysis, retain a contribution to the observed signal \cite{Macpherson:2024zwu}. Furthermore, employing this formalism allows us to effectively disentangle anisotropy effects due to observer motion, from those effects due to geometric fluctuations. As a result, we present a novel method to estimate the relative velocity between the observer and the matter rest frame at the observer's position. Importantly, this approach is model-independent, meaning that it does not rely on the Cosmological Principle or the Einstein field equations.

We show how well our analytical approach works using simple yet realistic analytical models of the local Universe. These models are constructed to closely match the observed multipoles of the expansion rate fluctuation field, as found in \cite{kalbouneh_marinoni_bel_2023}, while keeping things neat and intuitive. We further use these models to predict the precision with which future data will constrain the multipoles of the covariant cosmographic parameters at our position.

The paper is organized as follows. 
We review the salient properties of the expansion rate fluctuation field $\eta$ in section \ref{sec_def_eta}, while in section\ref{sec_pred_eta} its evolution with redshift, up to $\mathcal{O}(z^3)$ in an arbitrary spacetime, is calculated. We show that the amplitude of $\eta$ depends critically on covariant cosmographic parameters. In section \ref{mupoex} we express the multipoles of the $\eta$ field as a function of the multipoles of covariant cosmographic parameters and show how the choice of the observer reference frame critically affects the mapping. 
In section \ref{sec_model}, we develop a simple yet realistic analytical  model which simulates a local perturbation in the Universe (on scales $R \lesssim 500$ Mpc). 
We estimate in a fully model-independent way the resulting expansion rate fluctuation field $\eta$ in section \ref{sec_mulmeas} and, by comparing it with the input simulated value, we demonstrate the effectiveness of the cosmographic expansion even in the presence of large mass perturbations in the nearby Universe. In section \ref{sec_forecas}, we forecast the precision with which future data sets, such as the ZTF (Zwicky Transient Facility) survey \cite{amenouche:tel-04165406}, will constrain multipoles of the covariant cosmographic parameters. 
Warnings and precautions concerning the use of cosmographic expansion in the local Universe are discussed in section \ref{sec_limits}. Finally, in section \ref{sec_multiv}, we show how, in the context of the standard FLRW metric, the multipoles of $\eta$ provide insights into the amplitude of the multipoles of the 
radial peculiar velocity field of galaxies and we comment on its relation to the bulk flow. Conclusions and outlook are presented in section \ref{sec_conclusion}.

Hereafter we adopt the Einstein summation convention for repeated indices, in Greek letters (from 0 to 3) and Latin letters (from 1 to 3). We use natural units ($c=1$) and the metric signature is $(-+++)$.

\section{The expansion rate fluctuation field}\label{sec_def_eta}

The expansion rate fluctuation field \cite{kalbouneh_marinoni_bel_2023} is constructed as the monopole-free decimal log of the ratio between the redshift $z$ and the luminosity distance $d_L$, 
\begin{eqnarray}
  \eta(z,\boldsymbol{n}) &\equiv & \log\left(\frac{{z}}{{d}_{L}(z,\boldsymbol{n})}\right)-\mathcal{M}(z)\,, \nonumber \\
   \mathcal{M}(z) & \equiv&\int_S  \frac{\mathrm{d}\Omega}{4\pi} \, \log\left(\frac{{z}}{{d}_{L}(z,\boldsymbol{n})}\right)\,,
\label{defeta1}
\end{eqnarray}
where $S$ is a 2D spherical surface of radius $z$ centered on the observer and $\boldsymbol{n}$ is a unit vector specifying the observer's line of sight. Redshifts and distances (measured without redshift information) of objects on $S$ are the only data needed to construct the $\eta$ indicator. 

In practice, however, due to the discrete and sparse nature of the data, $\eta$ at redshift $z$ is more conveniently estimated using a spherical shell of relatively small width $\delta z$. Note that we fix the radius of $S$ using redshift and not distances. There are several reasons for this choice. Firstly, the distance depends on a dimensional normalization that varies from sample to sample and hampers direct comparisons between different data sets. Secondly, from an observational point of view, distances are affected by greater uncertainties (statistical and systematic) than redshifts (which, for the purposes of our analysis, are effectively treated as error-free). Finally, from a theoretical point of view, it is much easier to express physical quantities (and expand them in Taylor series) as a function of redshift than of distances, which facilitates comparison of the signal with theoretical predictions. 

 By construction, the fluctuating field $\eta(z,\boldsymbol{n})$ averages to zero on $S$. Consequently, it is only sensitive to the angular structure of the fluctuating component and its possible evolution with redshift. It is only in the presence of angular anisotropies of the luminosity distance, at a given redshift, that $\eta$ deviates from zero. For example, in an idealised FLRW cosmological model, $\eta=0$ at every epoch.

An observable without a monopole ($\mathcal{M}$) has several advantages. 
First, $\eta$ is completely independent of the normalization of the redshift-distance relationship, a fact that allows comparison of the signal extracted from different distance datasets, regardless of their zero-point calibration. 
This does not mean that the $\eta$ observable is blind to the Hubble constant. Information about $H_0$ is locked in the normalization term $\mathcal{M}$ appearing in $\ref{defeta1}$.
The additional advantage is that any potential systematic errors in measurements of the luminosity distances as a function of redshift do not affect the signal of $\eta$. 
The expansion rate fluctuation field is also independent of the chosen measuring units. Their only effect is to rescale the value of the monopole term $\mathcal{M}$.
We fix this freedom by choosing to
express $d_L$ in Megaparsec units and the redshift in velocity units (km/s).

We express the observable as a ratio between geometrical quantities, redshift and luminosity distance, in order to ensure that the function is well-behaved in the surroundings of the observer.
If the standard model is assumed, in the limit  $z\rightarrow 0$ the ratio converges to the Hubble constant. Moreover, when exploring beyond the standard metric, characteristic quantities, such as for example the covariant Hubble function, which generalizes the Hubble constant, are always locally defined at the observer's position \cite{Maartens:2023tib,kristian_sachs_1966,ellis_2009,MacCallum_Ellis_1970,Clarkson_theses_2000,clarkson_maartens_2010,heinesen_2021}. 
The logarithmic mapping was chosen for statistical reasons. The errors are Gaussian only in the distance modulus and not in the luminosity distance $d_L$. In this way, the estimator of $\eta$, 
\begin{equation}
  \hat\eta=\log(z)+5-\frac{\mu}{5}-\mathcal{M}\,,
\end{equation}
is unbiased. 

Redshifts and distances (measured without using redshift information) are the basic observational data needed for 
estimating $\eta$. However, in order to interpret the $\eta$ signal, we need to specify in which reference frame they are measured.
At a given arbitrary point in spacetime, there is, among others, a characteristic observer of the large-scale structure of the Universe: the {\it matter-comoving} observer at the event of observation $o$, who shares the motion of the surrounding dust flow. 
We emphasize that neither the heliocentric frame nor the  frame at rest relative to the Local Group of galaxies represents the rest frame of the matter observer, although the latter provides a  closer, albeit misleading, approximation. A full characterization, requiring the establishment of the covariant cosmography framework, will be addressed in  section \ref{sec_pred_eta}.

In the following, adopting the notation of \cite{Maartens:2023tib} we use a tilde (i.e. ${\tilde d}({\tilde z}), \tilde{z}$, \dots) to explicitly emphasize 
when relevant quantities are quoted as if they were measured in an arbitrary frame boosted with respect to the `natural' frame that is defined by matter. 
Without loss of generality for our conclusions, we assume that the boosted frame represents an observer at rest in a reference frame in which the CMB dipole -- whatever its nature (local or cosmological) -- disappears. We will refer to this boosted frame as the CMB frame. 
 We emphasize that the Doppler boosting of directly measured data (e.g., in the heliocentric frame) into the CMB frame does not require knowledge of the cosmological model, the spacetime metric, or the rest frame of the cosmic fluid. It only necessitates knowledge of the Sun's velocity relative to the CMB, which is well-constrained by CMB experiments \cite{Planck:2018nkj}.

Note that the CMB and the matter frames
 are indistinguishable in a uniform Universe described by the standard cosmological metric. On the other hand, these observers provide different estimates of the redshift and distance of the same source in a spacetime with arbitrary geometry.
Therefore, although in the following we will show that $\eta$ provides insight into the fluctuations of the space metric in a completely nonperturbative way, thus allowing investigation beyond the standard cosmological scenario in a model-independent way, the amplitude of these fluctuations is, fundamentally, observer-dependent.

\section{Expansion rate fluctuation field in an arbitrary spacetime} \label{sec_pred_eta}

Combining purely geometrical observables, essentially distances and time, 
the expansion rate fluctuation $\eta$  contains information about the structure of  the spacetime surrounding the observer. 

We begin by characterizing spacetime based on its matter content, assuming that this consists of a pressure-free perfect fluid, commonly referred to as 'dust',  which moves along geodesic paths. This assumption ensures that the worldlines traced by the fluid elements are smooth and allow for the definition of the matter 4-velocity. One can then demonstrate that the redshift $z$ and the angular diameter distance (and therefore the luminosity distance)  are well defined functions along each lightlike geodesic \cite{Maartens:2023tib}. Interestingly, the nature of the local metric can be investigated in a fully model-independent way by Taylor expanding the luminosity distance of a light source at a redshift $z$ along the line of sight direction specified by the unit vector  
$\boldsymbol{n}$ 
\begin{equation}
d_L(z,\boldsymbol{n})=d_{L}^{(1)}(\boldsymbol{n}) z + d_{L}^{(2)}(\boldsymbol{n}) z^2 + d_{L}^{(3)}(\boldsymbol{n}) z^3 + \mathcal{O}(z^4)\,.
\label{defdL0}
\end{equation}
This expansion applies under the assumption that the distance function at the observer's position is well-defined. It also assumes that the observer measuring the redshift $z$  and luminosity distance $d_L$ is stationary relative to the surrounding matter fluid (we refer to this observer as the matter observer).
This last choice incorporates the requirement that $z$ be zero when the distance vanishes, {\it i.e.} there is no zeroth-order term in the Taylor expansion: the ratio $z/d_L$ converges to $1/d_{L}^{(1)}$ in this limit.

The expansion coefficients $d_L^{(i)}$ in general depend on the position of the observer, the time of observation, and also the line of sight $\boldsymbol{n}$. More importantly, 
they provide glimpses into the underlying metric. In fact, as shown by  \cite{kristian_sachs_1966,ellis_2009, MacCallum_Ellis_1970,Clarkson_theses_2000,clarkson_maartens_2010,heinesen_2021, Maartens:2023tib}, they can be related to the matter frame {\it covariant cosmographic parameters,} $\mathbb{H}_o(\boldsymbol{n})$ (Hubble), $\mathbb{Q}_o(\boldsymbol{n})$ (deceleration), $\mathbb{J}_o(\boldsymbol{n})$ (jerk) and $\mathbb{R}_o(\boldsymbol{n})$ (curvature):  
\begin{equation}
  d_{L}^{(1)}(\boldsymbol{n})=\frac{1}{\mathbb{H}_o(\boldsymbol{n})},
  \qquad
  d_{L}^{(2)}(\boldsymbol{n})=\frac{1-\mathbb{Q}_o(\boldsymbol{n})}{2\mathbb{H}_o(\boldsymbol{n})},
 \qquad
  d_{L}^{(3)}(\boldsymbol{n})=\frac{\mathbb{Q}_o(\boldsymbol{n})-\mathbb{J}_o(\boldsymbol{n})+\mathbb{R}_o(\boldsymbol{n})+3\mathbb{Q}_o^2(\boldsymbol{n})-1}{6\mathbb{H}_o(\boldsymbol{n})}\,,
  \label{defdL123_1}
\end{equation}
where
\begin{equation}
  \mathbb{H}\circeq K^\mu K^\nu \Theta_{\mu\nu}\,,
\label{Heff10}
\end{equation}
\begin{equation}
  \mathbb{Q}\circeq-3+\frac{K^\mu K^\nu K^\alpha \nabla_\alpha\Theta_{\mu\nu}}{\mathbb{H}^2}\,,
\label{Qeff10}
\end{equation}
\begin{equation}
  \mathbb{R} \circeq 1+\mathbb{Q}-\frac{K^\mu K^\nu R_{\mu\nu}}{2\mathbb{H}^2}\,,
\label{Reff10}
\end{equation}
\begin{equation}
\mathbb{J} \circeq -10\mathbb{Q}-15+\frac{K^\mu K^\nu K^\alpha K^\beta \nabla_\alpha\nabla_\beta\Theta_{\mu\nu}}{\mathbb{H}^3}\,.
\label{Jeff10}
\end{equation}
Here $ \circeq $ indicates that all the quantities are evaluated at the event of observation $o$ (in the following, to simplify notation, we will omit the subscript $o$ from the cosmographic parameters), $\Theta_{\mu\nu}\equiv\nabla_\mu u_\nu$
(where $u^\nu$ is the 4-velocity vector field of the matter fluid)
and
\begin{equation}
  K^\mu\equiv\frac{k^\mu}{k^\nu u_\nu} \,,
\label{Kmu_0}
\end{equation}
with $k^\mu$ the past-pointing photon 4-wavevector ($K^{\mu}$ 
is normalized so that $K^\mu u_\mu=1$).

The covariant cosmographic parameters are defined so that they converge to the expansion coefficients for the luminosity distance in FLRW spacetime \cite{Visser_2004}:
\begin{subequations}
\begin{align}
 \mathbb{H}& \xrightarrow{FLRW} H_0\equiv\frac{\dot{a}_0}{a_0} & & \text{Hubble, } \\
 \mathbb{Q}& \xrightarrow{FLRW} q_0\equiv-\frac{\ddot{a}_0}{a_0 H_{0}^2}=\frac{1}{2}\Omega_{m0}-\Omega_{\Lambda 0} & &\text{Deceleration,} \\ 
 \mathbb{J} &\xrightarrow{FLRW}j_0\equiv\frac{\dddot{a}_{\!\!0}}{a_0 H_{0}^3}=\Omega_{m0}+\Omega_{\Lambda 0} & &\text{Jerk,} \\ 
 \mathbb{R}& \xrightarrow{FLRW} \Omega_{k0}\equiv 1-\Omega_{m0}-\Omega_{\Lambda 0} & &\text{Curvature,}
\end{align}
\end{subequations}
where the overdot denotes the derivative with respect to cosmic time and $\Omega_{i0}={8\pi G \rho_{i0}}/({3H_{0}^2})$ is the density parameter (we ignore radiation).

It is now straightforward to expand $\eta$ in power series of the redshift and predict its dependence on the covariant cosmographic parameters:
\begin{equation}
  \eta (z,\boldsymbol{n}) = \log \mathbb{H(\boldsymbol{n})} -\frac{1-\mathbb{Q(\boldsymbol{n})}}{2 \ln 10}\, z+\frac{7-\mathbb{Q(\boldsymbol{n}}) \big[10+9\mathbb{Q(\boldsymbol{n})}\big] +4 \big[\mathbb{J(\boldsymbol{n})}-\mathbb{R(\boldsymbol{n})}\big]}{24\ln 10}\, z^2
  -\mathcal{M} + \mathcal{O}\big(z^3\big)\,. 
\label{eta_exp_1}
\end{equation}
The amplitude of $\eta$ depends on the frame in which data are measured and will differ from what is predicted by eq. (\ref{eta_exp_1}) if the observer moves with respect to the matter frame \cite{Maartens:2023tib}. Note that at the event of observation, and in the limit of small changes $\delta \mathbb{H}\equiv \mathbb{H}-\langle \mathbb{H} \rangle_S$ in the covariant Hubble function, 
\begin{align}
\eta(\boldsymbol{n}) \circeq  \log \frac{\mathbb{H}(\boldsymbol{n})} { \;\;\langle \mathbb{H}\rangle_S} \propto \frac{\delta \mathbb{H}(\boldsymbol{n})}{\;\; \langle \mathbb{H} \rangle_S}  \,,  
\end{align} 
where $\langle \mathbb{H} \rangle_S$ is the monopole, as in eq. (\ref{defeta1}).
From this it can be seen that the observable allows us to directly quantify fluctuations in the expansion rate at the observer's position in the same way as $\delta \rho/\rho$ and $\delta T/T$ do for the density of matter and the radiation temperature respectively.  A nonzero value of $\eta_o$  indicates that the expansion rate has a more complex structure than that predicted by the standard model (a constant value).
A second thing worth noting is that truncating the expansion to $\mathcal{O}(z^3)$ generates a degeneracy between the jerk and the curvature. In principle, this degeneracy can be solved by increasing the accuracy of the approximation, but we will see in section section \ref{sec_forecas} how to tackle this problem with additional arguments based on the amplitude of the two functions.
 
Although theoretically clear and computationally advantageous, the matter frame is operationally 
difficult to define, in the context of the  fluid approximation hypothesis.
In other words, it is not easy to prescribe an observational procedure for identifying the observer that comoves with the surrounding matter fluid. 
In the fluid approximation, the matter observer’s worldline is determined by averaging over many nearby galaxies so that an intrinsically discrete point-like distribution, such as that of galaxies, can be considered with good approximation as a continuous fluid with sufficiently regular behavior. The
number of galaxies, or the scale of this averaging, can only be determined a posteriori by analyzing the data.

By contrast, the frame with respect to which the CMB dipole vanishes (the CMB frame) is operationally well-defined. It is thus much more practical to estimate the expansion rate fluctuations in the CMB frame, and then exploit theory to deduce the amplitude of the cosmographic parameters that would be inferred by a matter-comoving observer. 

In order to predict the amplitude of the $\eta$ signal in the CMB frame, we need to know how the relevant observables transform. This is derived in detail in \cite{Maartens:2023tib}. The redshift of an object in the CMB frame is related to that measured in the matter frame as
\begin{equation}
  \tilde{z}=z-\tilde{\boldsymbol{v}}_o\cdot\boldsymbol{n}\, (1+z) \,,
\label{ztrasf}  
\end{equation}
where we assume that the CMB frame has a physical velocity $\tilde v_o \equiv |\tilde{\boldsymbol{v}}|_o \ll 1$ with respect to the matter observer at the event of observation $o$.
Note that at linear order in velocity, $\tilde{\boldsymbol{v}}_o\cdot\boldsymbol{n} \approx \tilde{\boldsymbol{v}}_o\cdot \tilde{\boldsymbol{n}} $ 
where $ \tilde{\boldsymbol{n}} $ defines the line of sight of the boosted observer. The luminosity distance, instead, transforms between different frames as \cite{Maartens:2023tib,hui_greene_2006,Bonvin_Durrer_Gasparini_2006} 
\begin{equation}
  \tilde{d}_L(\tilde{z},\boldsymbol{n})=(1-\tilde{\boldsymbol{v}}_o\cdot\boldsymbol{n})\, d_{L}(z(\tilde{z}),\boldsymbol{n}) \,.
  \label{dLtrans1}
\end{equation}
We expand this expression up to third order in redshift, 
\begin{equation}
\tilde{d}_L=
\tilde d_L^{(0)}
+
\tilde d_L^{(1)}\tilde z 
+
\tilde d_L^{(2)}\tilde{z}^2
+
\tilde d_L^{(3)}\tilde{z}^3
+ \mathcal{O}(\tilde{z}^4),
\label{dLbar_exp1}
\end{equation}
and relate the expansion coefficients to those calculated in the matter frame: 
\begin{align}
&\tilde d_L^{(0)}= d_L^{(1)}\tilde{\boldsymbol{v}}_o\cdot\boldsymbol{n}
\;,\hspace{2.9cm}
\tilde d_L^{(1)}= d_L^{(1)}+2d_L^{(2)}\tilde{\boldsymbol{v}}_o\cdot\boldsymbol{n}
\;,
\notag \\
&\tilde d_L^{(2)}= d_L^{(2)}+\left(d_L^{(2)}+3d_L^{(3)}\right)\tilde{\boldsymbol{v}}_o\cdot\boldsymbol{n}\;,\quad
\tilde d_L^{(3)}= d_L^{(3)}+2(d_L^{(3)}+2d_L^{(4)})\tilde{\boldsymbol{v}}_o\cdot\boldsymbol{n} \,.
\end{align}
The expansion rate fluctuations measured in the boosted frame, in our case the CMB frame, is 
\begin{equation}
  \tilde \eta \approx -\log\left(\frac{\tilde d_L^{(0)}}{\tilde{z}}+\tilde d_L^{(1)}
+
\tilde d_L^{(2)}\tilde{z}
+
\tilde d_L^{(3)}\tilde{z}^2\right)-\tilde{\mathcal{M}}
\label{eta_bar_exp_1} \,,
\end{equation}
or equivalently, 
\begin{eqnarray}
  \tilde \eta (\tilde z,\boldsymbol{n}) & = &\eta (\tilde z,\boldsymbol{n}) -\log\left[1+\frac{\tilde{\boldsymbol{v}}_o\cdot\boldsymbol{n}}{\tilde z}(1+\tilde z)\right]+\mathcal{M}(\tilde z)-\tilde{\mathcal{M}}(\tilde z)  \nonumber
  \\
  & & {}+\tilde{\boldsymbol{v}}_o\cdot\boldsymbol{n}\left\{
  \frac{1}{\ln10}-\frac{1-\mathbb{Q}(\boldsymbol{n})}{2\ln10}+\frac{1-\mathbb{Q}(\boldsymbol{n})\big[4+9\mathbb{Q}(\boldsymbol{n})\big]+4\big[\mathbb{J}(\boldsymbol{n})-\mathbb{R}(\boldsymbol{n})\big]}{12\ln10}\,\tilde z \right\} + \mathcal{O}(\tilde z^3)\,. ~~~
\label{eta_bar_exp_3}
\end{eqnarray}

In order to obtain a real output, the logarithm must be applied only to positive numbers. This gives the condition $1+\tilde{\boldsymbol{v}}_o\cdot \boldsymbol{n}(1+\tilde z)/\tilde z>0$, which implies $\tilde z>\tilde{\boldsymbol{v}}_o\cdot \boldsymbol{n}$. It also implies that terms of order $\tilde z^2\times\tilde v_o$ can be safely neglected.
The above relationship gives the expansion rate fluctuations measured in the CMB frame as a function of the covariant cosmographic parameters. Note, however, that the values of the latter are expressed in the matter frame, which is the most natural choice for cosmological interpretations if no specific symmetry is attributed a priori to the cosmic line element. 
By setting $\tilde{\boldsymbol{v}}_o=0$ in this expression we re-obtain eq. (\ref{eta_exp_1}), the expansion rate fluctuations measured in the matter frame.

Observing the large-scale structure in different frames results in a different pattern of expansion rate fluctuations. At small redshifts, the difference is mainly due to the second term on the first line of eq. (\ref{eta_bar_exp_3}) and depends on the direction of the line of sight. Another feature of the expansion rate fluctuation field is worth noticing. In order to achieve second-order accuracy in powers of $z$, an expansion to the next higher order in the distance-redshift relationship is required. This means that, already at second-order in redshift, $\eta$ becomes sensitive to the jerk $\mathbb{J}$ and curvature $\mathbb{R}$ parameters (cf. eq. (\ref{eta_bar_exp_1})).

\section{\boldmath Multipolar expansion of $\eta$ }\label{mupoex}

The finer structure of the expansion rate fluctuation field is best appreciated by expressing  it by means of its spherical harmonic components
\begin{equation}
\eta(z,\theta,\phi)=\sum_{\ell=0}^{\infty}\sum_{m=-\ell}^{\ell} \eta_{\ell m}(z) Y_{\ell m} (\theta,\phi).
\end{equation}	
By decomposing the signal into an orthonormal basis, we ensure that the contributions from different angular scales are statistically independent and that statistical likelihood analyses are minimally affected by correlations between 
different modes.

In the following, we deal only with the case in which $\eta$ shows an axial symmetry around a given privileged direction in the local Universe. 
This choice is motivated by preliminary evidence found in \cite{kalbouneh_marinoni_bel_2023}, by reconstructing the expansion rate fluctuations 
up to $z<0.05$ using Cosmicflows-3 galaxy data \cite{Tully_2016} or the Pantheon sample of supernovae \cite{Scolnic2018}. As a bonus, this axisymmetric assumption allows us to simplify the formalism, by reducing the degrees of freedom of the fluctuation model, and to highlight its physical content, without sacrificing the generality of the conclusions. 

Any fluctuating quantity $f(z,\boldsymbol{n})$, including $\eta$, will depend only on the angle $\theta$ between the line of sight and the axis of symmetry ($Z$-axis). Then $f$ can be expanded in Legendre polynomials $P_\ell$. In this case the expansion coefficients $f_{\ell m}$ are zero for $m \neq 0$, and $f_{\ell 0}=[4\pi/(2\ell+1)]^{1/2}f_\ell$, where 	
\begin{equation}
f_\ell(z)=\frac{2\ell+1}{2} \int_{-1}^{1}f(z, \cos\theta) P_\ell(\cos\theta)\;\mathrm{d}(\cos\theta)\,.
\end{equation}
In this convention, 
$f_{2\ell+1}>0$
means that there is a maximum at $\theta=0$ and a minimum at $\theta=\pi$, and the opposite for 
$f_{2\ell+1}<0$.
For $f_{2\ell}>0$, the field has a maximum at $\theta=0$ and at $\theta=\pi$, while $f_{2\ell}<0$ means that there is a minimum in both directions.

 \subsection{\boldmath Relation between  $\eta$ multipoles and multipoles of the covariant cosmographic parameters}

There is also a theoretical motivation for decomposing the expansion rate fluctuation field into spherical harmonics. This approach parallels a similar decomposition of the covariant cosmographic parameters, which is of physical significance. Unlike standard functional parameterizations that involve infinite degrees of freedom, the degrees of freedom for each covariant cosmographic parameter are finite and are represented by their multipoles \cite{Hasse:1999,Clarkson_theses_2000,clarkson_maartens_2010,Umeh:2013,heinesen_2021,Maartens:2023tib}. By fitting a finite set of these multipoles to data, we can fully reconstruct the functional form of the Hubble, Deceleration, Jerk, and Curvature parameters. 

Moreover, the multipoles of the covariant cosmographic parameters offer a systematic method to inspect and classify anisotropies in the distance-redshift relationship, helping to identify potential symmetries. Each multipole carries independent information about the geometry of the geodesic congruence, providing powerful guidelines for determining the form of the line element that describes spacetime in our surroundings.
The task now is to establish the link between the multipoles of the expansion rate fluctuations $\eta$  and the multipoles of the covariant cosmographic parameters in an axisymmetric field configuration.

 By construction, since $\eta$ is a fluctuating variable with zero average inside the redshift shell  $S$ 
in which it is reconstructed, its monopole $\eta_0$ vanishes. Moreover, the covariant Hubble parameter $\mathbb{H}$ at the position of the observer displays only a monopole $\mathbb{H}_0=\langle \mathbb{H} \rangle_S$ and a quadrupolar distortion $\mathbb{H}_2$ (see \cite{Maartens:2023tib})
\begin{align}
 \mathbb{H}(\theta)=\mathbb{H}_0+\mathbb{H}_2 P_2(\cos \theta).  
\end{align}
Note that the subscript $0$ denotes the monopole $\ell=0$, and it should not be confused with $o$ which denotes instead  the event of observation.
From eq. (\ref{eta_exp_1}), in the limit of small redshift $(z\ll 1)$, and neglecting terms of order $(\mathbb{H}_2/\mathbb{H}_0)^2$, 
which contribute only at sub-percent level ($\sim0.01\%$, see Table \ref{tab_al}),  we deduce that 
\begin{align}\label{monom}
\mathcal{M}(z)&\approx\log \mathbb{H}_0-\frac{1-\mathbb{Q}_0}{2\ln10}\, z 
\\ \notag &~~~
+\frac{245
-35\mathbb{Q}_0 \big(10+9\mathbb{Q}_0\big)
-105(\mathbb{Q}_1)^2
-63(\mathbb{Q}_2)^2
-45(\mathbb{Q}_3)^2
+140\big(\mathbb{J}_0-\mathbb{R}_0 \big)
}{840\ln10}\,z^2. 
\end{align}

One would naively expect that only the monopoles of the covariant
cosmographic parameters contribute to the amplitude of the normalization term $\mathcal{M}$. This is not the case because the monopole of $\mathbb{Q}^2$ (see eq. (\ref{eta_exp_1})) is not the square of the monopole. To calculate it, we need to expand the deceleration parameter into
its multipolar components. By counting the degrees of freedom of the deceleration parameter (see eqs. (A22)--(A25) in \cite{Maartens:2023tib}), 
and assuming that $(\mathbb{H}_2/\mathbb{H}_0)^2\ll 1$, we deduce that only multipoles $\mathbb{Q}_\ell$ up to $\ell=5$ are nonzero, and that only multipoles up to $\ell=3$ are independent (while $\mathbb{Q}_4=-(36/35)\mathbb{Q}_2 \mathbb{H}_2/\mathbb{H}_0$ and $\mathbb{Q}_5=-(20/21)\mathbb{Q}_3 \mathbb{H}_2/\mathbb{H}_0$). 
As a consequence, while at low $z$ the normalization factor $\mathcal{M}$ depends only on the monopoles of $\mathbb{H}$
and $\mathbb{Q}$, at higher redshift it becomes sensitive to higher multipolar contamination, such as from the dipole, quadrupole and octupole of $\mathbb{Q}$. 

The first multipole of $\eta$ that in principle does not vanish, unless it is calculated 
at $z=0$, is the dipole. From eq. (\ref{eta_exp_1}), and under the same approximations as above, we get, up to second order in redshift,
\begin{equation}\label{dipe}
\eta_1(z) \approx \frac{\mathbb{Q}_1}{2 \ln10}\, z -\frac{49 \left(45 \mathbb{Q}_0+18 \mathbb{Q}_2+25\right) \mathbb{Q}_1+3 \mathbb{Q}_2 \mathbb{Q}_3\left( 189-144 \mathbb{H}_2/\mathbb{H}_0\right)-490(\mathbb{J}_1- \mathbb{R}_1)}{2940 \ln 10}\,z^2.
\end{equation}
A multipole of $\eta$ which might be different from zero, even when estimated in the limit $z\rightarrow 0$, is the quadrupole. Its dependence on the multipoles of the cosmographic parameters is
\begin{align}\label{quade}  
 \eta_2(z)&\approx \frac{\mathbb{H}_2}{\mathbb{H}_0\ln10}+\frac{\mathbb{Q}_2}{2\ln10}\, z \\ \notag
 &~~~ +
\frac{1}{84 \ln 10}\bigg\{-7 \left(9 \mathbb{Q}_0+5\right) \mathbb{Q}_2-9 
(\mathbb{Q}_2)^2-27 \mathbb{Q}_1 \mathbb{Q}_3-21(\mathbb{Q}_1)^2-6 (\mathbb{Q}_3)^2
\\ \notag &~~~~~~~~~~~~~~~~~~~~~~~~+\frac{8 }{385}\left[891 (\mathbb{Q}_2)^2+625 (\mathbb{Q}_3)^2\right]\frac{\mathbb{H}_2}{\mathbb{H}_0} +14( \mathbb{J}_2-\mathbb{R}_2)\bigg\}\,z^2 \,.
\end{align}
At the observer's position, the quadrupole of $\eta$ is mainly sourced by the ratio of the first two nonzero multipoles of the covariant
Hubble parameter. As the redshift increases, the quadrupole of higher-order cosmographic parameters also contributes to the signal. 
The relation between the octupole of the expansion rate fluctuation field measured in the matter rest frame and the covariant
cosmographic parameters is similar to that displayed by the dipole,
\begin{align}\label{octe}
\eta_3(z) &\approx \frac{\mathbb{Q}_3}{2 \ln10}\, z 
+\frac{1}{60 \ln 10} \bigg[-
\left(45 \mathbb{Q}_0+12 \mathbb{Q}_2+25\right) \mathbb{Q}_3-27 \mathbb{Q}_1 \mathbb{Q}_2
\\ \notag &\qquad\qquad\qquad\qquad\qquad
+\frac{16}{77}\,\mathbb{Q}_2\left(99\mathbb{Q}_1+103 \mathbb{Q}_3\right)\frac{\mathbb{H}_2}{\mathbb{H}_0}
+10(\mathbb{J}_3- \mathbb{R}_3)
\bigg]\, z^2.
\end{align}
Finally,  the hexadecapole,
\begin{align}\label{hexe}  
& \eta_4(z)\approx \frac{\mathbb{Q}_4}{2 \ln10}\, z +
\frac{1}{60060 \ln 10}\bigg\{-117\Big[
99\mathbb{Q}_2^2+5\mathbb{Q}_3(44\mathbb{Q}_1+9\mathbb{Q}_3)\Big]
+\frac{12}{7}\Big[3003\mathbb{Q}_2(5+9\mathbb{Q}_0)
\\ \nonumber & \qquad\qquad\qquad
\qquad\qquad\qquad\quad
+7020(\mathbb{Q}_2)^2+125\mathbb{Q}_3(91\mathbb{Q}_1+36\mathbb{Q}_3)\Big]
\frac{\mathbb{H}_2}{\mathbb{H}_0}+10010 (\mathbb{J}_4 - \mathbb{R}_4)\bigg\}\,z^2 \,,
\end{align}
is the  highest order moment of $\eta$ considered in the analysis. 
The linear term, driven by the amplitude of  $\mathbb{Q}_4$, makes a negligible contribution 
to the amplitude of $\eta_4$  since the hexadecapole of the deceleration parameter, being proportional to  $\mathbb{Q}_2 \mathbb{H}_2/\mathbb{H}_0$ is second-order small in perturbations. Note that, as will be justified in the section \ref{sec_forecas}, 
the leading contribution to $\eta_4$ comes instead from the single term $\mathbb{J}_4\, z^2$.
The hexadecapole is the highest order moment considered in the analysis 
since, at order $\mathcal O(z^3)$, nor $\mathbb{Q}_5$ (which is 
proportional to $\mathbb{Q}_3 \mathbb{H}_2/\mathbb{H}_0$) nor the term $\mathbb{J}_5 \, z^2 $ (as will be again discussed in section section \ref{sec_forecas}) contribute in a significant way to its amplitude.

In general, the leading contribution to $\mathcal{M}$ and $\eta_2$ comes from the monopole and quadrupole of $\mathbb{H}$ respectively, while the other multipoles of $\eta$ are dominated by the corresponding multipoles of $\mathbb{Q}$. The farther away from the observer the $\eta$ multipoles are estimated, the larger the contribution to the signal of the higher-order terms in the redshift-distance relation. 
In the local Universe, this contribution is all the more remarkable as the nonlinearities present in the distance-redshift relationship are significant. 
Indeed, as will be discussed later, some of the high-order multipoles of the 
covariant 
cosmographic parameters become increasingly important to compensate for the smallness of the progressively increasing powers of the $z$ variable.

\subsection{Motion of the matter frame relative to the CMB observer}\label{v_obs}

The advantage of estimating the expansion rate fluctuations $\eta$ in the CMB frame is that, in addition to cosmographic parameters, we can obtain also critical information about 
the velocity $\boldsymbol{v}_o$ of the matter comoving observer relative to the CMB frame ($\boldsymbol{v}_o= -\tilde{\boldsymbol{v}}_o$). 
This fact is all the more important if one considers that the estimation of the observer's velocity can be obtained in a completely model-independent manner, without making any assumptions about the metric (FLRW), nor requiring information about the peculiar velocity field of galaxies, the background cosmology or the power spectrum of matter perturbations (for model-dependent approaches,  and a discussion of the systematics induced by an improper definition of the background matter rest frame in that type of analysis, see for example  \cite{Horstmann:2021jjg, Sorrenti:2022zat}).

Note  that the velocity $\boldsymbol{v}_o$ of the matter frame relative to the CMB frame should not be confused  with the concept of bulk velocity in standard model studies. This bulk velocity characterizes the average of peculiar velocities over volumes in the FLRW background metric. 
In contrast, the nonperturbative and model-independent approach has no notion of peculiar motions of galaxies \cite{Maartens:2023tib}. Nevertheless, we can  make a comparison with the standard perturbative approach: in section \ref{sec_multiv} we show how  the standard bulk velocity relates to the multipoles of the expansion rate fluctuation field in the CMB frame.

In order to constrain  $\boldsymbol{v}_o$ we proceed as follows. In the limit $v_o\ll z\ll 1$, which well encompasses the range of available data, the relation between the expansion rate fluctuations measured in the matter frame ($\eta$) and in the CMB frame ($\tilde\eta$) is (cf. eq. (\ref{eta_bar_exp_3})) 
\begin{equation}
\tilde{\eta} -\eta \approx 
  -\log\left(1-V_o\cos \theta \right) +\mathcal{M}-\tilde{\mathcal{M}}\quad \mbox{where}\quad V_o= \frac{(1+\tilde{z})}{\tilde{z}}\, v_o \,.
\end{equation}
As a consequence, the mapping between the lowest multipoles ($\ell \leq 4$) measured in both frames is fairly well approximated by
\begin{eqnarray}
\tilde{\mathcal{M}}(\tilde{z}) & \approx & \mathcal{M}(\tilde{z})+\frac{V_o^2}{6\ln10} \,, \nonumber \\
\tilde \eta_\ell(\tilde{z}) & \approx & \eta_\ell(\tilde{z})+\frac{V_o}{\ln10} \left [ \delta_{1\ell}+\frac{V_o}{3} \delta_{2\ell}+\frac{V_o^2}{15}\left(3\delta_{1\ell}+2\delta_{3\ell}\right)\right]\,, \label{alm}
\end{eqnarray}
where $\delta_{\ell \ell'}$ is the Kronecker delta. The difference between the full $\tilde \eta_\ell$ and the approximated one is less than $10\%$ of the typical error of $\tilde \eta_\ell$.
Although each multipole of $\tilde{\eta}$ shows a specific dependence on $v_o$, 
in practice, i.e. in the redshift range where data are available for analysis ($z>0.01$) \cite{kalbouneh_marinoni_bel_2023},
the motion of the observer has a measurable effect only on the dipole $\tilde{\eta}_1$.
As eq. (\ref{alm}) shows, $\mathcal{M}$, $\tilde{\eta}_2$
and $\tilde{\eta}_3$ becomes sensitive to $v_o$ only at very low redshifts ($z<0.01$), in a regime where data have no constraining power. As the equation \ref{dipe} makes clear, the dipole of the deceleration parameter, which is naturally defined in the matter frame, will be systematically different if it is estimated in different reference frames, without explicitly factoring out the observer's motion. The discrepant estimates obtained by various authors \cite{Colin:2019ulu, Rubin:2019ywt,
Dhawan_Borderies_Macpherson_Heinesen_2022, Cowell_Dhawan_Macpherson_2023} are not induced by the different statistical inference schemes implemented in the analyses nor by systematic errors in the different measurements. It is a real, physically expected bias induced by the motion of the observer relative to the matter frame, but that can simply be eliminated by proper signal analysis. 

Our method indeed offers the first model-independent estimate of the relative velocity between the CMB observer and the matter frame at the observer’s position. It requires no assumptions about the nature of the CMB or the origin of its dipole, nor does it rely on references to the cosmological
principle or Einstein Field Equations. Furthermore, our formalism allows for a clear separation
between anisotropies induced by the kinematics of the observer (with respect to the matter frame)
and pure metric fluctuations, i.e. intrinsic anisotropies of the congruence of (fluid) matter geodesics.

Applying the method to real data holds the promise of potentially solving the dipole tension or at least clarifying the nature of any anisotropic signals measured at high redshift. This will be addressed in a future analysis. In the following, we show that although the observer's motion is degenerate with cosmographic parameters, a combined analysis of the low-order multipoles of  $\eta$   allows us to disentangle metric and kinematic effects and, in particular, to constrain $\boldsymbol{v}_o$.

\section{Analytical model for testing the cosmographic reconstruction scheme} \label{sec_model}

The effectiveness of the cosmographic approach for model-independent reconstruction of the metric structure of the local Universe ($z<0.1$) is examined by simulating linear (scalar) perturbations in an otherwise smooth FLRW background. This latter is chosen to be the Einstein de Sitter (EdS) Universe, a flat matter-dominated model that has simple analytical properties and thus perfectly suits illustrative purposes. The scale factor of the metric evolves as $a(t)=(t/t_0)^{2/3}$ (where $t_0$ is the age of the Universe today) and the background Hubble expansion rate is simply $H(t)=2/(3t)$.

\begin{figure*}
  \centering
  \includegraphics[scale=0.8]{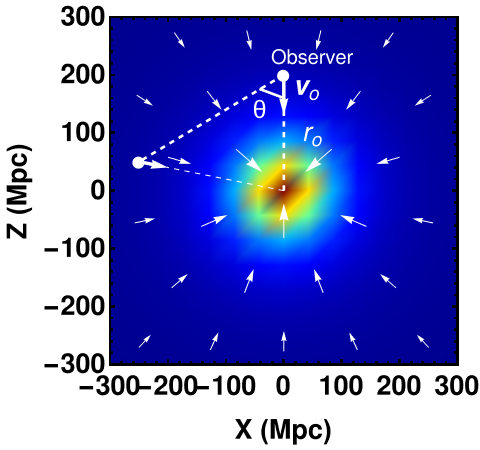}
  \caption{ Analytical model’s spherical density contrast projected onto a plane. We  show the observer's position relative to the attractor and the peculiar velocity perturbation caused by the attractor. Model parameters ($\theta$, $r_o$, $\boldsymbol{v}_o$) are also indicated.}
  \label{plt1}
\end{figure*}

By neglecting anisotropic stresses (we assume that Einstein’s field equations hold) and adopting the Newtonian gauge, the perturbed metric can be written as
\begin{equation}
\mathrm{d}s^2=-\left(1+2\Phi\right)\mathrm{d}t^2+a^2(t)(1-2\Phi)\left[\mathrm{d}r^2+ r^2(\mathrm{d}\psi^2+\sin^2\psi \,\mathrm{d} \phi^2)\right].
\label{metric1}
\end{equation}
We assume that the background/perturbation split is carried out, at linear order, in a smooth coordinate system where the CMB is at rest. This is not in general, and not in our specific perturbation model, the coordinate system in which matter is also at rest.

We then assume that there is only a single massive spherically symmetric structure (see Figure \ref{plt1}) that sources the time-independent gravitational field and model it as 
\begin{equation}
\Phi(r)=\Phi_c \frac{R_s}{r}\sinh^{-1}\left(\frac{r}{R_s}\right)\,,
\label{phi_r}
\end{equation}
where $R_s$ is the typical radial extension of the perturbation. This means that the relevant cosmological quantities measured by an off-center observer display an axially symmetric modulation. 
The potential $\Phi$ exactly solves, in Newtonian approximation and assuming that only the growing mode is of interest, the Poisson equation for an overdensity contrast 
$\delta\equiv{\rho}/{\rho_{\rm b}}-1$ ($\rho_{\rm b}$ is the time-dependent background density of the EdS model) of the form 
\begin{equation}
\delta(r,t_0)=\delta_c\left[1+\left(\frac{r}{R_s}\right)^2\right]^{-3/2}.
\label{delta_r}
\end{equation}
This profile fairly characterizes large matter overdensities such as, for example, the Shapley Supercluster \cite{marinoni_monaco_giuricin_costantini_1998}.
Here $\delta_c$ represents the central density contrast, which is related to the normalization of the potential as $\Phi_c= -{3}H^2_0 R_s^2 \delta_c/2$.

The peculiar velocity generated by the local gravitational field has only a radial component \cite{peebles_1980}:
\begin{equation}
v(r,t)=-\frac{2}{3 a H}\frac{\partial \Phi}{\partial r}=\frac{2 R_s \Phi_c}{ 3 H_0 r^2}\left(\frac{t}{t_0}\right)^{1/3}\left[\sinh^{-1}\left(\frac{r}{R_s}\right)-\frac{r}{\sqrt{R_s^2+r^2}}\right]\,.
\label{pec_vel}
\end{equation}
This physical velocity is measured by observers located at the positions of the 
matter particles and with fixed coordinates in the Newtonian gauge coordinate system, i.e. at rest in the CMB frame (cf. \cite{Maartens:2023tib}). Indeed they do
not see any CMB dipole since fluctuations are induced only by $\Phi$, and are thus negligible, being of second-order amplitude in velocity. These observers have 4-velocity $\tilde u^\mu$ and
the matter particles have 4-velocity $u^\mu$, where
\begin{equation}
\tilde u^\mu=(1-\Phi,\boldsymbol{0}), \quad
u^\mu=\left(1-\Phi, \frac{v}{a},0,0 \right)\,,
\end{equation}
in the given coordinates. 
 
Despite its simplicity, this analytical model is flexible enough to closely match the observed multipoles of the expansion rate fluctuation field, as estimated using real data, as we will show in section  \ref{sec_mulmeas}. Moreover it addresses two critical issues. First, it provides the ideal testbed for two different methods of reconstructing luminosity distances: the cosmographic method and the linear perturbation method. It is therefore instrumental in comparing the effectiveness of the model-independent approach in recovering fluctuations in the local expansion rate. 

By randomly sampling the continuous analytical model, moreover, we can also simulate catalogs of distances measured without using redshift. Analysis of these catalogs allows us to forecast the precision with which future datasets will be able to constrain the multipoles of covariant cosmographic parameters, as well as the velocity of the matter frame at the observer's position.

\subsection{Luminosity distance of analytical model: cosmographic approach}\label{dlexp}

The amplitude of the covariant cosmographic parameters
$\mathbb{H}, \mathbb{Q}, \mathbb{J}$ and $\mathbb{R}$ can be calculated in an arbitrary spacetime, and thus in our specific analytical model, using the eqs. (\ref{Heff10}), (\ref{Qeff10}), (\ref{Reff10}) and (\ref{Jeff10}). The metric and the normalized photon 4-momentum $K^\mu$ at the observation event are all that are needed.

If the observer is not exactly at the center of the symmetry of the model (the overdensity peak), the expansion rate field manifests an axial symmetry around the line connecting the observer with the density peak. Because of this, we can assume, without loss of generality, that $k^3=0$ and choose this axis of symmetry as the $Z$-axis. Since $k^\mu k_\mu=0$, two degrees of freedom are sufficient to fix $k^\mu$. We choose to make the affine parameter coincide with the physical distance measured by the observer $\tilde u^\mu$, so $k^\mu \tilde u_\mu \circeq 1$ \cite{Fleury:2015hgz}, which gives $k^0 \circeq -1+\Phi$ (using $a_0=1$).
We also introduce the angle $\theta$ which measures the separation between the direction of reception of the light ray and the direction of the center of symmetry as measured by the observer $\tilde u^\mu$, which is chosen so that when $\theta=0$ the observer line of sight is toward the center of the mass overdensity (see Figure \ref{plt1}). We thus obtain  
\begin{equation}
k^\mu \circeq \left(-1+\Phi,-(1+\Phi)\cos \theta,\frac{1}{r}(1+\Phi)\sin \theta, 0 \right)\,,
\end{equation}
and by using eq. (\ref{Kmu_0}),
\begin{equation}
K^\mu \circeq \left(-1+\Phi-v_o \cos\theta,-(1+\Phi+v_o \cos\theta)\cos \theta,\frac{1}{r}(1+\Phi+v_o \cos\theta)\sin \theta,0 \right)\,.
\label{Kvec1}
\end{equation}
As a check, note that $K^\mu u_\mu\circeq 1$ to leading order.
By linearizing eq. (\ref{Heff10}) around the observer's position (using
that the matter 4-acceleration is zero), we find 
\begin{equation}\label{loc_hub_par}
\mathbb{H} \circeq H\left(1-\frac{\dot{\Phi}}{H}-\Phi\right) + n^i n^j \partial_i v_j \,,
\end{equation}
where all derivatives are computed at the event of observation. 
An observer comoving with matter measures no dipolar component for the generalized Hubble parameter, only a monopole and a quadrupole \cite{Maartens:2023tib}. The latter, for our specific analytical model, are 
\begin{align}
\mathbb{H}_0 &=
H_0\left[1-\frac{1}{3}\delta_c \: \xi_o\left(\frac{ \xi_o^2}{(1+ \xi_o^2)^{3/2}}-\frac{9}{2} \xi_H^2\csch^{-1}( \xi_o)\right)\right]\,,
\\
\mathbb{H}_2 &=
\frac{2\delta_c H_0 \xi_o^3}{3(1+ \xi_o)^{3/2}}\left[-4-3 \xi_o^2+3(1+ \xi_o^2)^{3/2}\csch^{-1}( \xi_o)\right]\,,
\end{align}
where $ \xi_o\equiv R_s/r_o$, $\, \xi_H\equiv R_s/R_H =R_s H_0$, and $r_o$ is the radial coordinate of the observer.

The analytical (linearized) expressions of the nonzero multipoles of the higher-order cosmographic parameters ($\mathbb{Q}$, $\mathbb{J}$ and $\mathbb{R}$) as a function of the structural parameters of the analytical model are given in Appendix \ref{app_alX_eqns}. We remark that, at linear order in the perturbation parameter $\delta_c$, the multipoles  $\ell\geq 5$ vanish.

\begin{figure*}
  \centering
  \includegraphics[scale=0.6]{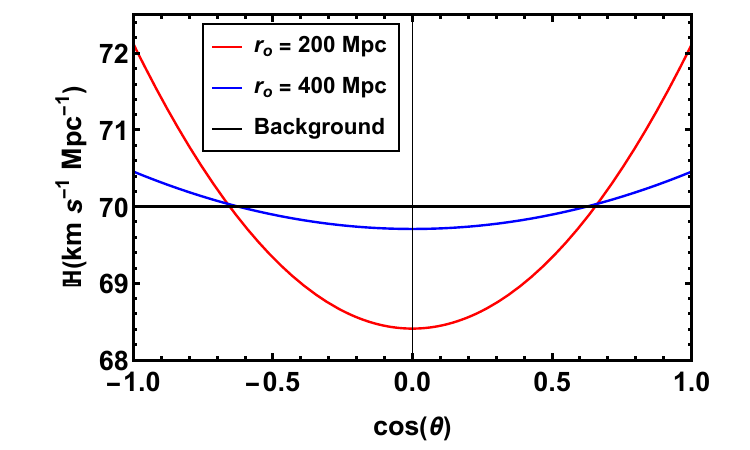}
  \includegraphics[scale=0.6]{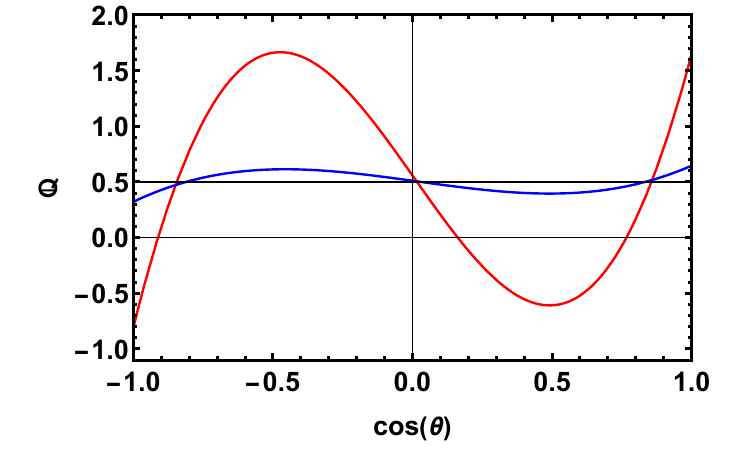}
  \includegraphics[scale=0.6]{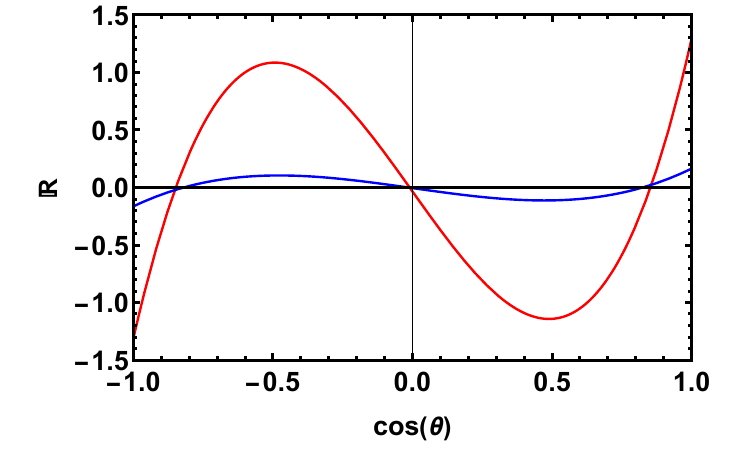}
  \includegraphics[scale=0.6]{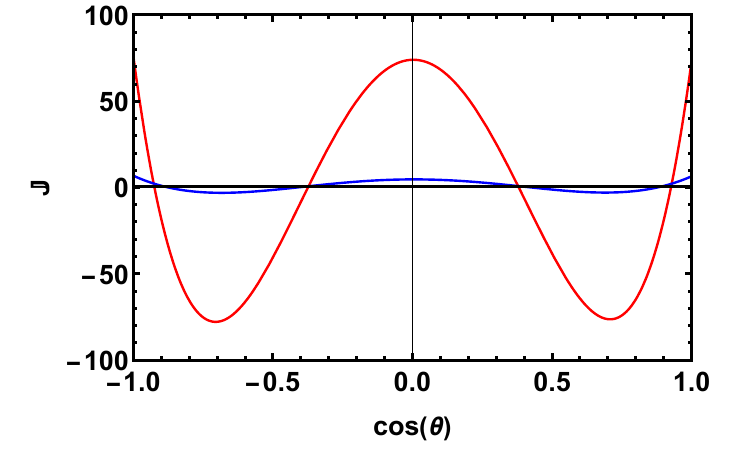}
  \caption{The lowest order covariant cosmographic parameters $\mathbb{H}$, $ \mathbb{Q}$, $\mathbb{R}$ and $\mathbb{J}$ at the event of observation are shown as a function of $\cos\theta$, the angular separation from the symmetry axis. The red curves represent local measurements for an observer $200\,$Mpc away from the overdensity, and the blue for an observer at a distance of $400\,$Mpc. The same model is assumed, with parameters $H_0 = 70\,$km/(s Mpc), $\delta_c = 2.5$ and $R_s=37.4\,$Mpc.}
  \label{HABC_1}
\end{figure*}

In Figure \ref{HABC_1} we show the angular dependence of the covariant cosmographic parameters 
$\mathbb{H}, \mathbb{Q}, \mathbb{J}$ and $\mathbb{R}$, 
as functions of $\cos\theta$ for two observers at different distances, $r_o=200$ and $400$ Mpc, from the high density peak.
We also show, for comparison, the input values of the Hubble, deceleration, jerk and curvature for the smooth background 
of the analytical model (the EdS model), whose value is, by construction, independent of direction. 
The amplitude of the anisotropies in each parameter becomes weaker as the distance of the observer from the
density peak increases. As the observer moves away from the center, the 
covariant cosmographic parameters asymptotically converge to the background values.

\subsection{Luminosity distance of analytical model: linear perturbation theory}\label{sec_dLpert}

We can determine the luminosity distance of objects in the single-attractor analytical model also through the linear cosmological perturbations. 
This independent determination, although based on knowledge of the line element (\ref{metric1}), and thus model-dependent, provides a useful comparative tool for assessing the validity of the third-order cosmographic expansion (cf. eq. (\ref{dLbar_exp1})) in the local perturbed Universe.

In a linearly perturbed EdS background, the redshift measured by a matter-comoving observer is \cite{durrer_2021}
\begin{equation}
1+z=(1+z_c)\big[1+\boldsymbol{n}\cdot(\boldsymbol{v}-\boldsymbol{v}_o)-(\Phi-\Phi_o)\big]\,,
\label{z_corr}
\end{equation}
where $z_c$ is the background cosmological redshift
\begin{equation}
  1+z_c=\frac{a_0}{a}\,.
\end{equation}
For an observer in the matter frame, 
the luminosity distance to a galaxy with peculiar velocity $\boldsymbol{v}$  is
\cite{hui_greene_2006,Bonvin_Durrer_Gasparini_2006} 
\begin{equation}\label{ddl} d_L(z)=d_L^c(z_c(z))\,\big(1+2\boldsymbol{n}\cdot\boldsymbol{v}-\boldsymbol{n}\cdot\boldsymbol{v}_o\big)\,.
\end{equation}
Here $\boldsymbol{v}_o$ is the velocity of the matter-comoving observer
and $d_L^c(z_c)$ is the luminosity distance between the events of emission and absorption, assuming that the Universe is perfectly smooth. In the EdS model, this takes the form 
\begin{equation}
  d_L^c(z_c)=\frac{2}{H_0}\left(1+z_c-\sqrt{1+z_c}\right)\,.
\end{equation}
Note that the perturbation of the luminosity distance which is induced by the gravitational potential $\Phi$ is not taken into account. Although these provide a first-order correction to the redshift expression (cf. eq. (\ref{z_corr}), at small redshift ($z\ll 1$),  their relative contribution to the distance expression is negligible. Both the redshift and the luminosity distance measured in the CMB frame can be obtained by simply setting $\boldsymbol{v}_o=0$. 

All the quantities in the preceding expressions ($z_c$, $\boldsymbol{v}$, $\Phi$, $\boldsymbol{n}$) depend on the coordinates $(t, r,\psi)$, at which the photon reaching the observer at the event ($t_0,r_o,\psi_o$) was emitted. We compute them by solving the geodesic equation of the photon in the EdS background for an off-center observer ($r_o\neq 0$) as detailed in Appendix (\ref{app_solgeod}). This, in turn, allows us to calculate, numerically, the expression for the luminosity distance of the sources as a function of the redshift and the line-of-sight direction and, therefore, the value of $\eta(z,\theta)$.

As a consistency test, we have checked that the cosmographic and the linear perturbation formalisms consistently provide the same expression of the Hubble parameter as measured in the matter frame. The Hubble parameter in a generic spacetime is defined as 
\begin{equation}
\mathbb{H} \circeq \frac{\mathrm{d}\, z}{\mathrm{d} \, d}\,. 
\end{equation}
In \cite{Maartens:2023tib}, we demonstrated that when calculated in the matter frame, this relation holds whatever is the operational definition of the distance $d$. 
Therefore, with $d=d_L$ and using eqs. (\ref{z_corr}) and (\ref{ddl}), we obtain
\begin{eqnarray}\label{acca0}
\mathbb{H} & \circeq &
H \left(1-\boldsymbol{n}\cdot\boldsymbol{v}+\boldsymbol{n}\cdot\frac{\mathrm{d}\boldsymbol{v}}{\mathrm{d}z_c} -\frac{\mathrm{d}\Phi}{\mathrm{d}z_c} \right) \nonumber \\
 & \circeq & H \left[1-n^i v_{i} +
\frac{n^j\big(-\dot v_j+n^i\partial_i v_j \big)-n^i \partial_i \Phi}{H_0}
\right]\,,
\label{eq_H_pert1}
\end{eqnarray}
where we neglected $\dot\Phi$.
Since a dust fluid element moves along time-like geodesics, then at ($t_0,r_o$), when the scale factor is $a=1$, its velocity must satisfy the equation of motion
\begin{equation}
\dot v_{i} +H v_i \circeq -\partial_i \Phi \,.
\end{equation}
Once inserted in eq. (\ref{eq_H_pert1}), this allows 
us to consistently retrieve the formula for the local Hubble parameter given in eq. (\ref{loc_hub_par}) if we ignore the terms in order $\Phi$. However, the terms containing the gravitational potential could also be found if the contribution of $\Phi$ was also taken into account in the expression of $d_L$ (according to the formula C21 in \cite{hui_greene_2006}).
We also note, incidentally, that if the 4-acceleration of the fluid element does not vanish, then the terms which have one $n^i$ in eq. (\ref{eq_H_pert1}) will not cancel. As a consequence, the generalized Hubble parameter measured by a matter-comoving observer will display also a dipolar component. 

\section{Expansion rate fluctuation field of the analytical model}\label{sec_mulmeas}

\begin{figure*}[!ht]
  \centering
  \includegraphics[scale=0.44]{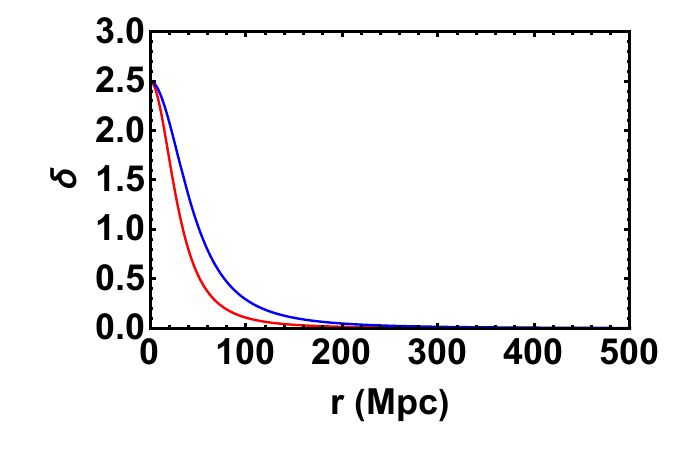}
  \includegraphics[scale=0.44]{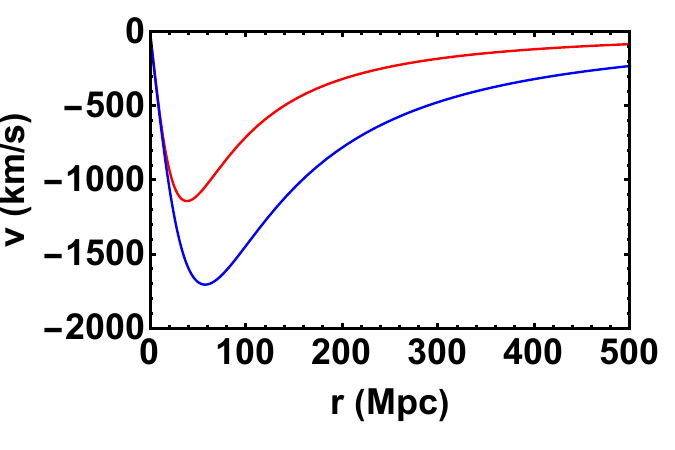}
   \includegraphics[scale=0.44]{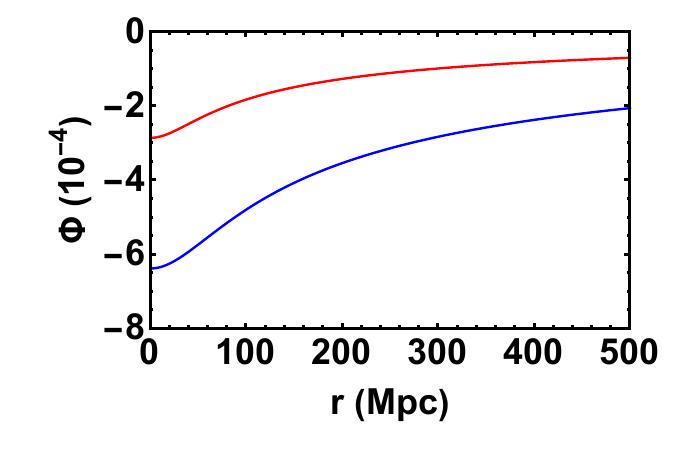}
  \caption{The density contrast, the radial component of the peculiar velocity, and the gravitational potential ($\Phi/c^2$), all evaluated at present time, as a function of the radial coordinate $r$ for the $M1$ (red) and $M2$ (blue) models.}
  \label{vel_phi_r}
\end{figure*}

\begin{figure*}
  \centering
  \includegraphics[scale=0.48]{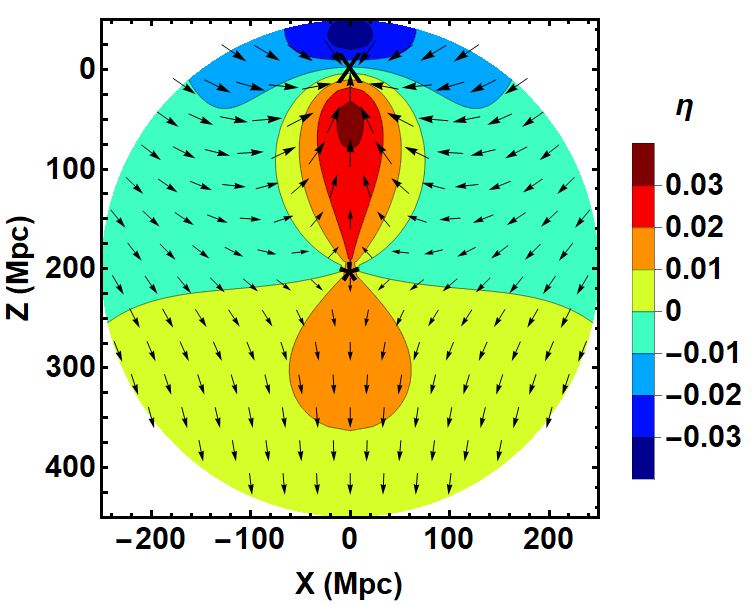}
  \includegraphics[scale=0.48]{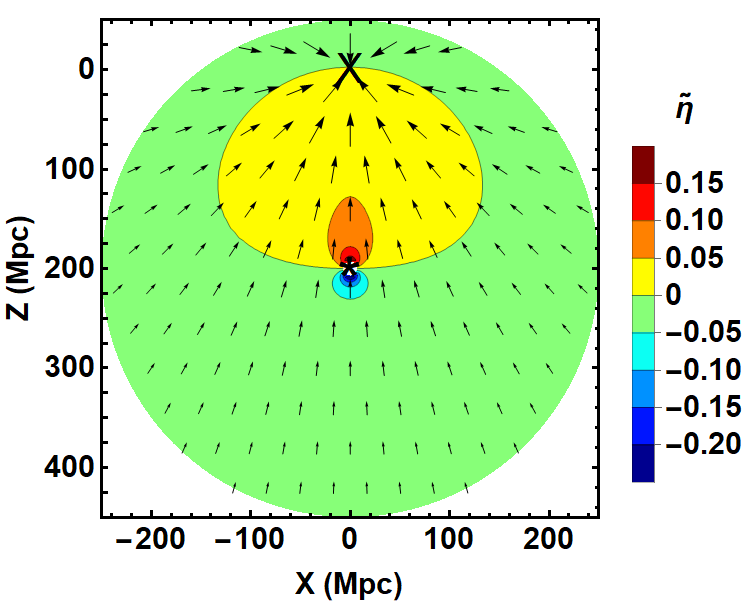}
  \caption{The expansion rate fluctuation field in the matter frame (left) and in the CMB frame (right) for the model $M1$ is portrayed in  the $(X,Z)$-plane containing the axis between the center of the mass overdensity, at $(X,Z)=(0,0)$, and the observer, at (0,200). Arrows display the peculiar velocity (vector) field, while color maps and isocontours display the amplitude of the expansion rate fluctuation (scalar) field.}
  \label{eta_explain}
\end{figure*}

As well as serving as a test bed for the limits of the cosmographic approach in the local Universe, the analytical model is designed to roughly reproduce the expansion rate fluctuations measured with real data by \cite{kalbouneh_marinoni_bel_2023}, thus enabling a better understanding of their physical origin. It also allows us to compare and contrast the structure and characteristics of the expansion rate fluctuation field, measured in two different reference systems, 
the matter-comoving and the CMB frames.

\begin{figure*}
  \centering
  \includegraphics[scale=0.23]{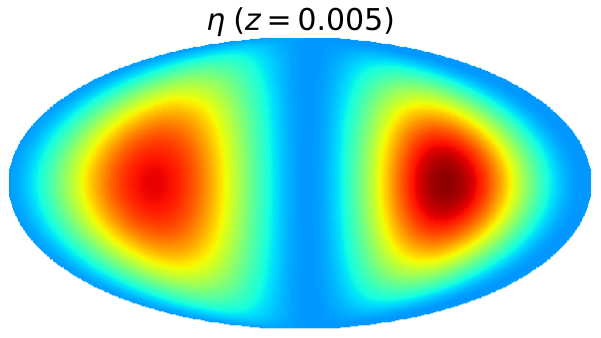}
  \includegraphics[scale=0.23]{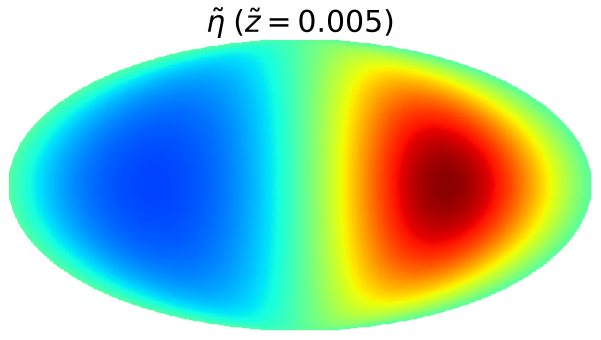}
  \includegraphics[scale=0.23]{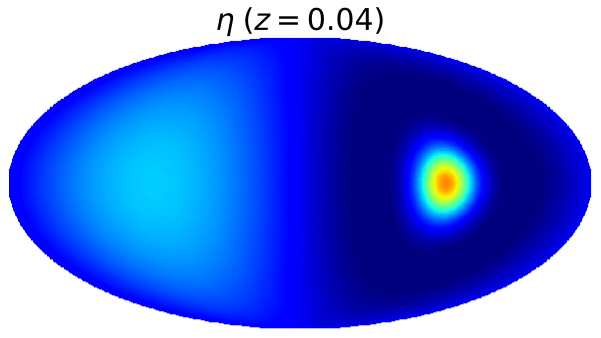}
  \includegraphics[scale=0.23]{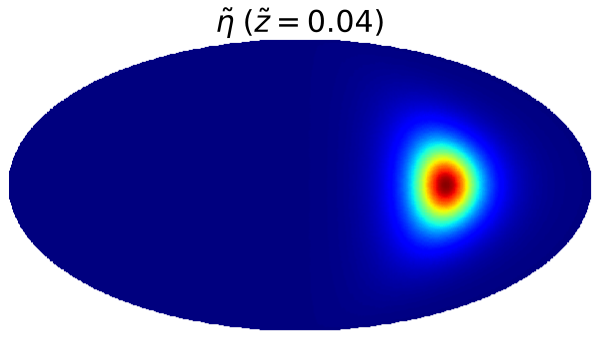}  \\ 
  \includegraphics[scale=0.35]{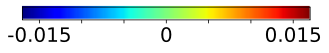} 
  \hspace{0.4cm}
  \includegraphics[scale=0.35]{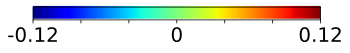} 
  \hspace{0.6cm}
  \includegraphics[scale=0.35]{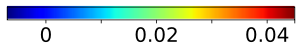} 
  \hspace{0.8cm}  
  \includegraphics[scale=0.35]{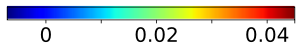}
  \\
  \includegraphics[scale=0.23]{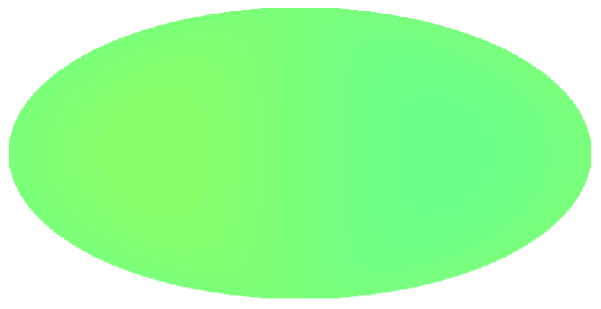}
  \includegraphics[scale=0.23]{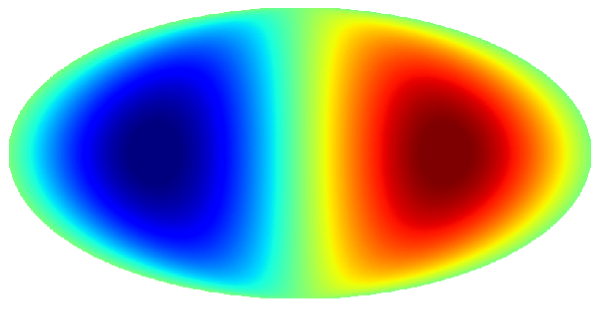}
  \includegraphics[scale=0.23]{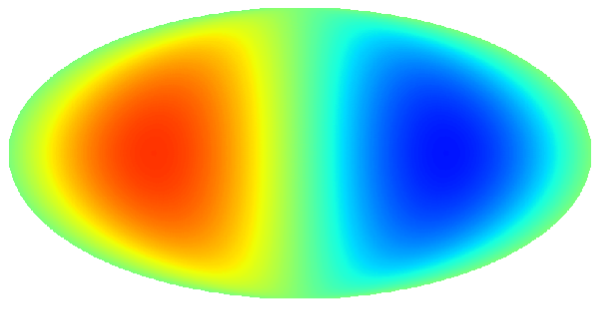}
  \includegraphics[scale=0.23]{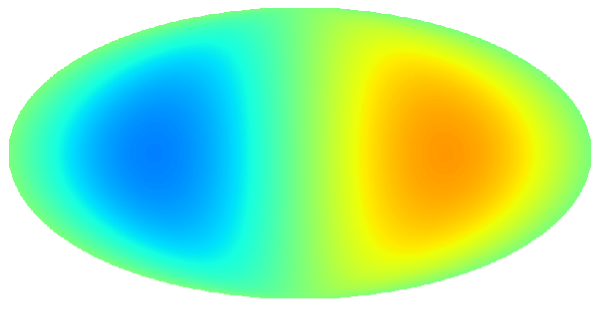}
  \\
  \includegraphics[scale=0.23]{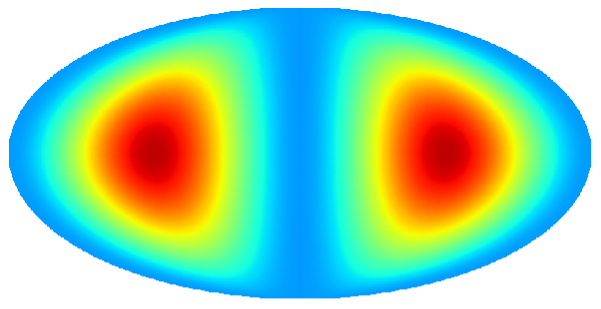}
   \includegraphics[scale=0.23]{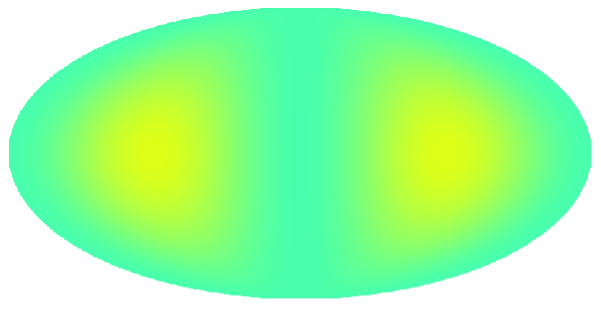}
  \includegraphics[scale=0.23]{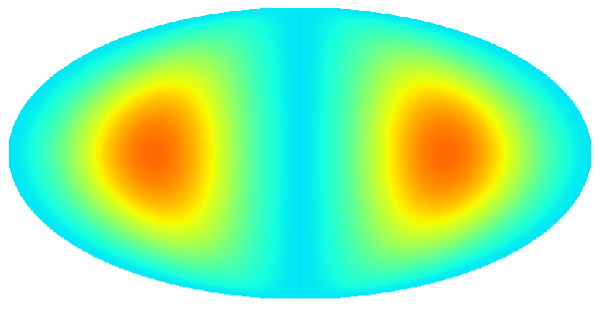}
  \includegraphics[scale=0.23]{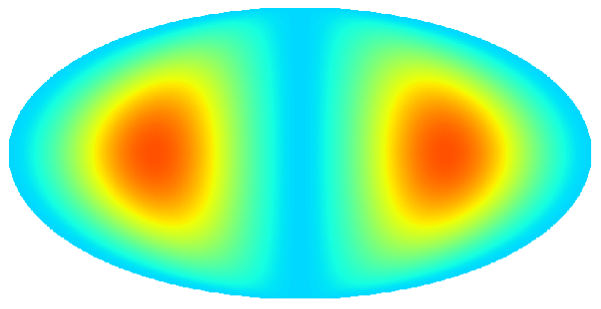}
  \\
  \includegraphics[scale=0.23]{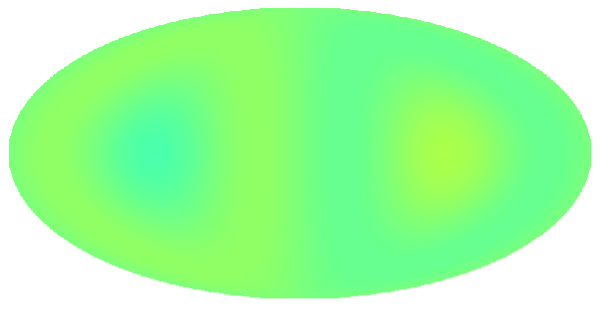}
   \includegraphics[scale=0.23]{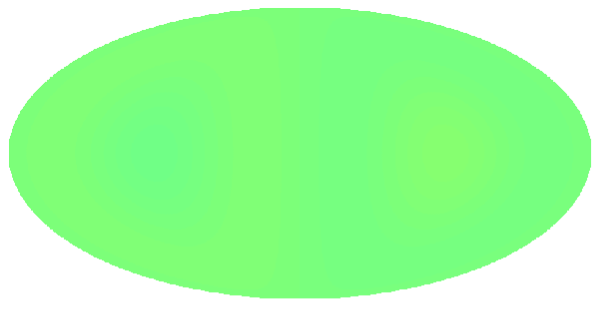}
  \includegraphics[scale=0.23]{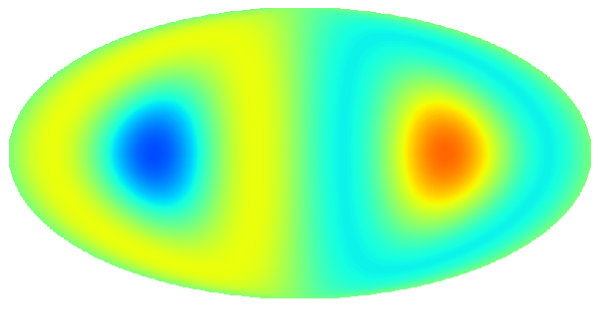}
  \includegraphics[scale=0.23]{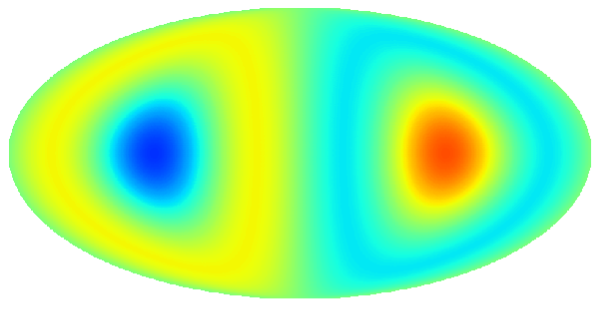}
  \\
  \includegraphics[scale=0.23]{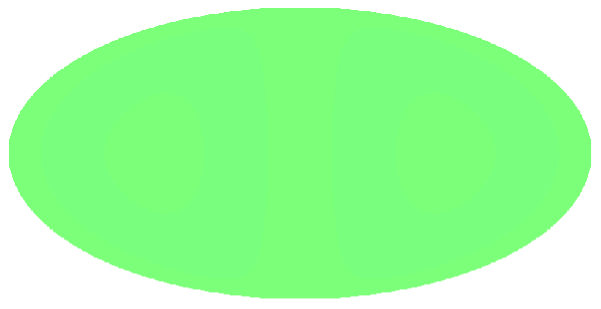}
   \includegraphics[scale=0.23]{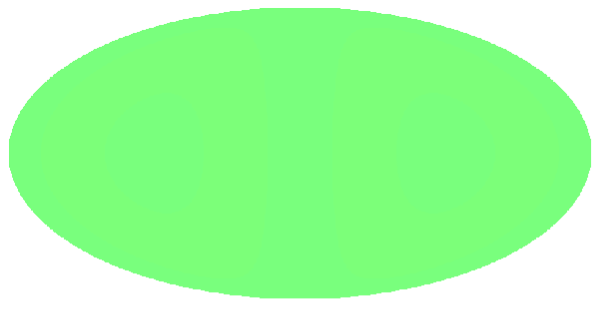}
  \includegraphics[scale=0.23]{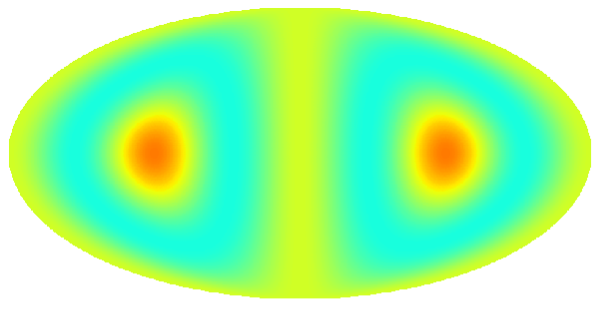}
  \includegraphics[scale=0.23]{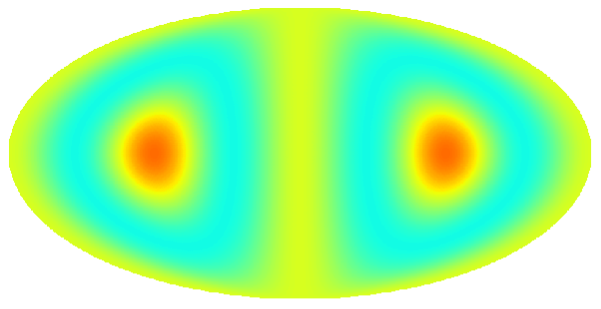}
  \\
  \hspace{0.2cm}
  \includegraphics[scale=0.35]{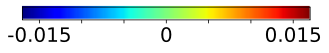} 
  \hspace{0.6cm}
  \includegraphics[scale=0.35]{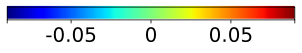} 
  \hspace{0.5cm}
  \includegraphics[scale=0.35]{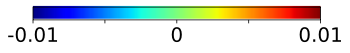}
  \hspace{0.5cm}
  \includegraphics[scale=0.35]{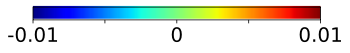} 
  \caption{All-sky maps of the expansion rate fluctuation field generated by the $M1$ model in matter and CMB frames, at two redshifts (first row, 4 columns). The rows show the multipolar components: dipole (row 2), quadrupole (row 3), octupole (row 4), and hexadecapole (row 5). The first and second columns
   show $\eta$ (matter frame) and $\tilde\eta$ (CMB frame)
   at $z=0.005$. Columns 3 and 4 show the same fields at a deeper redshift ($z=0.04$). The position of the overdense region is chosen to lie on the equator of the Mollweide projection, $90^\circ$ away from the center.}
  \label{eta_m_mult}
\end{figure*}

We set the parameters of the analytical model so as to describe two competing and antagonistic scenarios often invoked to explain the kinematics of the local Universe within the standard model of cosmology. One model (hereafter $M1$) has a peculiar velocity field which is sourced by a relatively local mass overdensity and thus results in a bulk
motion which vanishes on small averaging scales. In the opposite scenario, dubbed $M2$, the attractor mass is larger and further away from the observer, thus a coherent bulk motion signal is still present in a large volume with typical size $R \sim 500$Mpc.

The small-scale bulk model $M1$ is given by using
\begin{align}
M1:\quad r_o=200\,{\rm Mpc}\,,~~  \delta_c=2.5\,, ~~ R_s=37.4\,{\rm Mpc}\,,
\end{align}
in eq. (\ref{delta_r}). 
This choice implies $\Phi_c =-2.9\times10^{-4}$. 
These values are suggested by studies of the peculiar velocity field in the local Universe \cite{marinoni_monaco_giuricin_costantini_1998} and fairly characterize the scenario in which such a field is generated by a single attractor like the Shapley supercluster. The values we choose for the parameters $\delta_c$ and $R_s$ are consistent with \cite{marinoni_monaco_giuricin_costantini_1998}, but they are fixed so that the observer (here assumed to be 200 Mpc far from Shapley) moves relative to the CMB with a velocity of $320$ km/s. This value is the component of our Local Group which is generated by a matter distribution at distance $>38\,$Mpc, as measured by \cite{Tully_Pomarède_Graziani_Courtois_Hoffman_Shaya_2019}. 
The large-scale bulk model $M2$ is given by
\begin{align}
M2:\quad r_o=400\,{\rm Mpc}\,,~~  \delta_c=2.5\,, ~~ R_s=56\,{\rm Mpc} \,,
\end{align}
which generates the same observer's velocity ($\sim 320$ km/s). 

The background EdS geometry is common to the two models, with $H_0 = 70$ $\mathrm{km \; s^{-1} Mpc^{-1}}$,  $q_0=0.5$, $\Omega_{k0}=0$, and $j_0=1$. 
In Figure \ref{vel_phi_r}, the density contrast, the radial component of the velocity profile today, and the gravitational potential for these two scenarios are shown as a function of the radial coordinate $r$.

In Figure \ref{eta_explain} we display a planar section of the $M1$ analytical model of the Universe. This plane contains the center of the Shapley overdensity contrast and the observer, at a distance of 200 Mpc. We also show, superimposed, the peculiar velocity (vector) field generated by the mass overdensity as well as the isocontours of the expansion rate fluctuation (scalar) field. Both fields show different configurations, depending on whether they are measured in the matter or CMB frames respectively. 

Regardless of the observer, a common feature that emerges from visual inspection is the characteristic anisotropic structure of the expansion rate fluctuation field induced by metric perturbations. In addition, when looking along a fixed direction of the observer's line of sight, one can also visually appreciate the nonlinearities in the distance-redshift relationships, which are most evident in the direction of the supercluster.

\begin{figure*}
  \centering
  \includegraphics[scale=0.37]{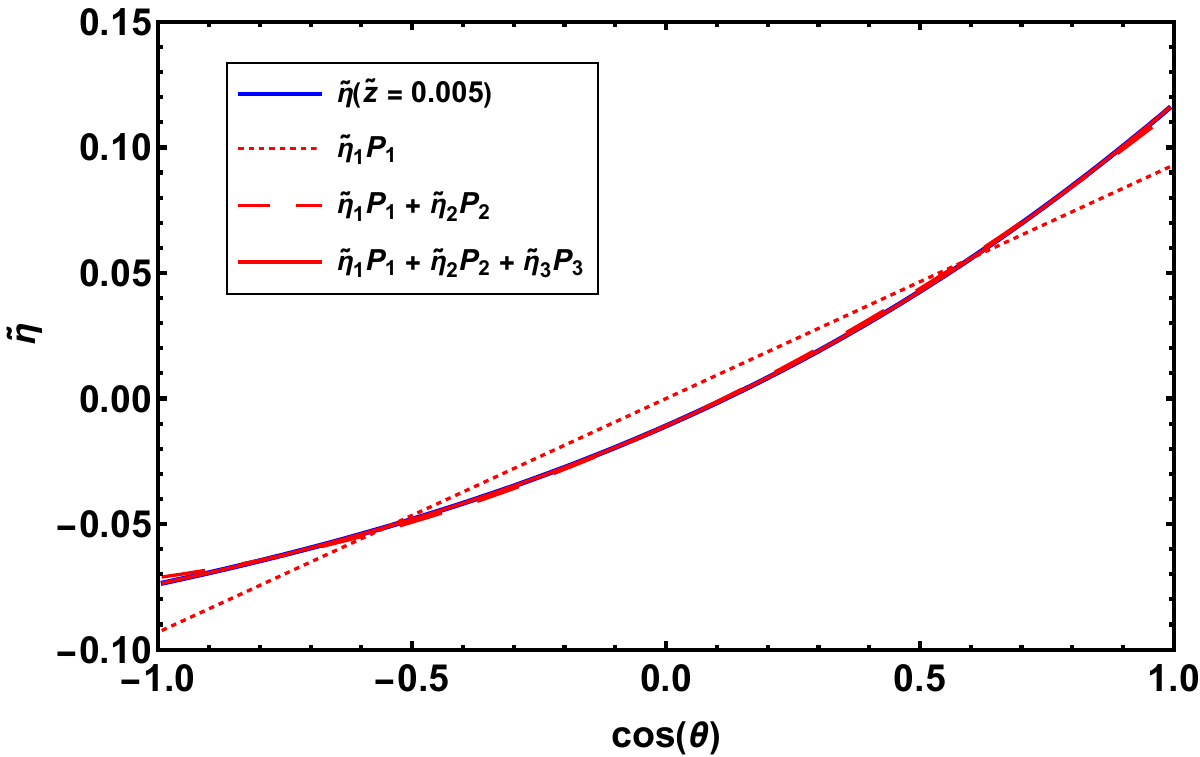}
  \includegraphics[scale=0.37]{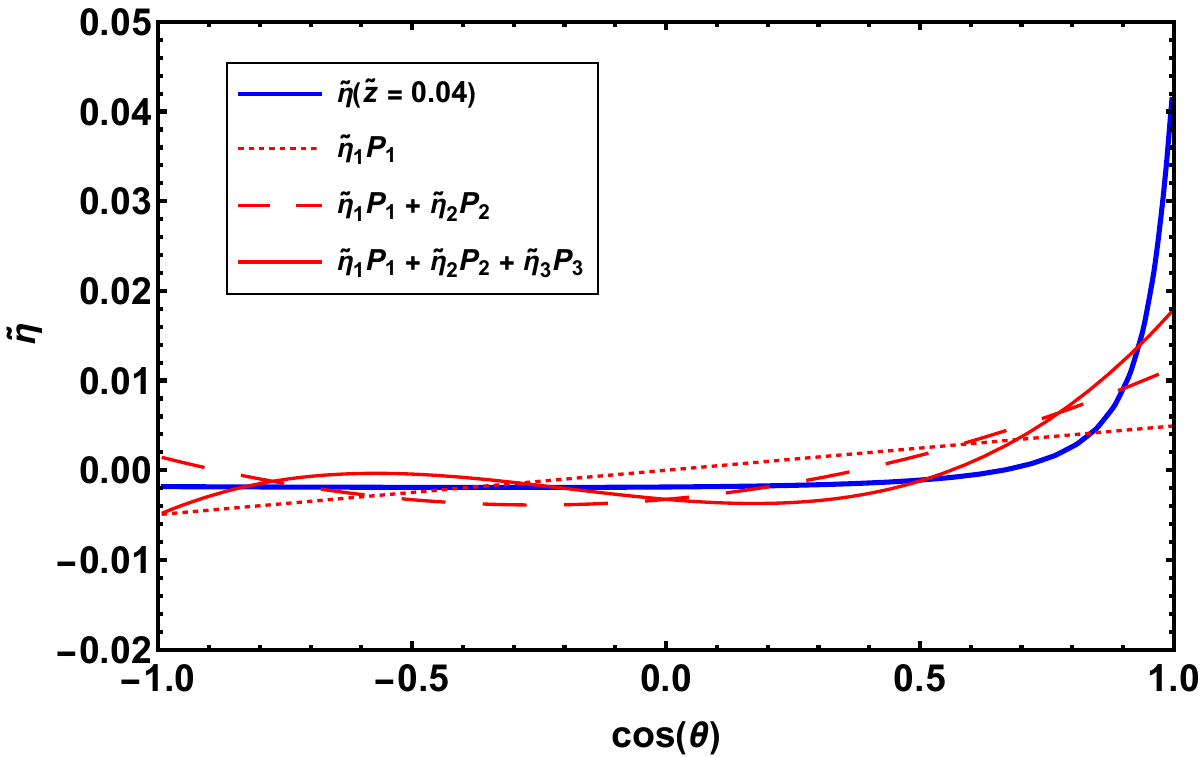}
  \includegraphics[scale=0.37]{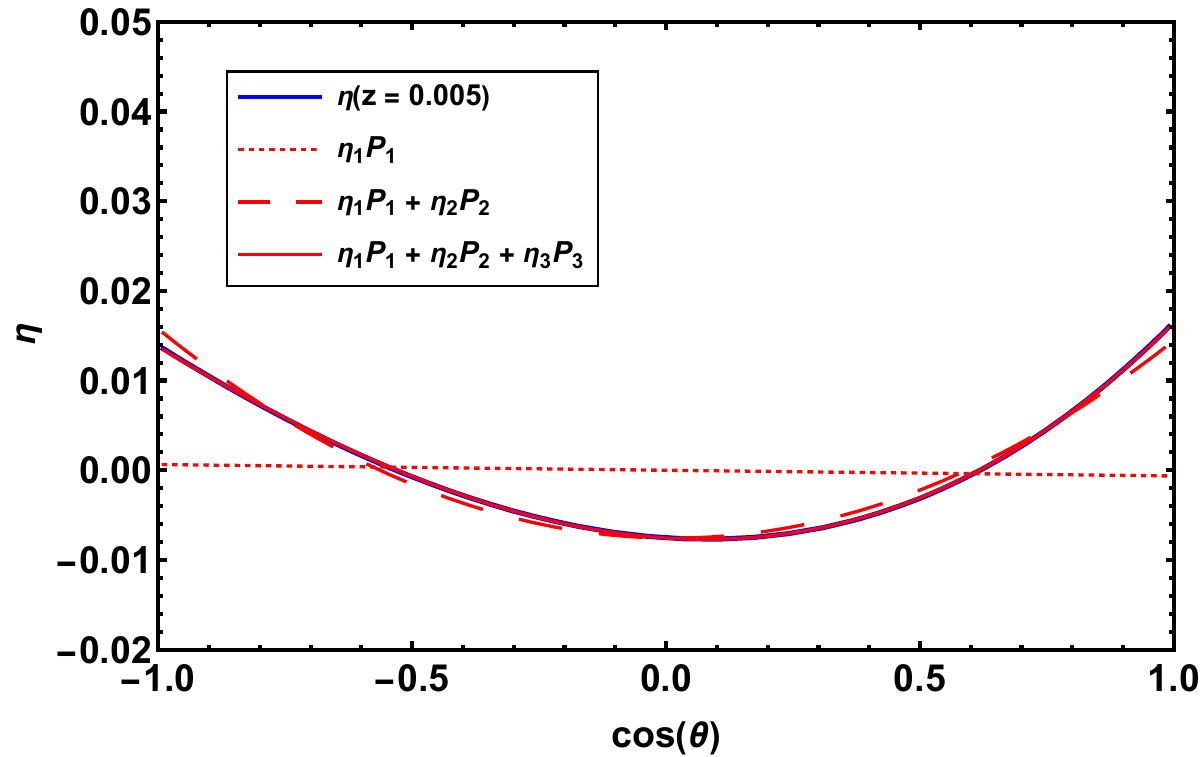}
  \includegraphics[scale=0.37]{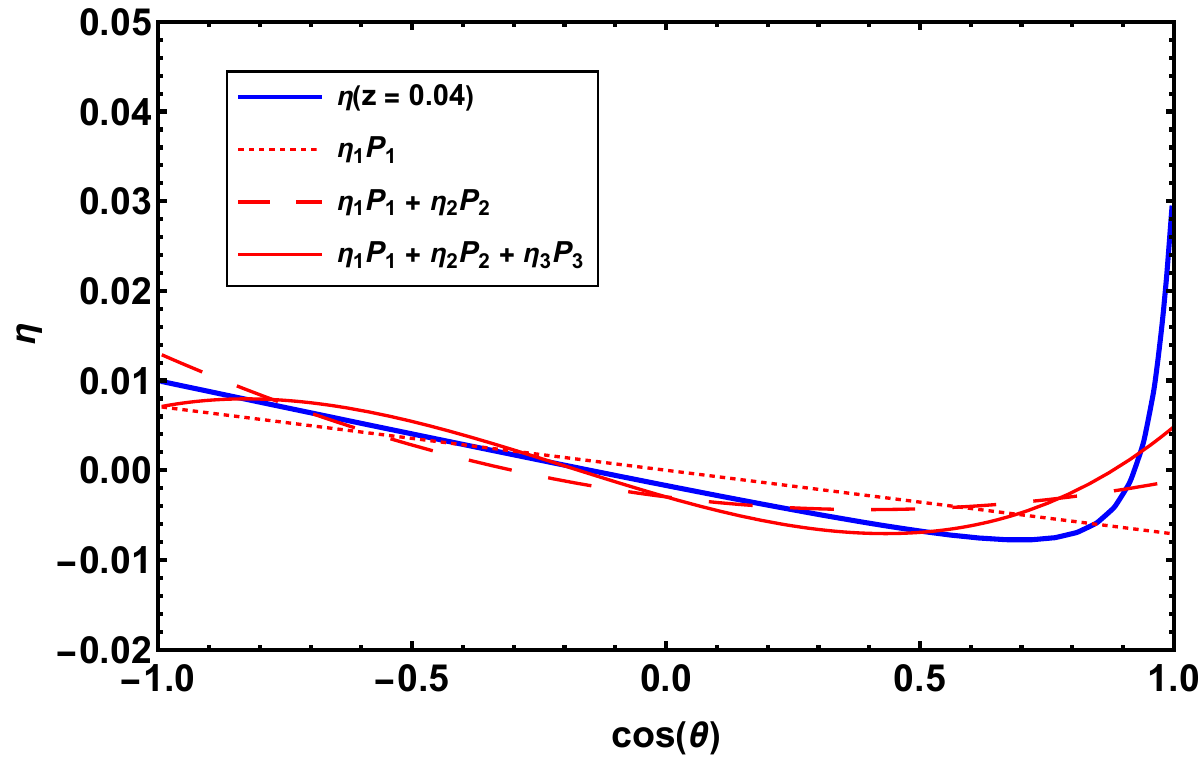}
  \caption{The expansion rate fluctuation field of the $M1$ model in the CMB frame (upper panels) and matter frame (lower panels), shown as a function of $\cos\theta$ at redshift $z=0.005$ (left panels) and $z=0.04$ (right panels). We also show how the signal is reconstructed by different order Legendre expansions, up to $\ell=3$.}
  \label{eta_cos}
\end{figure*}

\begin{figure}
  \centering
  \includegraphics[scale=0.54]{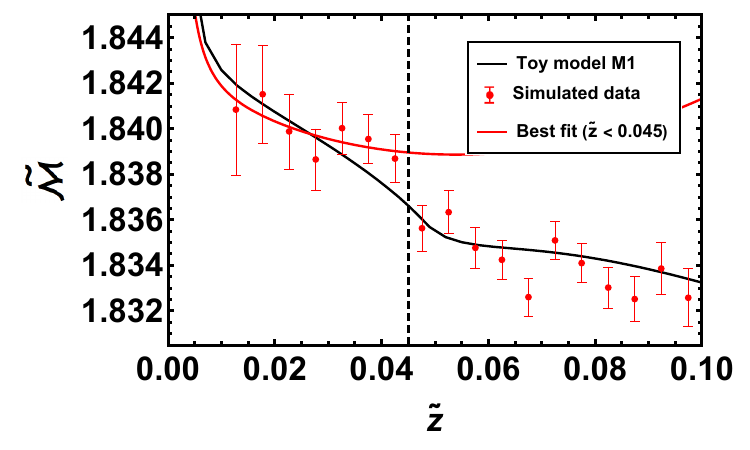}
  \includegraphics[scale=0.54]{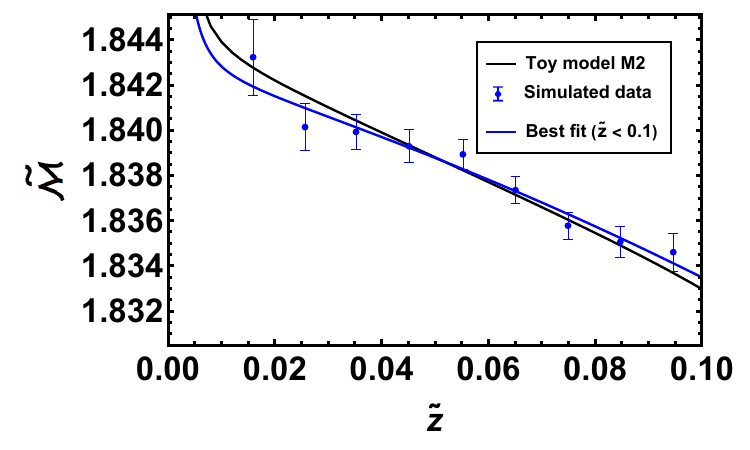}
  \includegraphics[scale=0.54]{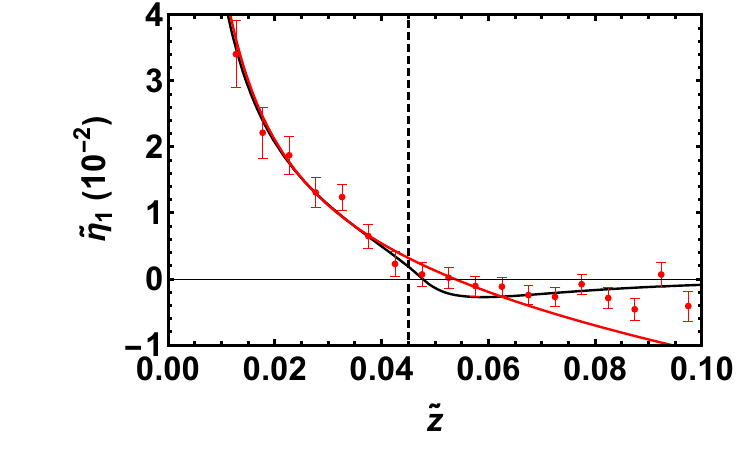}
  \includegraphics[scale=0.54]{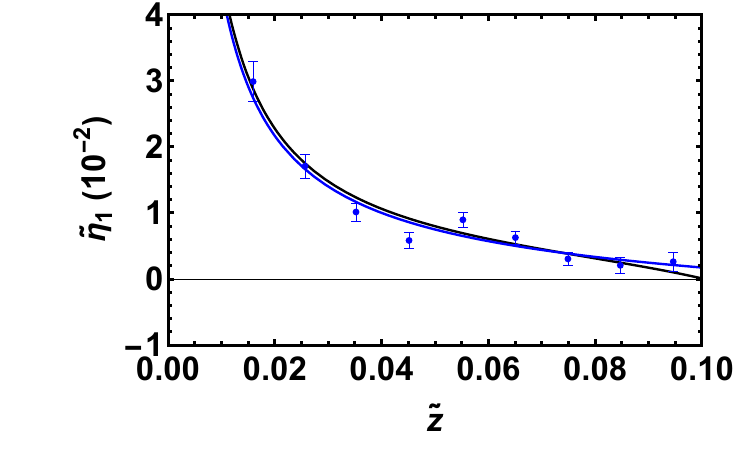}
  \includegraphics[scale=0.54]{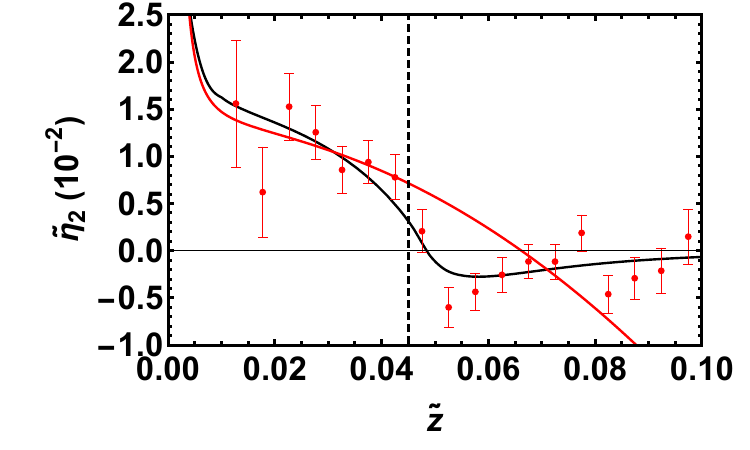}
  \includegraphics[scale=0.54]{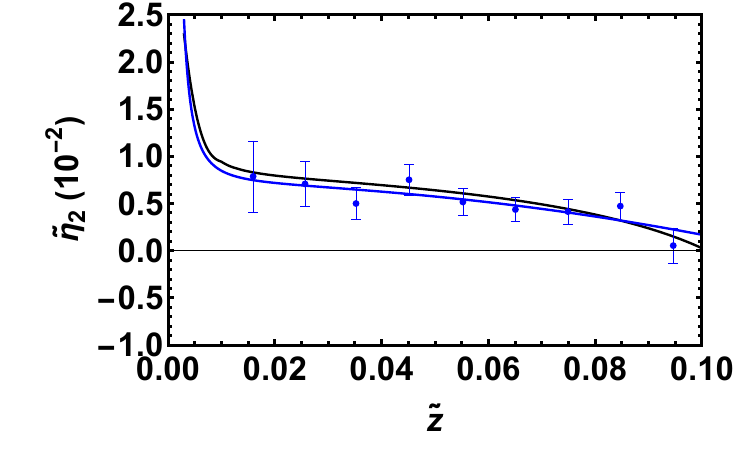}
  \includegraphics[scale=0.54]{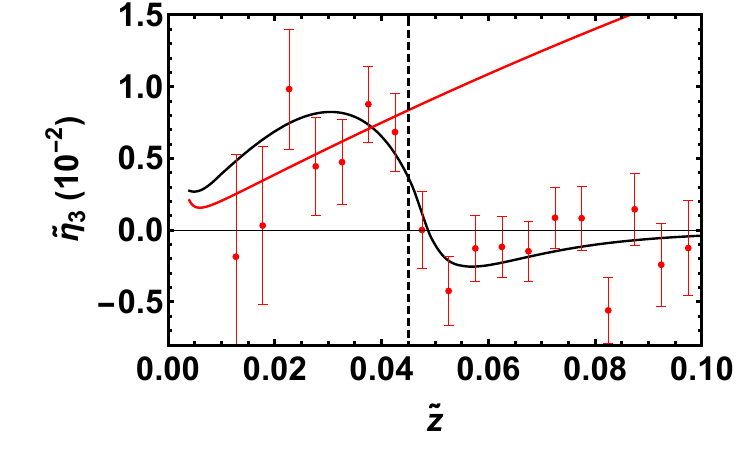}
  \includegraphics[scale=0.54]{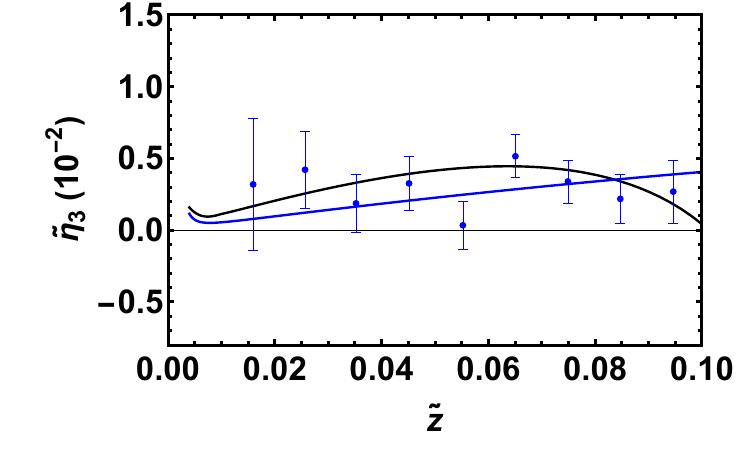}
  \includegraphics[scale=0.54]{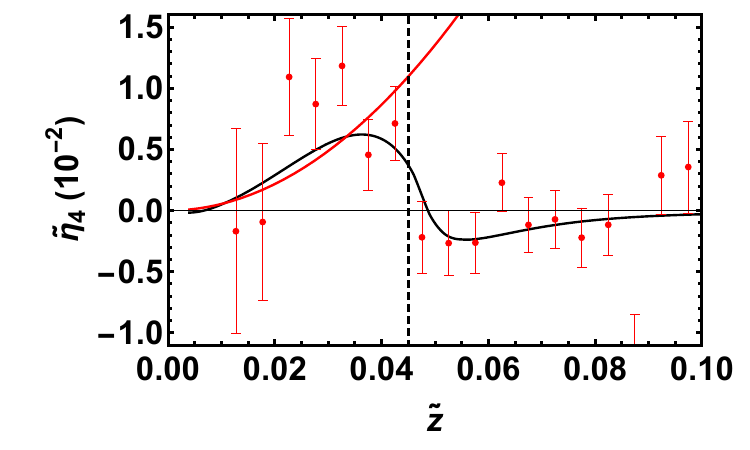}
  \includegraphics[scale=0.54]{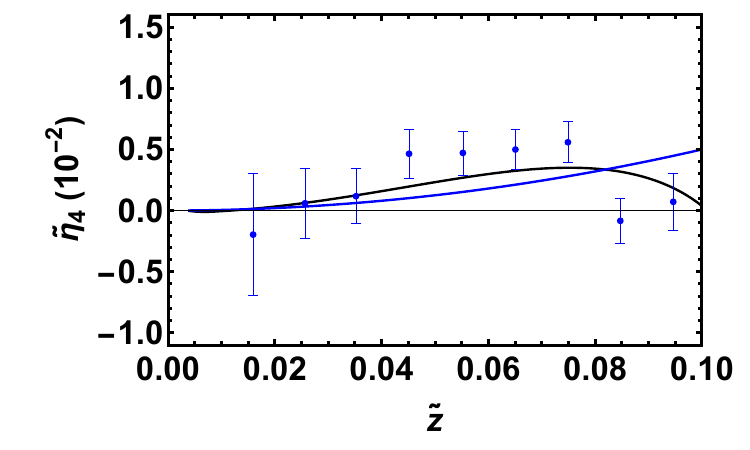}
    \captionsetup{font=footnotesize}
   \caption{Left panels:
   The normalization $\mathcal{M}$, dipole, quadrupole and octupole of the expansion rate fluctuation field in the CMB frame ($\tilde\eta_\ell$), reconstructed using $M1$ simulated galaxy data in different redshift shells (red points). These are compared to the input continuous value (black solid line) and to the best-fitting theoretical predictions (red curves). The latter are obtained by expanding $\tilde{\eta}$ using the cosmographic approach, truncated up to $\mathcal{O}(z^3)$, and by fitting only the points with $\tilde z<0.045$. Right panels: as on the left,  but for the $M2$ simulated data (blue points). The best theoretical models (blue curves) are obtained by fitting all the available measurements up to $z=0.1$.}
  \label{sim_m_d_q_M12}
\end{figure}

The scalar nature of the $\eta$ observable allows the multipolar structure of the signal to be captured even if only visually. At low redshift (close to the observer), the amplitude of $\eta$ measured in the matter frame is dominated by a quadrupolar component, while that inferred in the CMB frame shows a typical dipolar pattern. 
The fact that, by definition, the velocity field vanishes around the matter-comoving observer (see left panel of Figure \ref{eta_explain}), means that the fluctuations of the expansion velocity field are dominated by a shear component. In other words, the expansion rate is similarly perturbed in antipodal directions around the observer. When the distance increases, the peculiar velocity field reveals more complex characteristics. As a consequence, the expansion rate fluctuation field also acquires non-trivial distortions. 
In the CMB frame (right panel of Figure \ref{eta_explain}), the expansion rate field is locally dominated by a dipolar component, induced by the motion of the observer relative to the chosen reference frame. Its amplitude increases closer to the observer.

 A quantitative estimation of the multipolar components of $\eta$ for $M1$ is presented in 
 Figure \ref{eta_m_mult}, where we show the multipoles of the expansion rate fluctuation measured by the matter-comoving and CMB-comoving observers. The multipole decomposition is shown at two characteristic redshifts, $z=0.005$ -- assumed to represent very local measurements, and $z=0.04$ -- representing regions where the peak density of the analytical model severely distorts the distance-redshift relationship. The nonzero value of a multipole provides information on the structure of the anisotropy in the expansion rate, while its evolution with redshift reflects the level of nonlinearity in the distance-redshift relation. 
 The qualitative similarity with the signal extracted by \cite{kalbouneh_marinoni_bel_2023} (in the CMB frame) from Cosmicflows-3 and Pantheon samples is apparent: the axes of the multipoles are aligned, the geometry is axisymmetrical, and the amplitude and redshift scaling of each multipole roughly reproduce what was observed by \cite{kalbouneh_marinoni_bel_2023} (compare their Figure 4).

The expansion rate fluctuations in the CMB frame are enhanced relative to the signal measured by the matter-frame observer. This increase is essentially contributed by the dipolar component, which, as expected on the basis of eq. (\ref{alm}), is the multipole that is most sensitive to observer velocity effects. Furthermore, as eq. (\ref{eta_bar_exp_3}) makes clear, the smaller the volume surrounding the observer in which the dipole signal is estimated, the greater the increase in the signal. On the contrary, in the same limit $z\rightarrow 0$, the dipole of $\eta$ in the matter frame is expected to vanish (cf. eq. (\ref{dipe})), as shown in Figure \ref{eta_m_mult}. The distortions induced by the motion of the observer in the dipole are of order $v_o/\tilde z$. As one pushes the analysis to higher $\tilde z$, the amplitude of $\tilde{\eta}_1$ is suppressed inversely proportional to redshift. Consequently, the signal becomes more sensitive to metric contributions, particularly to the dipole of the deceleration parameter, $\mathbb{Q}_1$. The increasingly important effects induced by $\mathbb{Q}_1$ can be seen by comparing the low- and high-redshift dipolar components measured in the matter frame.

Visual inspection of Figure \ref{eta_m_mult} shows that 
the relative motion of the CMB and matter frames only affects the dipole of $\tilde{\eta}$, while minimally affecting its quadrupole and octupole. 
In fact, already at very low redshift, the contribution of velocity-induced distortions to the signal is ${\cal O}(v_o^2/z^2)$ and ${\cal O}(v_o^3/z^3)$ respectively. 
Note that, contrary to the case of other multipoles, $\eta_2$ 
does not vanish for $z\rightarrow 0$, but in this limit becomes proportional to the amplitude of the quadrupole $\mathbb{H}_2$.

Similarly to the case of the dipole, as redshift increases, the differences between the reconstructed multipoles of $\eta$ in the matter and CMB frames, are reduced and the signal acquires specific dependencies on the higher-order multipoles of the covariant cosmographic parameters.
Figure \ref{eta_m_mult} shows that the quadrupole signal $\eta_2$ in the $M1$ model
weakens with redshift, because the quadrupole contributions of $\mathbb{Q}$ and $\mathbb{J}-\mathbb{R}$ -- which are negative in this model (see Table \ref{tab_al}) -- gradually become more important. The same applies to the octupole and the hexadecapole signals $\eta_3$ and $\eta_4$, whose low-$z$ amplitude becomes progressively more sensitive to a mixture of multipoles of higher-order cosmographic parameters. Although these are only ${\cal O}(z^2)$ corrections, large values of the covariant cosmographic parameters, as indeed is the case if the metric has large fluctuations (see Table \ref{tab_al}), compensate for the suppression induced by the powers of small $z$. 

In Figure \ref{eta_cos} we show how well the expansion rate fluctuations in the CMB and matter frames are reconstructed by the lower multipoles, at redshifts $z=0.005$ and $z=0.04$.
As expected, a multipolar expansion up to $\ell =2$
is enough to achieve percentage estimation precision at low redshift. However,
the higher the redshift, the more inaccurate the estimation is. Note that, given the different functional shapes of the expansion rate fluctuations in different observer frames, at an equal order of expansion, the Legendre approximation in the matter frame has a somewhat higher level of precision than in the CMB frame, with the accuracy increasing by roughly $5\%$.

\section{Forecasting precision on cosmographic parameter estimation}\label{sec_forecas}

\begin{table}
\scalebox{0.67}{
\begin{tabular}{|ccccccccccc|}
\hline
\multicolumn{1}{|c||}{Model}            & \multicolumn{5}{c||}{\begin{tabular}[c]{@{}c@{}}Small-Scale Bulk ($M1$)\\ ($\delta_c=2.5, R_s=37.4$ Mpc, $r_o=200$ Mpc)\end{tabular}}                         & \multicolumn{5}{c|}{\begin{tabular}[c]{@{}c@{}} Large-Scale Bulk ($M2$)\\ ($\delta_c=2.5, R_s=56$ Mpc, $r_o=400$ Mpc)\end{tabular}}              \\ \hline\hline
\multicolumn{1}{|c||}{$X$}             & \multicolumn{1}{c|}{$X_0$}   & \multicolumn{1}{c|}{$X_1$}   & \multicolumn{1}{c|}{$X_2$}   & \multicolumn{1}{c|}{$X_3$}   & \multicolumn{1}{c||}{$X_4$} & \multicolumn{1}{c|}{$X_0$}    & \multicolumn{1}{c|}{$X_1$}    & \multicolumn{1}{c|}{$X_2$}   & \multicolumn{1}{c|}{$X_3$}   & $X_4$ \\ \hline
\multicolumn{1}{|c||}{$\mathbb{H}$ (km/s/Mpc)}       & \multicolumn{1}{c|}{$69.65$}   & \multicolumn{1}{c|}{-}     & \multicolumn{1}{c|}{$2.47$}   & \multicolumn{1}{c|}{-}     & \multicolumn{1}{c||}{-}   & \multicolumn{1}{c|}{$69.86$}    & \multicolumn{1}{c|}{-}      & \multicolumn{1}{c|}{$1.29$}    & \multicolumn{1}{c|}{-}     & -   \\ \hline
\multicolumn{1}{|c||}{$\mathbb{Q}$}       & \multicolumn{1}{c|}{$0.51$}    & \multicolumn{1}{c|}{$-0.58$}   & \multicolumn{1}{c|}{$-0.09$}  & \multicolumn{1}{c|}{$1.88$}   & \multicolumn{1}{c||}{0}   & \multicolumn{1}{c|}{$0.505$}    & \multicolumn{1}{c|}{$-0.125$}   & \multicolumn{1}{c|}{$-0.05$}   & \multicolumn{1}{c|}{$0.51$}   & 0   \\ \hline
\multicolumn{1}{|c||}{$\mathbb{J}-\mathbb{R}$}       & \multicolumn{1}{c|}{$-6.53$}   & \multicolumn{1}{c|}{$1.16$}    & \multicolumn{1}{c|}{$-57.3$}   & \multicolumn{1}{c|}{$-3.76$}   & \multicolumn{1}{c||}{$138.5$} & \multicolumn{1}{c|}{$0.17$}     & \multicolumn{1}{c|}{$0.25$}    & \multicolumn{1}{c|}{$-6.37$}   & \multicolumn{1}{c|}{$-1.02$}   & $19.3$ \\ \hline
\multicolumn{1}{|c||}{$v_o$ (km/s)}            & \multicolumn{5}{c||}{$320$}                                                                         & \multicolumn{5}{c|}{$320$}                                                                 \\ \hline\hline
\multicolumn{11}{|c|}{Best fitting values}                                                                                                                                                                           \\ \hline\hline
\multicolumn{1}{|c||}{$X$}             & \multicolumn{1}{c|}{$X_0$}   & \multicolumn{1}{c|}{$X_1$}   & \multicolumn{1}{c|}{$X_2$}   & \multicolumn{1}{c|}{$X_3$}   & \multicolumn{1}{c||}{$X_4$} & \multicolumn{1}{c|}{$X_0$}    & \multicolumn{1}{c|}{$X_1$}    & \multicolumn{1}{c|}{$X_2$}   & \multicolumn{1}{c|}{$X_3$}   & $X_4$ \\ \hline
\multicolumn{1}{|c||}{$\mathbb{H}$ (km/s/Mpc)}       & \multicolumn{1}{c|}{$69.5\pm1.0$} & \multicolumn{1}{c|}{-}      & \multicolumn{1}{c|}{$2.1\pm0.5$} & \multicolumn{1}{c|}{-}      & \multicolumn{1}{c||}{-}    & \multicolumn{1}{c|}{$69.63\pm0.30$} & \multicolumn{1}{c|}{-}       & \multicolumn{1}{c|}{$1.1\pm0.2$} & \multicolumn{1}{c|}{-}      & -    \\ \hline
\multicolumn{1}{|c||}{$\mathbb{Q}$}       & \multicolumn{1}{c|}{$0.4\pm2.0$} & \multicolumn{1}{c|}{$-0.9\pm0.3$} & \multicolumn{1}{c|}{$0$} & \multicolumn{1}{c|}{$0.9\pm0.2$} & \multicolumn{1}{c||}{0}    & \multicolumn{1}{c|}{$0.7\pm0.3$} & \multicolumn{1}{c|}{$-0.18\pm0.05$} & \multicolumn{1}{c|}{$0$} & \multicolumn{1}{c|}{$0.23\pm0.05$} & 0    \\ \hline
\multicolumn{1}{|c||}{$\mathbb{J}-\mathbb{R}$} & \multicolumn{1}{c|}{$16\pm87$}  & \multicolumn{1}{c|}{$0$}  & \multicolumn{1}{c|}{$-42\pm32$} & \multicolumn{1}{c|}{$0$} & \multicolumn{1}{c||}{$73\pm16$}    & \multicolumn{1}{c|}{$-2.3\pm5.9$}    & \multicolumn{1}{c|}{$0$}   & \multicolumn{1}{c|}{$-7.4\pm3.1$}  & \multicolumn{1}{c|}{$0$}  & $6.8\pm1.9$    \\ \hline
\multicolumn{1}{|c||}{$v_o$ (km/s)}            & \multicolumn{5}{c||}{$336\pm32$}                                                                      & \multicolumn{5}{c|}{$303\pm20$}                                                              \\ \hline
\end{tabular}}
\captionsetup{font=small}
 \caption{ The upper table shows the Legendre multipoles $X_\ell$ ($\ell \leq 4$) of the covariant cosmographic parameters and the relative velocity $v_o$ between the matter and the CMB frames for the $M1$ (left) and $M2$ (right) models. These quantities are calculated using the equations presented in Appendix \ref{app_alX_eqns}. The lower table shows the best-fit values (and the $68\%$ confidence interval) deduced from the likelihood analysis performed using $15,000$ discrete objects sampling the analytical models and roughly reproducing the ZTF survey characteristics. A cell containing a dash indicates that the multipole vanishes by definition. A cell containing a zero means that the multipole is intrinsically small and its value is therefore set to zero by default. Note that for $\ell>0$, $\mathbb{J}_{\ell} -\mathbb{R}_\ell \approx \mathbb{J}_\ell$.}
\label{tab_al}
\end{table}

\begin{figure}
  \includegraphics[scale=0.41]{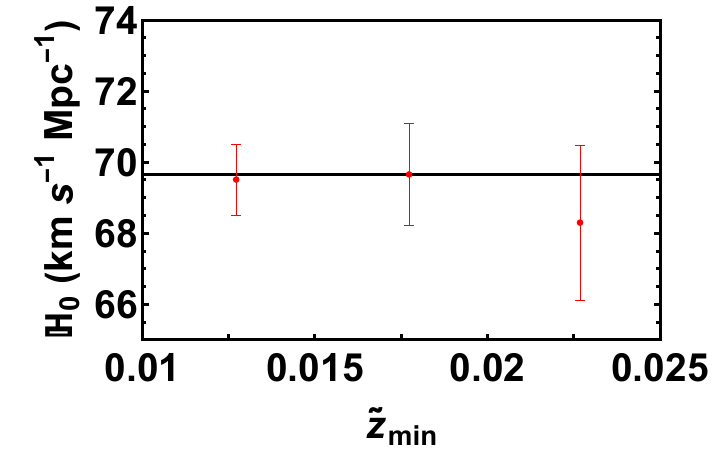}
  \includegraphics[scale=0.41]{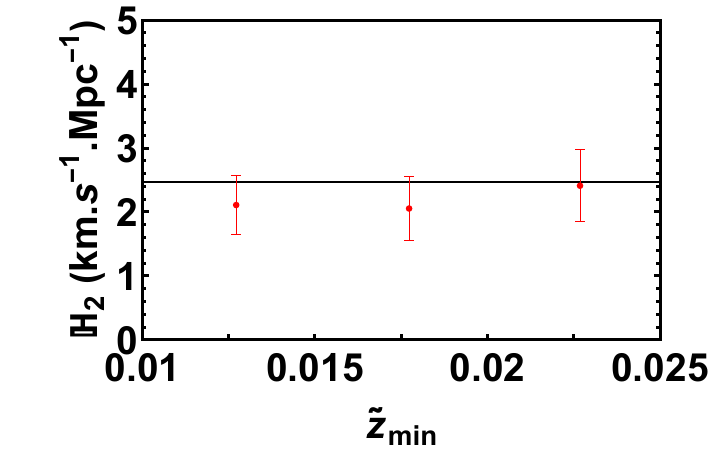}
  \includegraphics[scale=0.41]{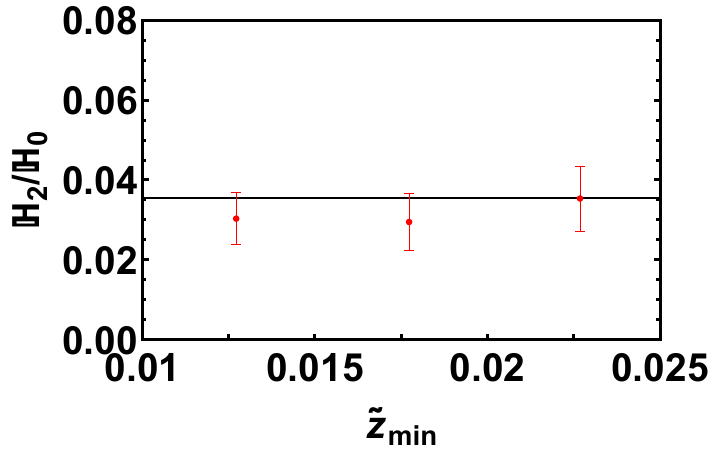}
  \\
  \includegraphics[scale=0.41]{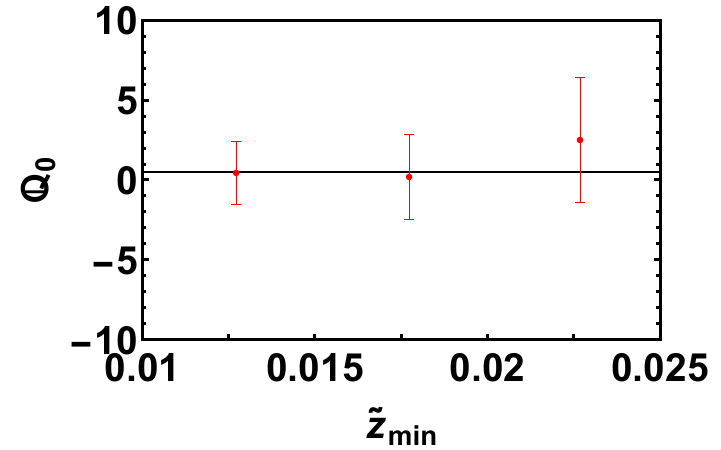}
  \includegraphics[scale=0.41]{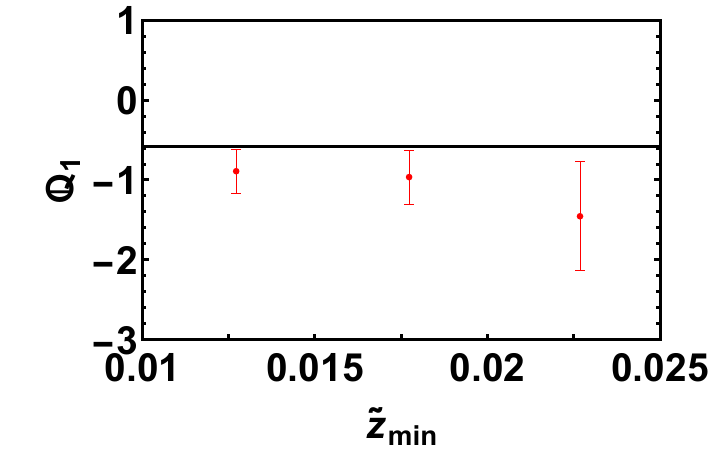}
  \includegraphics[scale=0.41]{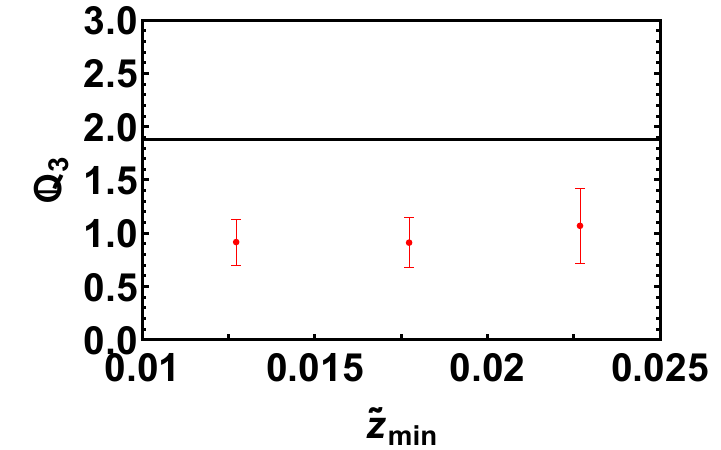}
  \\
  \includegraphics[scale=0.41]{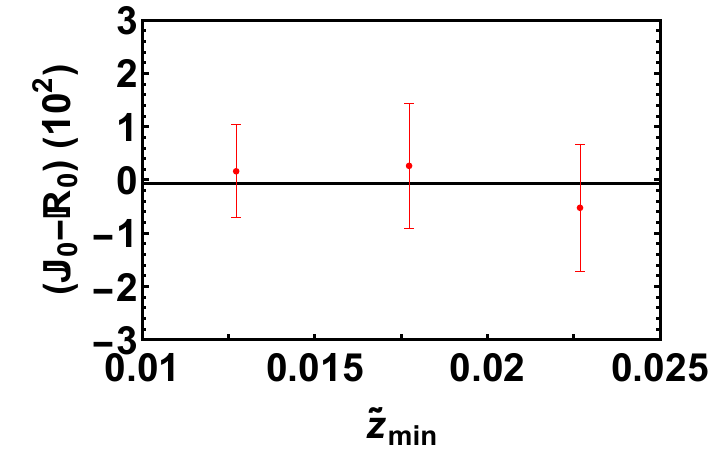}
  \includegraphics[scale=0.41]{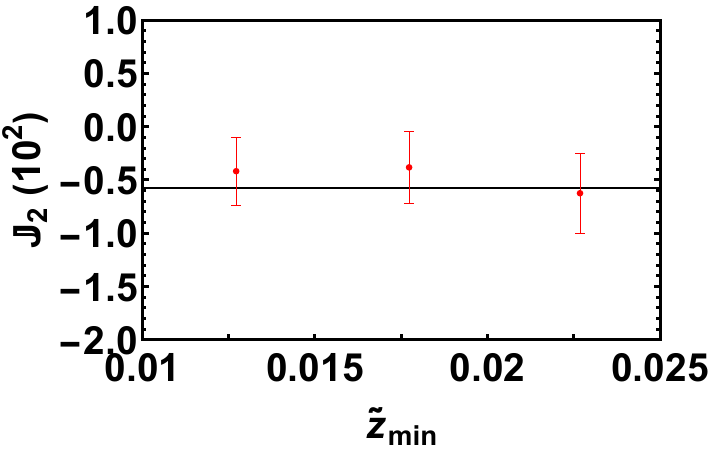}
  \includegraphics[scale=0.41]{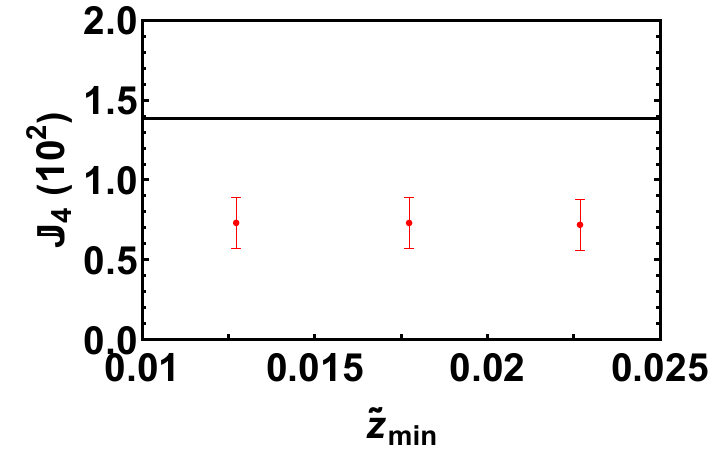}
   \caption{The best-fitting values (and the associated $1\sigma$ error bars) for the covariant cosmographic parameters extracted from the analysis of the $M1$ simulation in the shell $\tilde z_{\rm min}<\tilde z<0.045$, 
   are shown as a function of $\tilde z_{\rm min}$. The black lines show, for comparison, the true simulated input values (see Table \ref{tab_al}).}
  \label{cospar_M1}
\end{figure}

\begin{figure}
  \includegraphics[scale=0.41]{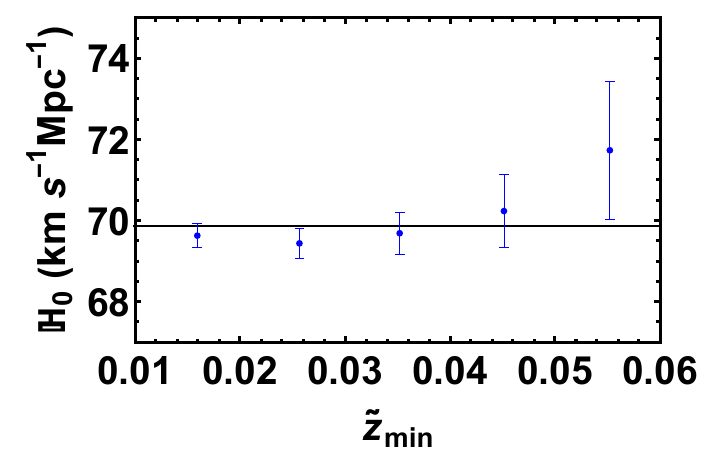}
  \includegraphics[scale=0.41]{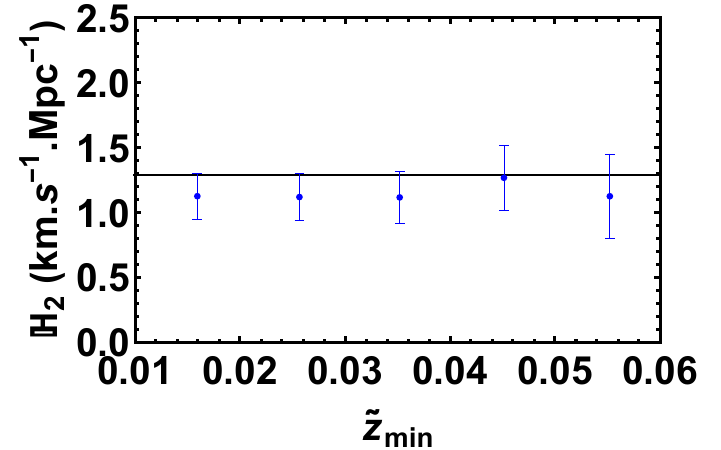}
  \includegraphics[scale=0.41]{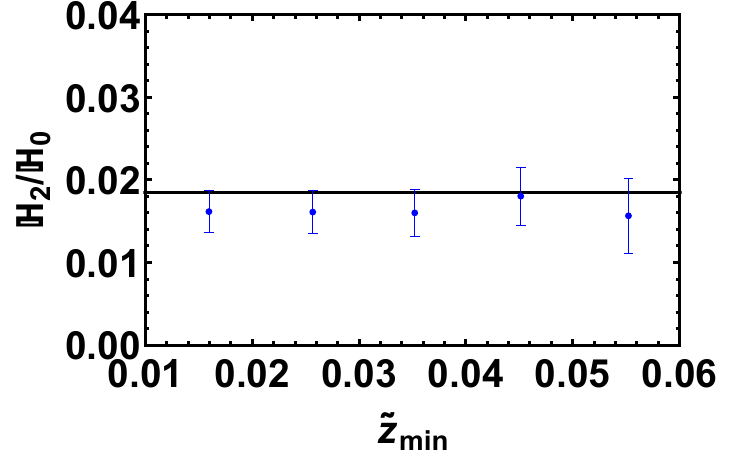}
  \\
  \includegraphics[scale=0.41]{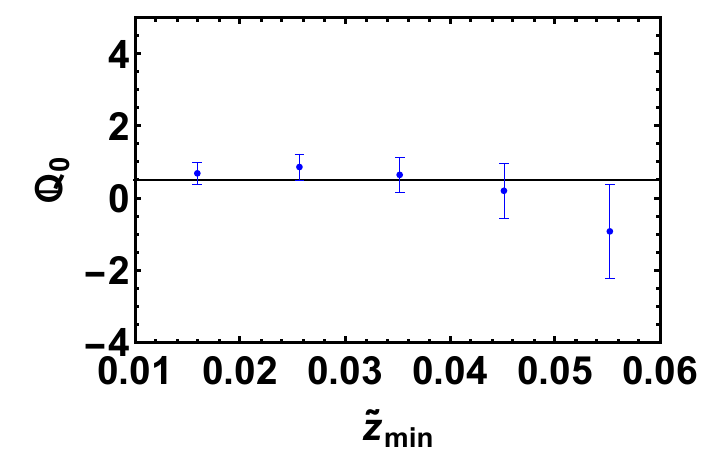}
  \includegraphics[scale=0.41]{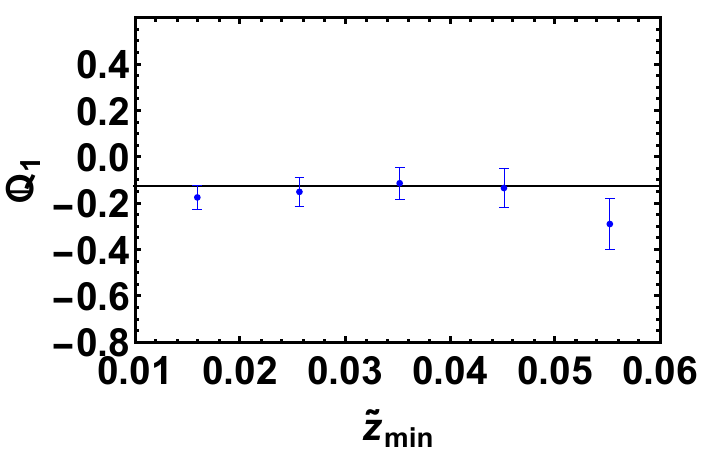}
  \includegraphics[scale=0.41]{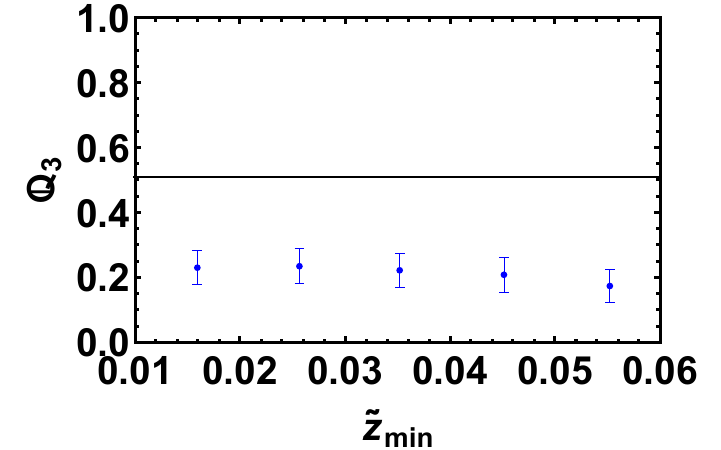}
  \\
  \includegraphics[scale=0.41]{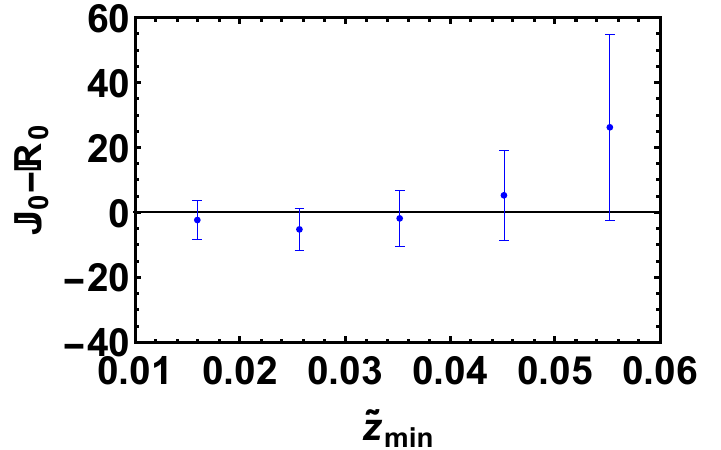}
  \includegraphics[scale=0.41]{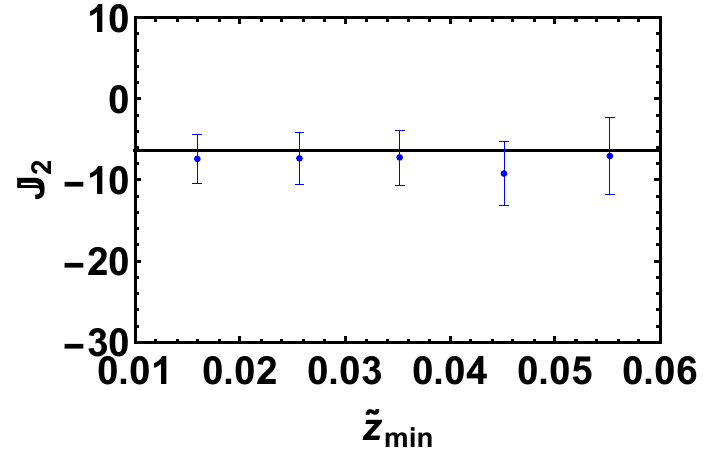}
  \includegraphics[scale=0.41]{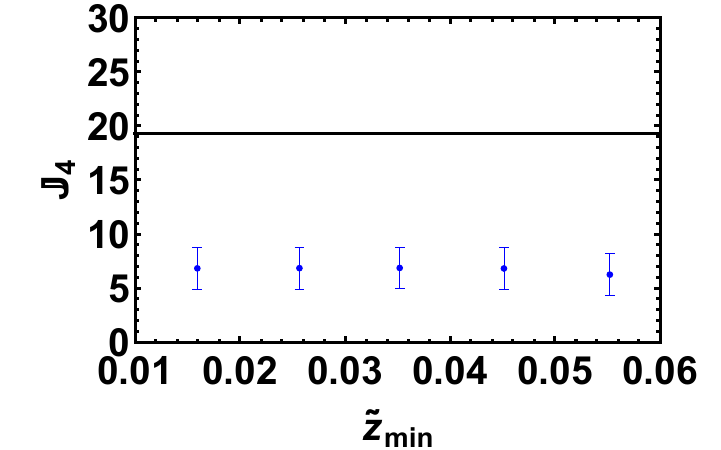}
  \caption{As in Figure \ref{cospar_M1}, but for the $M2$ simulation. The best-fitting parameters recovered in the shell $\tilde z_{\rm min}<\tilde z<0.1$ are shown as a function of $\tilde z_{\rm min}$.}
  \label{cospar_M2}
\end{figure}

By means of the analytical models, we can also assess the accuracy with which the (idealized) continuous $\eta$ field is reconstructed using a (real) discrete point process, and forecast the precision with which future data sets will constrain multipoles of the covariant cosmographic parameters in the local Universe ($z \lesssim 0.1)$.

To this end, we simulate distances and redshifts for $15\times 10^3$ galaxies, distributed in the redshift interval $0.01 < \tilde z <0.1$, by sampling the continuous analytical models $M1$ and $M2$ according to the redshift distribution expected from the ZTF survey \cite{amenouche:tel-04165406}.
In order to make these two simulated catalogs of distances (measured without using redshift information) as realistic as possible, we assume that the distance modulus of each object has a Gaussian distribution with standard deviation $\sigma_{\mu}=0.15$ \cite{Dhawan_2022}, and no correlation between the measurements.

We measure the expansion rate fluctuations (in the CMB frame) in spherical shells $S$ of thickness $\delta \tilde z=0.005$ for the $M1$ model and $\delta \tilde z=0.01$ for $M2$ model. 
In Figure \ref{sim_m_d_q_M12}, the amplitude of the multipoles $\tilde{\eta}_\ell$, inferred using the discrete catalogs $M1$ and $M2$, is compared to the input theoretical value characterizing the continuous matter density fields. The estimation is unbiased and its accuracy, for both the small- and large-scale bulk scenarios, is remarkable. However, extracting the value of covariant cosmographic parameters from these measurements is challenging. 

The covariant cosmographic parameters $\mathbb{H}$, $\mathbb{Q}$, $\mathbb{J}$ and 
$\mathbb{R}$ introduce  50 degrees of freedom in the analysis \cite{heinesen_2021,Kalbouneh_p3}. By fitting all independent multipoles to the data simultaneously, one is faced with statistical degeneracies that prevent accurate parameter determination \cite{Dhawan_Borderies_Macpherson_Heinesen_2022,Cowell_Dhawan_Macpherson_2023}.
Additional arguments, based on the symmetries of the expansion velocity field and the vanishing  of the fluid  4-acceleration, help circumvent this problem.  We start by recognizing the axisymmetric configurations for the expansion rate fluctuations, through the analysis of the multipolar structure of the $\eta$ signal. Then we neglect terms of order ${\cal O}[(\mathbb{H}_2/\mathbb{H}_0)^2]$ and ${\cal O}(z^3)$. Consequently, the fitting parameters are consistently reduced down to a total of 12: 2 d.o.f. related to $\mathbb{H}$, 4 to $\mathbb{Q}$ and 6 to $\mathbb{J}-\mathbb{R}$.  A second major simplification is induced by the fact that some multipoles of  the covariant cosmographic parameters (cf. equations (\ref{Qeff10},\ref{Reff10},\ref{Jeff10})) contain spatial gradients, as shown by  \cite{Heinesen_Macpherson_2022}.  The magnitude of those  multipoles is thus increased by factors inversely proportional to the typical spatial scales of the inhomogeneities. For example, we show in Appendix \ref{app_alX_eqns} that the odd multipoles of $\mathbb Q, \mathbb{R}$ and $\mathbb{J}$, containing a spatial derivative, are proportional to $1/\xi_H$ (with  $\xi_H \ll 1$) while the monopole, the quadrupole and the hexadecapole of $\mathbb J$, containing a second order spatial derivative,  scale as $1/\xi_H^2$. Because of this only terms containing  powers of  $(z/\xi_{H})$ contribute significantly in the expansions (\ref{monom},\ref{dipe},\ref{quade},\ref{octe},\ref{hexe}). 

The same argument also allows to resolve the degeneracy between $\mathbb{J}$ and $\mathbb{R}$ appearing when truncating the expansion of $\eta$ at order $\mathcal{O}(z^3)$: the contribution of the latter parameter being always much smaller with respect to that of the former, is always negligible for all $\ell \ge 1$ (in principle we never discard the monopoles  since we expect to recover the FLRW cosmographic parameters in the limit of small expansion rate perturbations). 
Summing up, for each covariant cosmographic  parameter,  on top of the monopoles,  we  consider, in the fitting scheme, only the multipoles $\ell \ge 1$ that  have a maximum number of spatial derivatives ({\it dominant} multipoles).   We can therefore further reduce the total number of free-fitting parameters to 9. Note that the velocity $v_o$ of the matter frame relative to the CMB frame is also counted in this number.

The optimal analysis scheme therefore consists in decomposing $\tilde\eta$ that would be measured in the CMB frame into spherical harmonics; then we compare each multipole with theoretical predictions via a likelihood analysis in order to constrain the free parameters of the model. These parameters are
$\boldsymbol{X}=\big(v_o, \mathbb{H}_0, \mathbb{H}_2, \mathbb{Q}_0, \mathbb{Q}_1, \mathbb{Q}_3, \mathbb{J}_0-\mathbb{R}_0, \mathbb{J}_2, \mathbb{J}_4\big)$.
This is achieved in practice by minimizing the $\chi^2$ statistic
\begin{equation}  \chi^2(\boldsymbol{X})=\sum_{i=1}^{N_{\rm sh}} \left[\frac{\tilde{\mathcal{M}}(\boldsymbol{X},\tilde z_i)-\tilde{\mathcal{M}}_i}{\sigma_{\tilde{\mathcal{M}}_i}}\right]^2+\sum_{\ell=1}^{\ell_{\rm max}} \sum_{i=1}^{N_{\rm sh}}\left[\frac{\tilde \eta_{\ell}(\boldsymbol{X},\tilde z_i)-(\tilde\eta_{\ell})_i}{\sigma_{(\tilde\eta_{\ell})_i}}\right]^2,
\label{chi2}
\end{equation}
where $N_{\rm sh}$ is the number of redshift shells, indexed by $i$, while $\ell$ labels each multipole ($\ell_{\rm max}=4$). The theoretical models for $\tilde{\mathcal{M}}$ and $\tilde \eta_{\ell}$ are obtained by combining equations (\ref{monom}), (\ref{dipe}), (\ref{quade}), (\ref{octe}) and (\ref{hexe}) with equations (\ref{alm}) -- and by setting $\mathbb Q_2=\mathbb J_1-\mathbb R_1=\mathbb J_3-\mathbb R_3=0$, as discussed above. The method for estimating the multipoles 
$\tilde{\mathcal{M}}_i$ and $(\tilde{\eta}_{\ell})_i$,
and their uncertainties in each shell $i$, is detailed in Appendix \ref{forleg}: see eqs. (\ref{a0err}) and (\ref{alerr}).

The accuracy of this inference scheme is tested using the catalogs $M1$ and $M2$. In Figure \ref{sim_m_d_q_M12} (left panels), we show the result of fitting the redshift evolution of the lowest multipoles of $\tilde{\eta}$, together with the normalizing factor $\mathcal{\tilde{M}}$, by means of the $M1$ dataset. 
Figure \ref{sim_m_d_q_M12} (right panels) presents similar results obtained from the analysis of the $M2$ catalog. In Table \ref{tab_al}, we quote the best-fitting values of the covariant cosmographic parameters and compare them to the true input values, for both models $M1$ and $M2$.

Figure \ref{sim_m_d_q_M12} shows how well the 
discrete objects simulating the ZTF survey trace the underlying continuous (exact) field and how well the
multipoles of $\eta$ are reconstructed using the expansion in covariant cosmographic parameters -- {\it i.e.,} by 
neglecting $\mathcal{O}(z^3)$ terms in eqs. (\ref{monom}), (\ref{dipe}), (\ref{quade}), (\ref{octe}) and $(\ref{hexe})$.
For the analytical model $M1$, we see that the goodness of the fit is satisfactory at low redshift, but breaks down at $z \approx 0.045$. This value is not an insignificant one: it is approximately the distance of the large density peak simulated in the model $M1$, the scale at which strong nonlinearities in the distance-redshift relationship emerge.
The causes of these nonlinear effects and their effect on cosmographic reconstruction are discussed in detail in the next section. We observe that the existence and value of this break-down scale can be easily recognized from the data, without any prior knowledge of the magnitude and size of matter fluctuations in the local Universe.

In the case of the large-scale bulk model $M2$, where the attractor lies on the edge of the region covered by the simulated data  ($0.01<\tilde z<0.1$), the cosmographic expansion is even more effective, describing the evolution of the signal over the entire redshift interval analyzed. As a consequence, the value of the $\chi^2$ statistic becomes highly significant. We find $\chi^2=49.2$ for $9\times5-9=36$ degrees of freedom, implying that the probability of finding a higher $\chi^2$ value is around $7\%$.

In Figures \ref{cospar_M1} and \ref{cospar_M2} we display the 
best-fitting values for the lowest multipoles of the covariant cosmographic parameters $\mathbb{H}$ (monopole and quadrupole), $\mathbb{Q}$ (monopole, dipole, and octupole), and $\mathbb{J}-\mathbb{R}$ (monopole, quadrupole, and hexadecapole). We also show how well these estimates approximate the simulated input values. Note that the cosmographic parameters characterize the structure of spacetime at the position of the observer. However, it is not feasible to constrain their amplitude using measurements of $\eta$ at $z=0$. 
Their value is extrapolated from $\eta$ measurements made in 
spherical shells centered on the observer and with variable width. The upper boundary is fixed at the maximum redshift (at which the fit is not rejected by the $\chi^2$ test), while the lower boundary of the shell, $\tilde z_{\rm min}$, is left free to vary. The smaller $\tilde z_{\rm min}$ becomes, the more data are used in the likelihood analysis, and the closer the best-fitting parameter is to the actual value simulated at $z=0$.
As a matter of fact, virtually all the estimates of the cosmographic parameters converge to the simulated values as $\tilde z_{\rm min}\to 0$. 
The statistical uncertainties of the reconstructed parameters decrease as the volume in which the expansion rate fluctuation is calculated grows larger. Note however that estimates for different values of $\tilde z_{\rm min}$ 
are correlated, given the cumulative nature of the binning strategy.

It is important that the likelihood analysis is performed also by fitting the normalization factor $\mathcal{M}$. 
This quantity could be biased in principle because it contains information on the zero-point calibration of brightness distances. It is also sensitive to possible systematic, distance-dependent errors in the estimation of the distance modulus. If, for conservative reasons, we exclude the $\mathcal{M}$ function from the likelihood analysis, we cannot any more provide an independent estimate of the amplitudes of the monopole and quadrupole of the generalized Hubble parameter $\mathbb{H}$. Only the ratio $\mathbb{H}_2/\mathbb{H}_0$ (shown in the upper right plots in Figure \ref{cospar_M1} and \ref{cospar_M2}) can be constrained in this case. By contrast, the accuracy in estimating the multipoles (with $\ell>0$) of $\mathbb Q$ and $\mathbb{J}$ is not affected by the analysis strategy. As a consequence, any systematic miscalibration of the normalization of galaxy distances will not have any impact in the reconstruction of the multipoles of $\mathbb{Q}$ and $\mathbb{J}$.

As shown in Table \ref{tab_al} and also in Figures  \ref{cospar_M1} and \ref{cospar_M2}, 
the  dipole and octupole of $\mathbb{Q}$ are recovered with a significant signal-to-noise ratio.  We note, however, that $\mathbb Q_3$ (and also $\mathbb J_4$) are  mis-recovered (for both  $M1$ and $M2$). As a  consequence of the subdominance of certain multipoles, the only term that contributes to the amplitude of $\eta_3$ (cf. eq. (\ref{octe})) is the linear one (i.e. $\mathbb Q_3$), while the amplitude of $\eta_4$ is effectively determined only by $\mathbb{J}_4$ (cf. eq. (\ref{hexe})). Therefore, the theoretical form of the octupole and hexadecapole of the expansion rate fluctuation field is controlled by only one monomial of the redshift, which is a severe limitation in the ability to reproduce the observed redshift evolution of the data. 
The optimal way to tackle this issue is to include in the analysis the fourth and fifth derivatives of distance (snap and crackle) in the expansion (\ref{defdL0}). Here, instead, we simply parametrize our ignorance by means of two nuisance parameters $\mathbb A$ and $\mathbb B$ over which we subsequently marginalize in the statistical analysis. Although in this way we  overall increase the statistical uncertainty in the best fitting parameters $\mathbb{Q}_3$
and $\mathbb{J}_4$, we  significantly reduce the systematic error. 
We thus add,  in accordance with an educated guess about the symmetry of the problem,  a term $\mathbb A\, z^3$ to eq. (\ref{octe}) and a term $\mathbb B\,z^4$ to eq. (\ref{hexe}) and re-do the likelihood analysis (for a more refined analysis see \cite{Kalbouneh_p3}).   
For $M1$, we obtain  $\mathbb Q_3= 1.0\pm0.6$ and $\mathbb J_4= 94\pm 69$. As expected the discrepancy between the recovered octupole of $\mathbb{Q}$
and its simulated value decreased from $4.9\sigma$ to $1.5\sigma$. As for the hexadecapole of $\mathbb{J}$, the deviation from the expectation reduces from $4.1\sigma$ to $0.7\sigma$ (the best fitting values for nuisance parameters are $\mathbb A=-19\pm89$ and $\mathbb B=(-1\pm3) \times 10^3$ respectively.
For $M2$, the new analysis returns  $\mathbb Q_3= 0.36\pm0.13$ and $\mathbb J_4= 32\pm 7$ and the tension decreases from $5.6\sigma$ to $1.2\sigma$, and from $6.6\sigma$ to $1.8\sigma$, respectively (the best fitting values for nuisance parameters are $\mathbb A=-4\pm4$ and $\mathbb B=-276\pm71$ respectively.

The velocity $v_o=320$ km/s of the matter relative to the CMB is reconstructed quite well from the analysis of both simulated catalogs $M1$ and $M2$. 

\begin{figure}
  \includegraphics[scale=0.6]{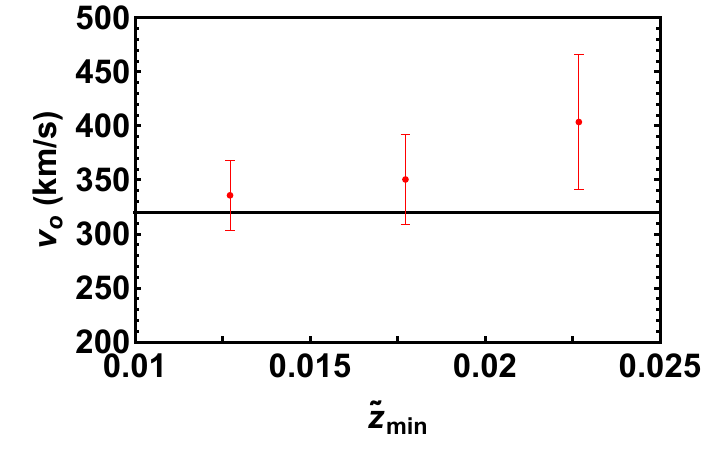}
  \includegraphics[scale=0.6]{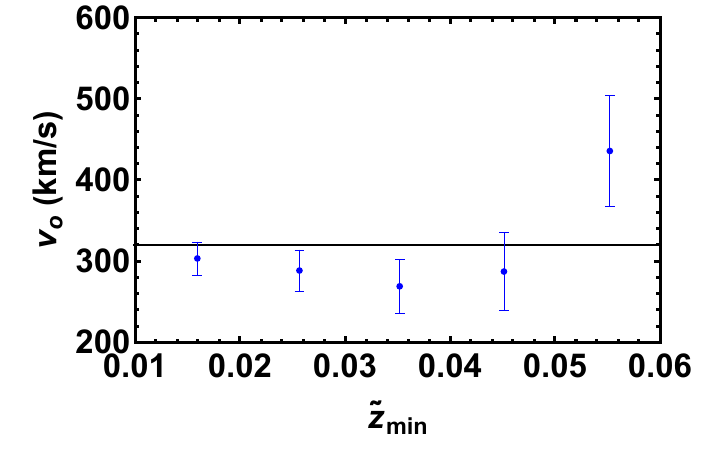}
   \caption{The best-fitting velocity (and the associated $1\sigma$ errorbar) of the matter frame relative to the CMB frame at the event of observation. {\it Left:} Results from the analysis of the $M1$ simulation in shells $\tilde z_{\rm min}<\tilde z<0.045$. {\it Right:} Results from the analysis of the $M2$ simulation in shells $\tilde z_{\rm min}<\tilde z<0.1$.}
  \label{vobs}
\end{figure}

\begin{figure}
  \centering
  \includegraphics[scale=0.45]{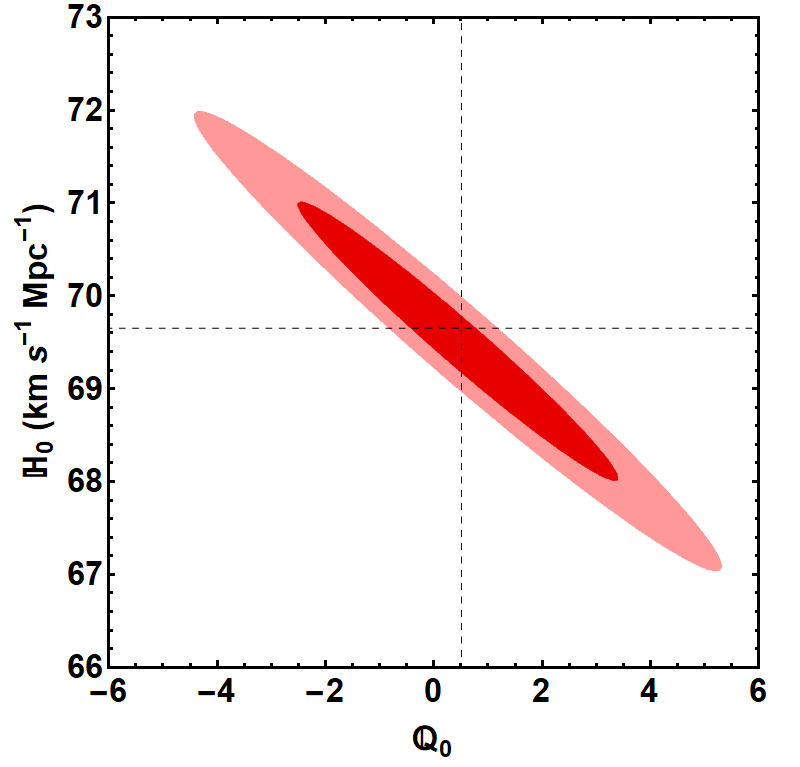}
  \includegraphics[scale=0.45]{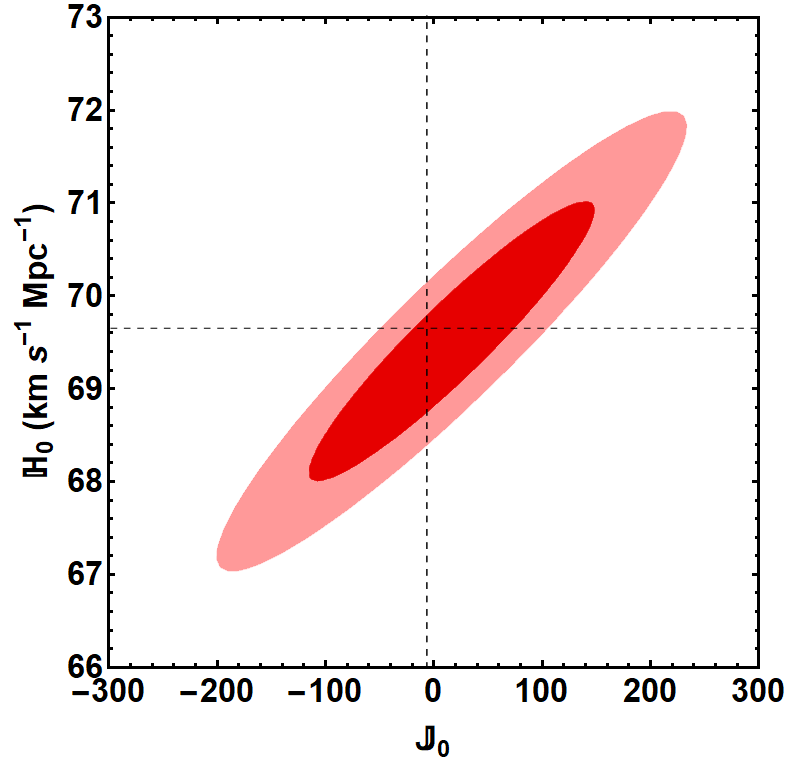}
  \includegraphics[scale=0.45]{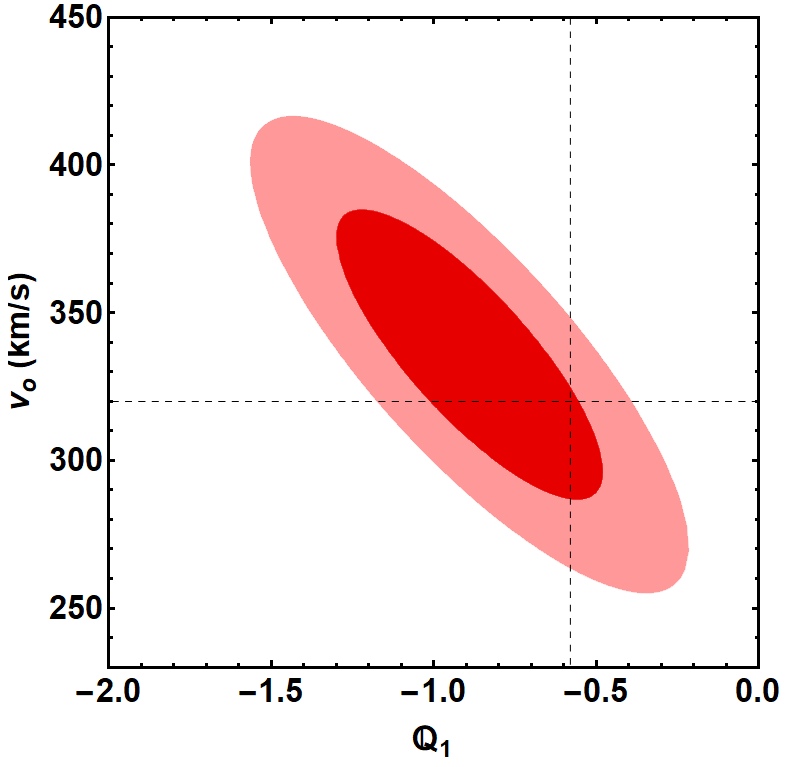}
  \includegraphics[scale=0.45]{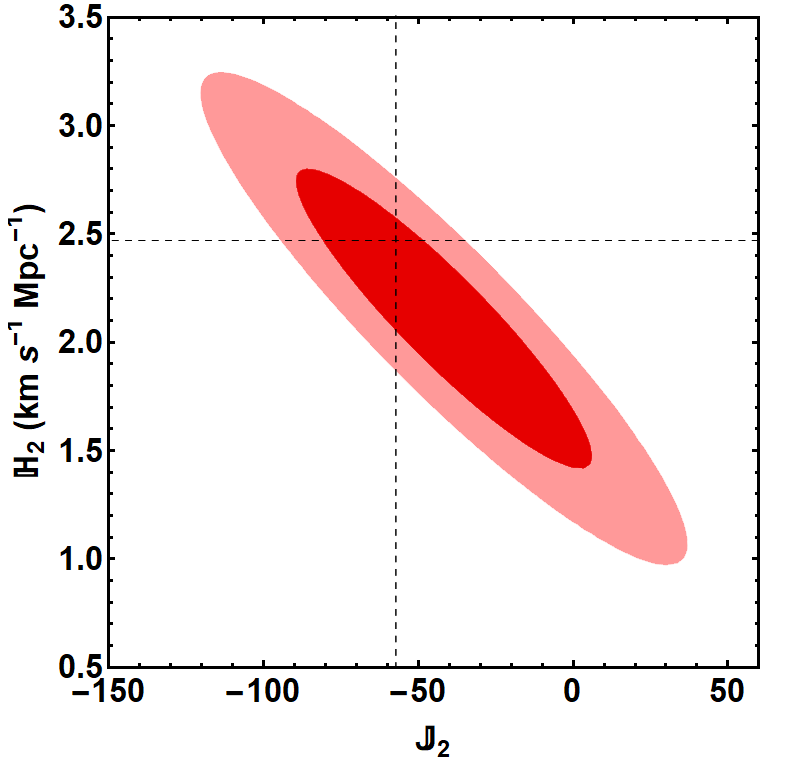}
  \caption{The $68\%$ and $95\%$ confidence level contours in various 2D parameter spaces obtained from the likelihood analysis of the $M1$ simulated data ($\tilde z<0.045$). The intersection of the two dashed lines indicates the input value.}
  \label{elps_err_M1}
\end{figure}

\begin{figure}
  \centering
  \includegraphics[scale=0.44]{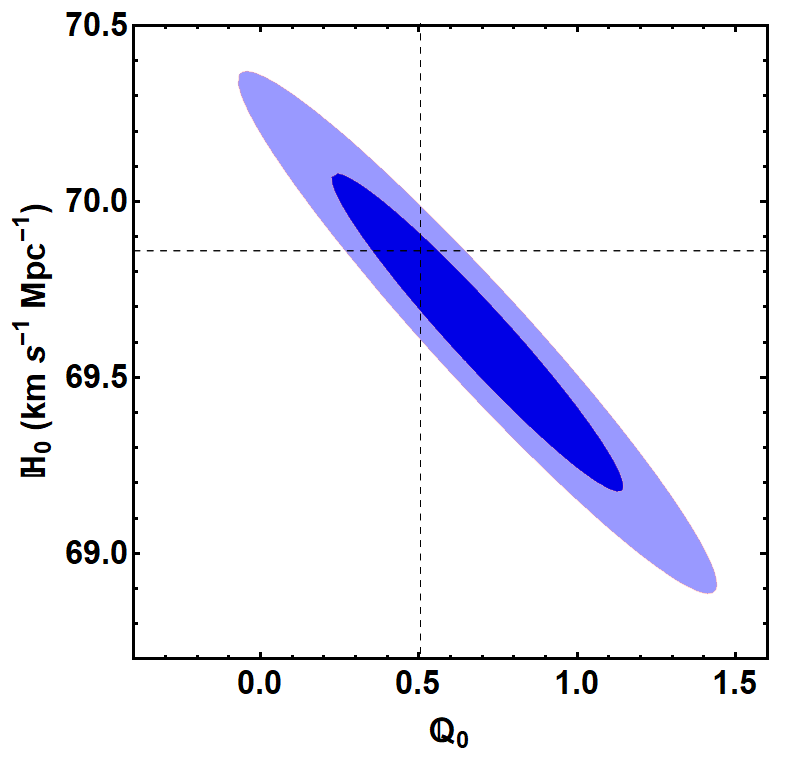}
  \includegraphics[scale=0.44]{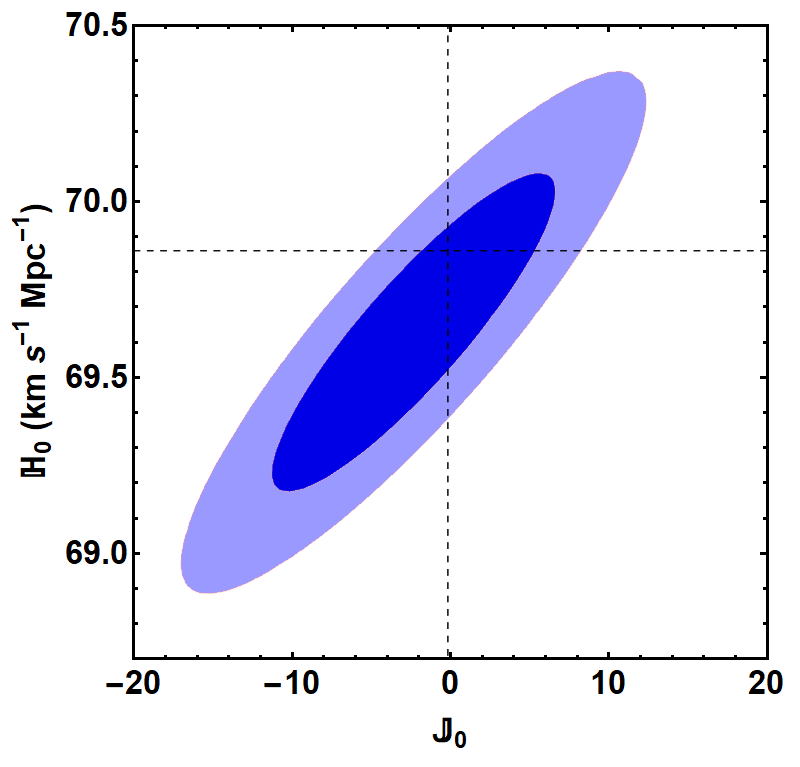}
  \includegraphics[scale=0.44]{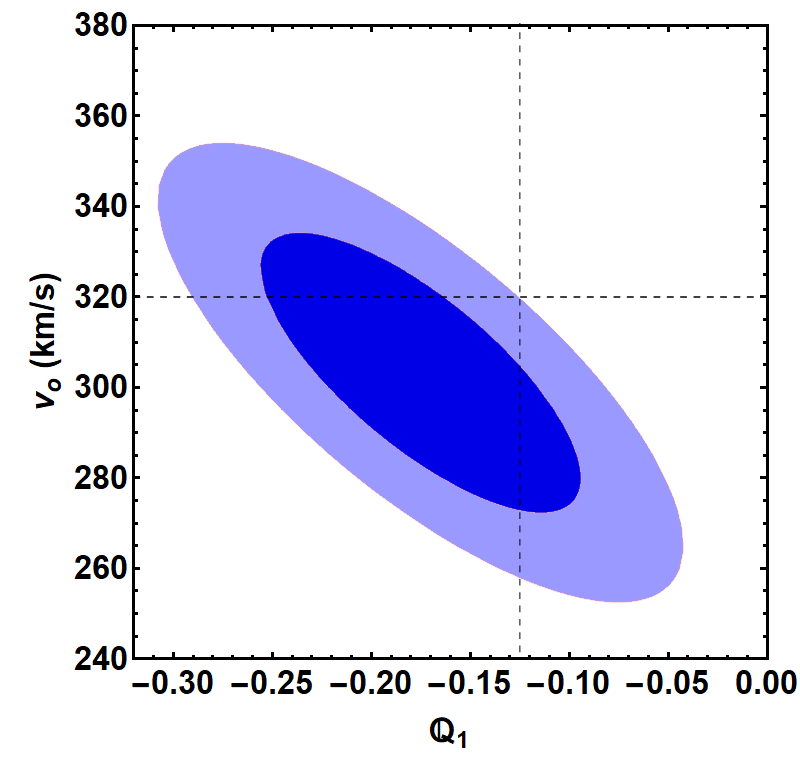}
  \includegraphics[scale=0.44]{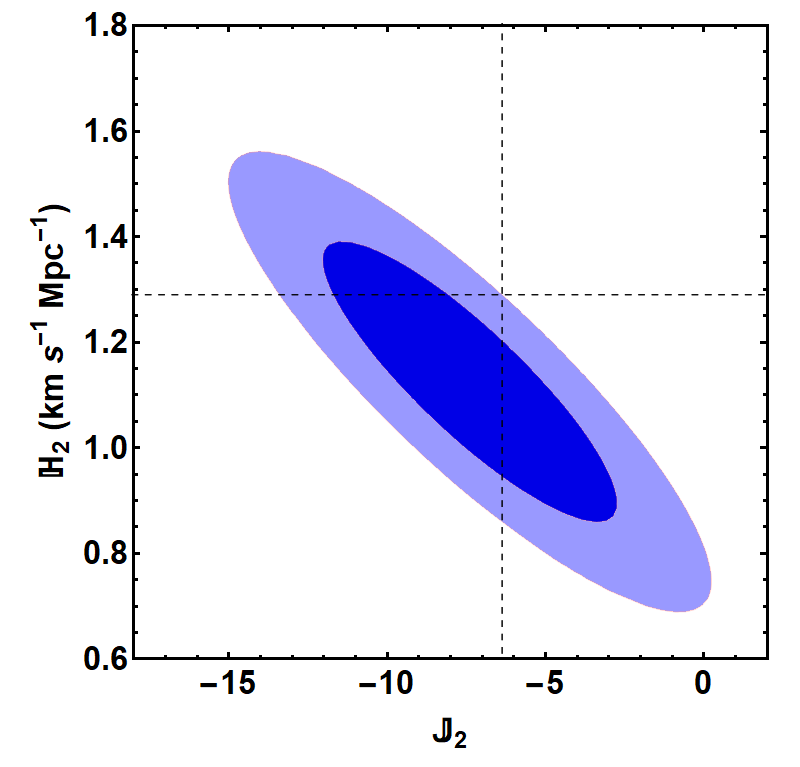}
  \caption{The $68\%$ and $95\%$ confidence level contours in various 2D parameter spaces obtained from the likelihood analysis of the $M2$ simulated data ($\tilde z<0.1$). The intersection of the two dashed lines indicates the input value.}
  \label{elps_err_M2}
\end{figure}

Figure \ref{vobs} shows that the relative error of the best-fitting velocity is about $10\%$ and $7\%$ for the $M1$ and $M2$ scenarios respectively. 
The same caveat as before applies. 
If $v_o$ is inferred from the analysis of  $\tilde\eta$ reconstructed in shells at large distance from the observer, the estimation is affected by large uncertainties. However, as soon as the lower boundary $\tilde z_{\rm min}$ of the shell in which $\tilde{\eta}$ is reconstructed approaches zero, the estimates of the local covariant cosmographic parameters stabilize and in the limit converge to the input simulated values. 

Parameter estimation is clearly affected by statistical degeneracies. These are visible in Figures~\ref{elps_err_M1} and \ref{elps_err_M2} which show the 2-dimensional $68\%$ and $95\%$ likelihood contours of relevant cosmographic parameters. The analysis of $M1$ and $M2$ suggests that the accuracy of statistical inference depends critically on the structure of the expansion rate fluctuations, rather than on the quality and precision of the distance measurements. 
The higher the redshift at which the cosmographic expansion breaks down, i.e., the wider the redshift interval in which the amplitude of $\eta$ can be approximated in a model-independent way by an expansion in powers of $z$, the smaller the uncertainties in the recovered cosmographic parameters. The smoother and more spread over a large volume are the perturbations in the local expansion rate (model $M2$), the better the precision of the recovered parameters will be -- compared to the case where large perturbations are close to the observer's position (model $M1$). 

Nevertheless, even in the latter `pessimistic scenario', data from a future survey like the ZTF are sufficient to constrain the relevant covariant cosmographic parameters with a fair degree of precision. The parameters that one expects to better reconstruct are the local value of the monopole of the covariant Hubble parameter $\mathbb{H}_0$ (with an optimistic/ pessimistic 
 resolution of $0.3/1.0$ km s$^{-1}$ Mpc$^{-1}$), 
the quadrupole of $\mathbb{H}$ ($\sim 4 \sigma$ detection in the pessimistic scenario, i.e., a relative uncertainty of $24\%$), and the dipole in $\mathbb{Q}$, which is recovered with nearly a $33\%$ precision under the most conservative assumptions. 

The degree of reliability of the predictions depends strictly on the degree of realism of the analytical model. The constraints on the multipoles of the covariant cosmographic parameters could deviate from expectations even if the specific parameters of future surveys were significantly different from those simplistically assumed in this study. Nevertheless, the results obtained are encouraging and suggest the possibility of obtaining fairly precise estimates of the covariant cosmographic parameters at scales $(z \lesssim 0.1)$ where the CP ceases to be applicable, and of characterising the expansion rate of the local universe in a more meaningful and complete way than is possible using the $H_0$ parameter of the Standard Model alone.

\section{Limits of the cosmographic expansion} \label{sec_limits}

The tensorial expansion of redshift as a function of distance, which lies at the heart of covariant cosmography, demands smoothness and single-valuedness. Therefore, the applicability of these assumptions in the local, nonlinearly perturbed universe must be critically analyzed.

\begin{figure}
  \centering
  \includegraphics[scale=0.6]{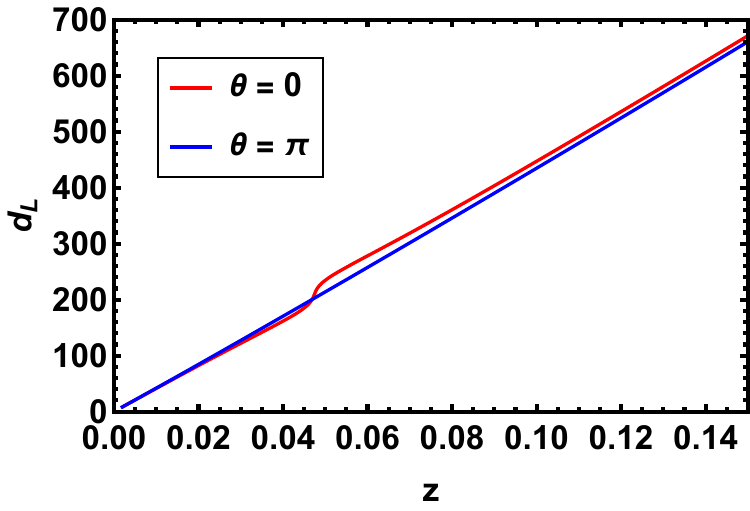}
  \includegraphics[scale=0.6]{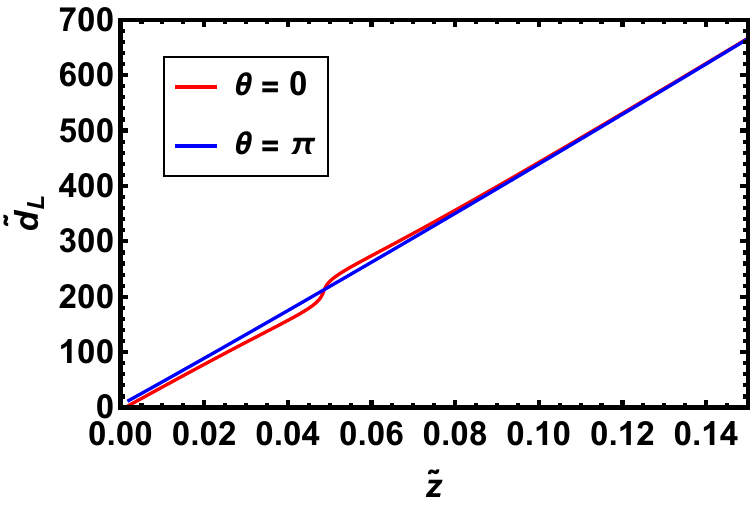}
  \caption{Luminosity distance in the $M1$ model, calculated using linear perturbation theory for a matter-comoving observer (left) and a CMB-comoving observer (right). The scaling with redshift along two different line-of-sight directions is shown: towards the center of the density peak ($\theta=0$, solid red line) and in the antipodal direction ($\theta=\pi$, solid blue line).}
  \label{dL_2r_1}
\end{figure}

The luminosity distance $d_L$ to a source at redshift $z$ depends on the direction of the line of sight, whether calculated in a model-independent cosmographic approach or in a linearly perturbed FLRW model. 
In Figure \ref{dL_2r_1} we display its redshift scaling in the direction of the center of the mass overdensity ($\theta=0$), and the opposite direction ($\theta=\pi$), for the matter- and CMB-comoving observers, at a distance of $r_o=200\,$Mpc from the density peak of the $M1$ model. 

The luminosity distance shows a characteristic elongated $S$-shaped trend along the observer's line of sight, bending approximately at the position of the density peak. This feature is induced by the radial component of the peculiar velocity of the emitters -- which increases/ decreases relative to the background cosmological value for the observed redshift of a source placed before/ after the density peak. If the density peak is sufficiently large (not far from our $\delta_c$), the luminosity distance becomes a multivariate function of $z$, and the validity of the cosmographic expansion breaks down. 
Even before reaching this pathological configuration, 
the more nonlinear the density field, the poorer becomes the reconstruction of the distance using only lower-order expansion terms.

\begin{figure*}
  \centering
  \includegraphics[scale=0.6]{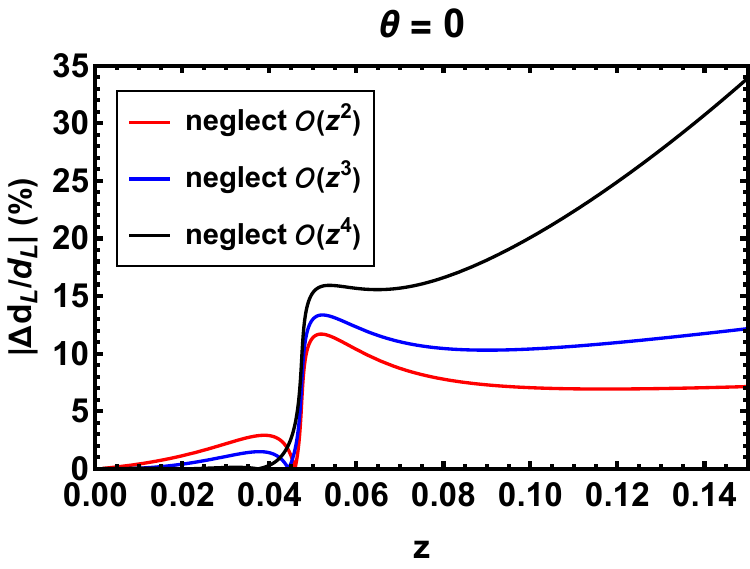}
  \includegraphics[scale=0.6]{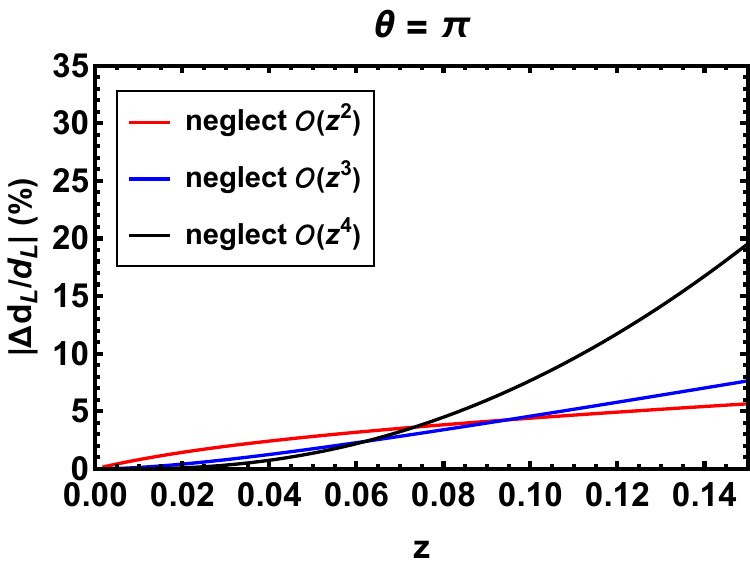}
  \caption{The relative discrepancy between different orders of expansion of the luminosity distance (measured by a matter comoving observer) and the true input value (model $M1$), in the direction of the centre of the density peak (left) and in the opposite direction (right).}
   \label{dLerr1}
\end{figure*}

Figure \ref{dLerr1} compares the luminosity distance reconstructed using the cosmographic approach against that reconstructed using linear perturbation theory. The relative discrepancy is shown along two line-of-sight directions ($\theta=0$ and $\theta=\pi$) and for different expansion orders in redshift.

The cosmographic expansion of $d_L$, eq. (\ref{defdL0}), works extremely well locally, around the observer, and along any line of sight. It improves the higher the order of expansion is. However, already at relatively low redshifts, $z \approx 0.05$ in the case
of the $M1$ model, its accuracy is severely limited if the field is dominated by a nearby large density peak. In this configuration, the luminosity distance deviates considerably from linearity and the cosmographic parameters acquire a large amplitude to allow an accurate description of these distortions in regimes where $z\ll 1$. 

In fact, a truncated polynomial expansion at a low order is ineffective in capturing these nonlinear features. It is therefore essential to know at what scales the covariant cosmographic expansion fails if it is to be successfully applied in the highly irregular regions of the local Universe. As expected, if the gravitational source is located far from the observer, as in the case of the $M2$ model, the redshift range in which cosmography can be safely applied increases accordingly. The combined analysis of the independent multipoles of $\eta$, as explained in the previous section, makes it easy to identify the scale at which the covariant cosmographic approach breaks down.

\section{Multipoles of the radial peculiar velocity field}\label{sec_multiv}

After showing how to recover model-independent information about the spacetime surrounding the observer, we move on to interpret $\tilde{\eta}$ anisotropies in a more conventional framework, as small deviations from the FLRW metric. 
In this scenario, fluctuations in $\tilde\eta$,
instead of providing information on the covariant cosmographic parameters, and thus on the geometric structure of the geodesic congruence representing the cosmic  fluid,  simply give an insight into the amplitude and distribution of the radial peculiar velocity field, $v_n \equiv \boldsymbol{v}\cdot \boldsymbol{n}$, of galaxies relative to an idealized matter frame -- which coincides with the CMB frame in perturbed FLRW. If we ignore the gravitational potential in the perturbed FLRW model, we have $\tilde z=z_c+v_n(1+z_c)$ and $\tilde d_L=d_{L}^{c}(1+2v_n)$. As a consequence, we can express $\tilde\eta$ as
\begin{equation}
\tilde\eta=\log\left[\frac{z_c+v_n(1+z_c)}{d_L^c(1+2v_n)}\right]-\tilde{\mathcal{M}}=\log\left(\frac{z_c}{d_L^c}\right)-\tilde{ \mathcal{M}} + \log\left[1+\frac{v_n}{z_c}(1-z_c)\right].
\end{equation}

If $v_n\ll \tilde z$, the multipoles of the radial component of the peculiar velocity field are related to the multipoles of the expansion rate fluctuation field $\tilde\eta$ (for $\ell>0$) as
\begin{equation}
v_{n\ell}(\tilde z)=\tilde\eta_{\ell}(\tilde{z})\,\tilde{z}\,\ln10\,.
\end{equation}
The average of the multipoles of $v_{n}$ over a spherical volume of radius $R$ is
\begin{equation}
\big\langle v_{n\ell} \big\rangle_R =\frac{\ln10}{V(R)}\,{\int_0^{\tilde z(R)}\tilde\eta_\ell\; \tilde z \; \mathrm{d}V}\,,
\label{vl_th}
\end{equation}
where $\tilde z(R)\approx H_0 R$, and $V$ is the volume. Note that $\langle v_{n1}\rangle_R \equiv v_{\rm b}$ corresponds to what is traditionally called in the literature the bulk velocity
(cf. eq. (2) in \cite{Adi_Nusser_2015}) that is a measure of the common radial streaming velocity shared by all the fluid elements inside the given region. To avoid possible confusion, it is worth stressing that $v_o$, the observer velocity with respect to CMB (see section \ref{v_obs}), is not conceptually, and does not coincide numerically, with the bulk velocity of a spatial volume. Nevertheless, when the volume inside which the bulk velocity is reconstructed is decreased, i.e., when $R \rightarrow 0$, the bulk velocity of matter in that volume well approximates the motion of the matter frame relative to the CMB frame.

\begin{figure}
\centering
  \includegraphics[scale=0.62]{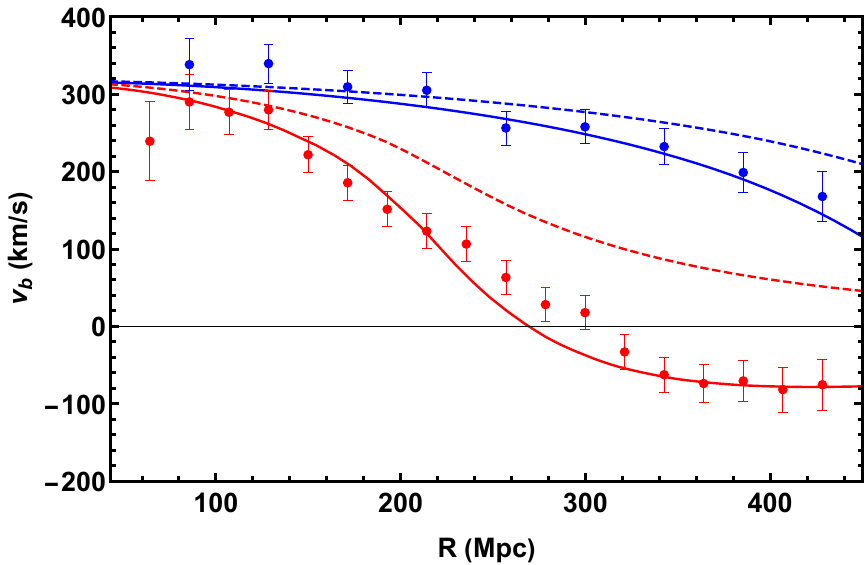}
  \\
  \includegraphics[scale=0.62]{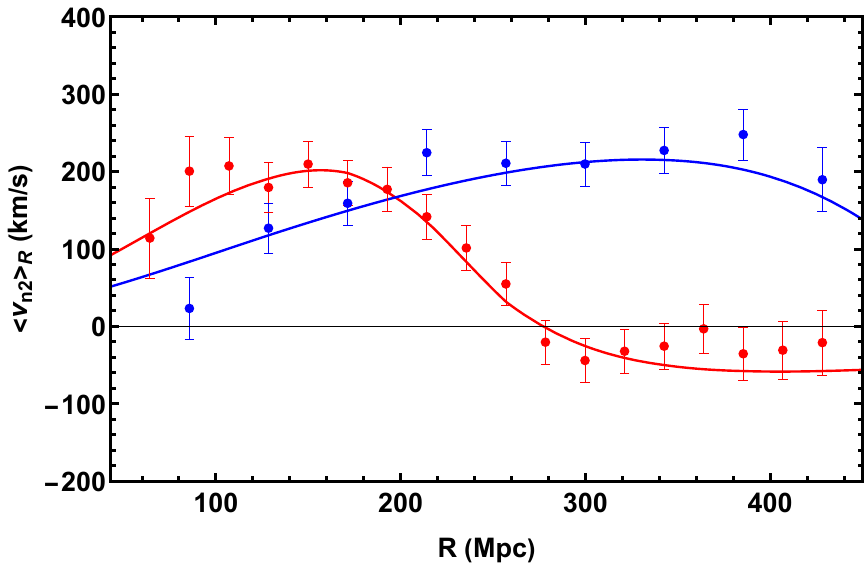}
  \\
  \includegraphics[scale=0.62]{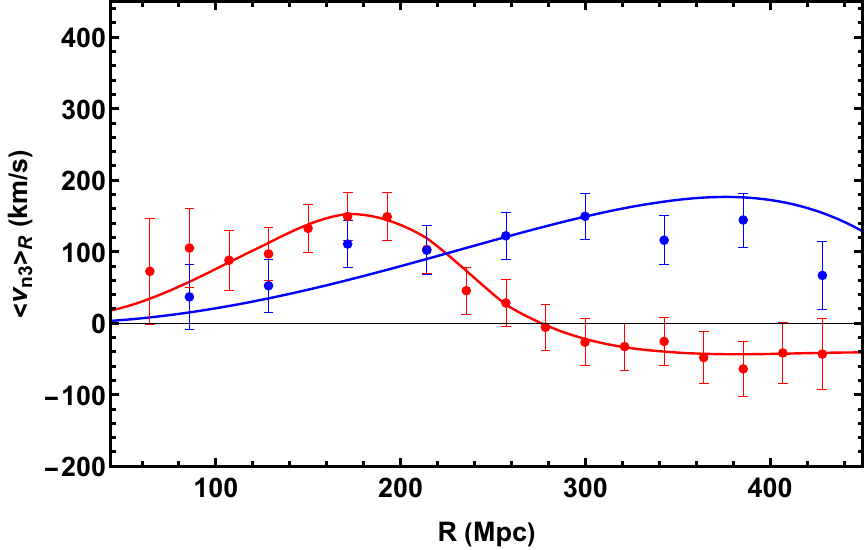}
   \caption{The bulk velocity $v_{\rm b}=\langle v_{n1}\rangle_R$ {\it {(top)}}, together with the mean value of the quadrupole {\it(centre)} and octupole {\it {(bottom)}} of the peculiar radial velocity field, estimated using eq. (\ref{vl_th}), are shown as a function of the radius $R$ of the averaging volume. Dots represent the value inferred from discrete sampling (15,000 objects), while the continuous lines represent the underlying continuous field value. 
   The red/ blue colors indicate results from the $M1/ M2$ simulations. The dashed curves in the left panel present the estimates of the bulk velocity reconstructed by averaging the $3D$ peculiar velocity field,
   $v_{\rm b}=\langle v \rangle_R$.} 
   \label{bulkvel}
\end{figure}

Figure \ref{bulkvel} shows how the reconstructed bulk velocity and the average higher multipoles of the radial velocity field, estimated using eq. (\ref{vl_th}), compare with the simulated values. The evolution, for both small-scale and large-scale velocity models, is shown as a function of the radius $R$ of the spherical volume over which data are averaged. 
Note that in model $M1$ the bulk velocity decreases sharply once the averaging volume becomes large enough to contain the density fluctuation generating the peculiar motions. The bulk velocity becomes negative when the radius of the volume in which it is reconstructed becomes larger than the distance (from the observer) of the dominant inhomogeneity generating the peculiar velocity field. 
A similar fast asymptotic transition also characterizes the volume average of the quadrupole and octupole of the radial velocity fields. These latter quantities vanish if the volume becomes sufficiently large to contain the dominant inhomogeneity. 

By contrast, in model $M2$ the slow decrease of the bulk velocity with the depth of the survey is accompanied by a systematic increase of the higher velocity moments. This is because the latter systematically peak at about the position of the large mass concentration that generates the streaming motions. 
Higher-order moments in the expansion rate fluctuation field $\eta$ are therefore instrumental in investigating the structure and extent of density inhomogeneities, making it possible to complement or corroborate evidence about the convergence (or not) of matter flows in large regions of the Universe.

Figure \ref{bulkvel} (upper panel) also shows the bulk velocity defined as the volume average of the three-dimensional peculiar velocity vectors (see eq. (1) in \cite{Adi_Nusser_2015}) as opposed to that reconstructed from the radial component alone.
 Interestingly, the two estimates converge up to the common scales where both the quadrupole and octupole of the radial velocity field stop growing in a monotonic manner and reverse. In volumes with radii larger than this scale, the estimate obtained from the radial component of the peculiar velocities systematically underestimates the bulk velocity reconstructed from 3D information.
 We defer to further analysis whether this is a generic conclusion or depends on the specific asymmetric configuration of the model considered.

\section{Conclusions}\label{sec_conclusion}

In local regions of the Universe, matter density fluctuations are so large in terms of amplitude and spatial scales that the application of the Cosmological Principle and the cosmological laws derived from it is problematic. Perturbation theory is certainly a powerful method for modelling and interpreting deviations from ideal uniform conditions.
However, this approach has the drawback of requiring the assumption of a background cosmological metric. Can inhomogeneities and anisotropies be quantified in a model-independent way? Are there ways to let the data speak for themselves without assuming a priori FLRW scenarios or splitting the observed signals into a background and a perturbed component? Can we access the physical content of the measurements in a fully covariant way (see \cite{Maartens:2023tib}) and yet determine the motion of the observer with respect to the surrounding matter frame? 

We address these challenges using the expansion rate fluctuation field $\eta$, a statistically unbiased, Gaussian, scalar field that is sensitive to deviations from isotropy of the redshift-distance relationship. In \cite{kalbouneh_marinoni_bel_2023} it is shown how to extract the multipole moments of this observable. By analyzing its structure, it is found that unexpected symmetries characterize the local spacetime. Here we show that, in addition to its simple statistical properties, this observable also has a distinctive theoretical appeal. It allows us to constrain relevant cosmological quantities in a fully model-independent way.

Our main conclusions are as follows. 
\begin{enumerate}
\item We provide a formula to predict, up to $\mathcal{O}(z^3)$, the redshift scaling of the expansion rate fluctuation field. Its amplitude depends on the lower-order covariant cosmographic parameters that characterize in a model-independent way the structure of a generic spacetime in the surroundings of the event of observation, namely $ \mathbb{H}$ (Hubble), $\mathbb{Q}$ (deceleration), $\mathbb{J}$ (jerk) and $\mathbb{R}$ (curvature). 

\item The formalism is developed in such a way that it is straightforward to adapt it to observations carried out in a reference frame which is comoving with matter or in an arbitrary frame boosted with respect to it. It is crucial to carefully characterize the frame in which data is observationally measured and the frame in which quantities are theoretically characterized. Indeed the covariant cosmographic parameters at the event of observation have, in general, different amplitudes depending on the observer's state of motion at that event \cite{Maartens:2023tib}. We show that it is practical to estimate the $\eta$ signal in the CMB frame, the frame of an observer who measures no CMB dipole, whatever the physical origin of that dipole is. At the same time, it is theoretically more convenient and natural to interpret the physical content of the signal in the matter frame, that of an observer at rest with respect to the surrounding matter fluid.

\item We have made explicit the dependence of the spherical harmonic components of $\eta$ on the multipoles of the covariant cosmographic parameters for a matter-comoving observer and in an axially symmetric configuration. According to the finding of \cite{kalbouneh_marinoni_bel_2023}, this axial symmetry fairly well describes available data in the local Universe. A schematic summary of these relations is presented in Figure \ref{summa}.

\begin{figure*}
  \centering
  \includegraphics[scale=0.38]{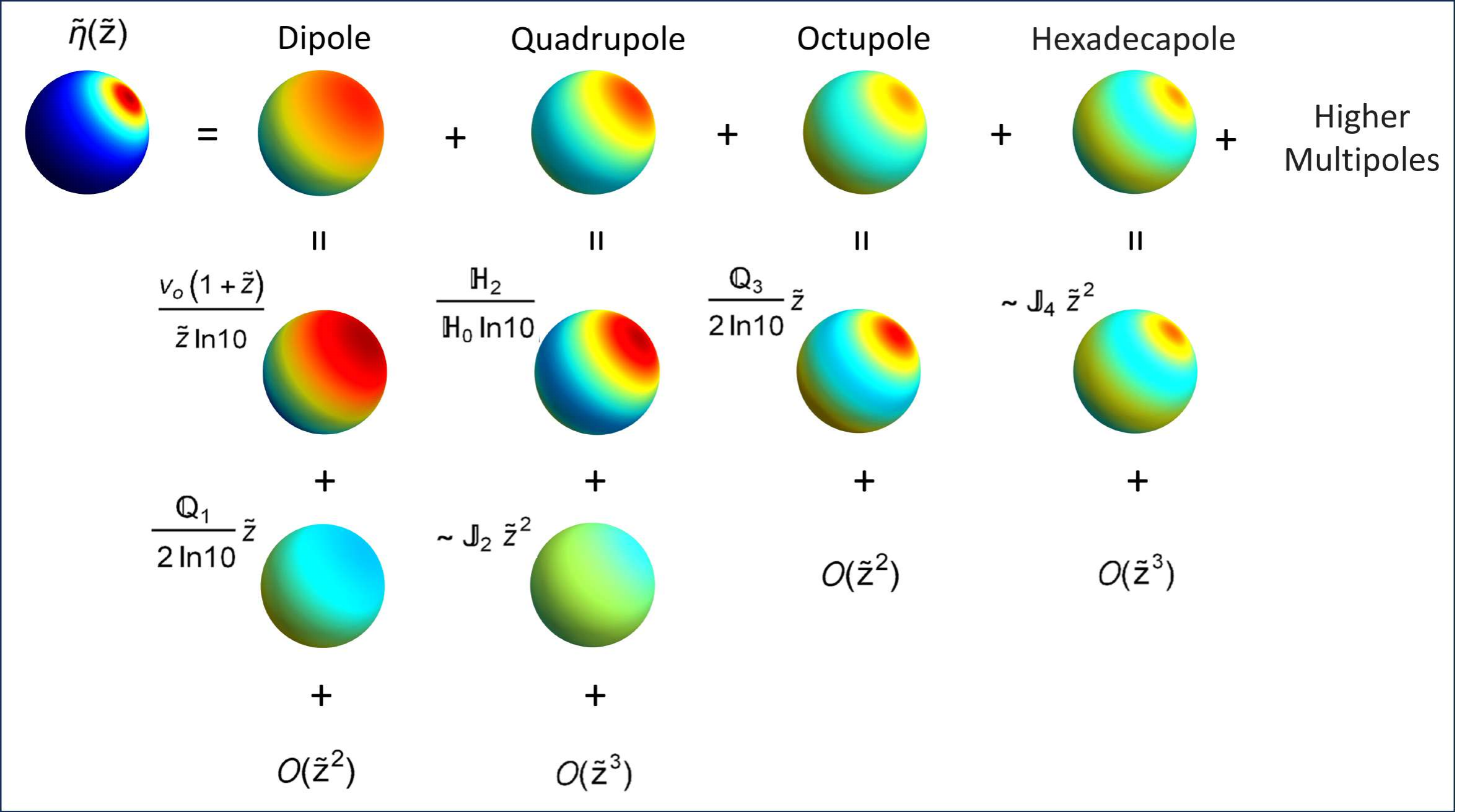}
  \caption{Graphic summary illustrating the decomposition of the expansion rate fluctuation field inferred in the CMB frame ($\tilde\eta$) in a given shell $S$ at redshift $\tilde z$, in terms of its axially-symmetric multipoles (horizontal direction). Each multipole is further related (vertical direction) to the dominant physical quantities contributing to its amplitude: these are the multipoles of the (lower-order) covariant cosmographic parameters and the relative velocity $v_o$ of the observer relative to the matter frame.}
  \label{summa}
\end{figure*}  

The dipole of $\eta$, when reconstructed in the CMB frame, provides unique information about the velocity of the observer relative to the matter frame. 
However, in order to infer the amplitude of this relative motion, it is necessary to disentangle it from the intrinsic distortions contributed by inhomogeneities in the metric. The latter effects are all the more important the deeper the volume analyzed. 
The next multipole of $\eta$, the quadrupole, is mainly sensitive to the ratio between the quadrupole and the monopole of the covariant Hubble parameter ($\mathbb{H}_2/\mathbb{H}_0$), although it also contains information about the quadrupole of the deceleration parameter. 
 
\item By exploiting a simple analytical model of metric perturbations, we highlight the virtues and also the limitations of a power series expansion of $\eta$ as a function of redshift. Surprisingly, we find that the cosmographic expansion works well even in the local Universe, where large deviations in the distance-redshift relation, induced by density fluctuations, are expected. In particular, we show how to use the redshift evolution of $\eta$ multipoles to identify the redshift intervals where the cosmographic expansion breaks down and those where a meaningful theoretical inference can be made. 

\item By randomly sampling the continuous analytical model, we simulate all-sky catalogs of distances (typically containing 15,000 entries in the redshift interval $0<z<0.1$ and with Gaussian error in the distance modulus $\sigma_{\mu} =0.15$). We use them to assess the precision and accuracy with which future datasets, such as that under collection by the ZTF survey, will constrain the velocity of the observer (relative to the surrounding matter frame) and the multipoles of the covariant cosmographic parameters.

\item We also explore the cosmological potential of a more conventional analysis, in which the $\eta$ signal is interpreted in the context of the standard cosmological model. We show that in this case, the multipoles of $\eta$ provide insights into the amplitude of the multipoles of the radial peculiar velocity field of galaxies. Notably, the dipole of $\eta$ extracted in a given volume $V$ is proportional to the amplitude of the bulk motion on that characteristic scale. 
\end{enumerate}

Looking ahead to the future, we aim to build on current work and extend it in complementary directions. From an empirical point of view, we plan to estimate the expansion rate fluctuations from 
updated datasets which have recently become available, i.e., the Cosmicflows-4 catalog \cite{Tully:2022rbj_CF4} and the Pantheon+ sample \cite{Brout:2022vxf_Pantheon+}. 
In addition, we aim to make more accurate predictions of the ZTF survey's potential to constrain covariant cosmographic parameters, taking into account possible covariances in data uncertainties and more accurate maps of the celestial distribution of sources.
From the theoretical side, we aim to exploit these insights into the structure of the local Universe in order to construct a spacetime metric that captures the essential features of the anisotropic $\eta$ signal. 
Finally, from the methodological side, we need to develop reconstruction techniques that allow us to estimate the expansion rate fluctuation field even from data with partial (and anisotropic) sky coverage. The purpose is to extend the multipolar analysis of the redshift-distance relation to the high redshift Universe. 
\vfill

\acknowledgments We would like to thank Julien Bel, Chris Clarkson, Asta Heinesen, Federico Piazza, Jessica Santiago for
useful discussions. BK and CM are supported by the {\it Programme National Cosmologie et Galaxies} (PNCG) and {\it Programme National Gravitation Références Astronomie Métrologie} (PNGRAM), of CNRS/INSU with INP and IN2P3, co-funded by CEA and CNES. They also received support from the {\it Agence Nationale de la Recherche} under the grant ANR-24-CE31-6963-01, and the French government under the France 2030 investment plan, as part of the Initiative d’Excellence d'Aix-Marseille Université -  A*MIDEX (AMX-19-IET-012).
RM is supported by the South African Radio Astronomy Observatory and the National Research Foundation (grant no. 75415).

\newpage

\appendix

\section{Light geodesics in Einstein de Sitter spacetime }\label{app_solgeod}

Consider the FLRW metric 
\begin{equation}
\mathrm{d}s^2=-\mathrm{d}t^2+a^2(t)\left[\mathrm{d}r^2+ r^2(\mathrm{d}\psi^2+\sin^2\psi\, \mathrm{d} \phi^2)\right],
\label{metric_BG}
\end{equation}
with an off-centre observer at a distance $r_o$ from the center of the coordinates (at $r=0$). We choose the $Z$-axis along this direction. 
The past pointing 4-wavevector of light is
\begin{equation}
k^\mu=\frac{\mathrm{d}x^\mu}{\mathrm{d}\lambda}\,,
\end{equation}
where $\lambda$ is an affine parameter. We determine the geodesic path by solving the geodesic equations 
\begin{equation}
\frac{\mathrm{d}^2 x^\mu}{\mathrm{d} \lambda^2}+\Gamma^{\mu}_{\;\;\nu\alpha}\frac{\mathrm{d}x^\nu}{\mathrm{d} \lambda}\frac{\mathrm{d}x^\alpha}{\mathrm{d} \lambda}=0 \,.
\label{geod_eq}
\end{equation}
For the off-center observer, $k^3=0$ by symmetry. We choose the affine parameter 
as the physical distance measured by an observer at rest in the smooth background, with 4-velocity $\tilde u^\mu=\delta^\mu_0$, so that 
$\lambda=0$ at the observer position and $(k^\mu \tilde u_\mu)_o=1$. The initial conditions are 
\begin{equation}
t(0)=t_0\,, \;\;\;\;\;\; r(0)=r_o\,, \;\;\;\;\;\; \psi(0)=0\,, \quad
%
\frac{\mathrm{d} t}{\mathrm{d} \lambda}\bigg|_0 = -1\,,
\;\;\;\;\;\; 
\frac{\mathrm{d}r}{\mathrm{d} \lambda}\bigg|_0 =-\cos\theta\,,
\;\;\;\;\;\; 
\frac{\mathrm{d} \psi}{\mathrm{d} \lambda}\bigg|_0 = \frac{1}{r_o}\sin\theta \,,
\label{kmu0}
\end{equation}
where $t_0$ is the time of observation (today), and 
$\theta$ is the angle between the $Z$-axis and the direction the light ray measured by the observer $\tilde u^\mu$. 
The photon trajectory is thus 
\begin{align}
  t(\lambda)&=t_0 \left(1-\frac{5}{3}\frac{\lambda}{t_0}\right)^{3/5}
  \\
r(\lambda)&=r_o\left\{ 1+6\cos{\theta}\frac{t_0}{r_o}\left[\left(1-\frac{5}{3}\frac{\lambda}{t_0}\right)^{1/5}-1\right]+9\frac{t_{0}^2}{r_{o}^2}\left[\left(1-\frac{5}{3}\frac{\lambda}{t_0}\right)^{1/5}-1\right]^2\right\}^{1/2}\,,
\\
\psi(\lambda)&=\tan^{-1}(\cot\theta)-\tan^{-1}\left\{\cot\theta+3\frac{t_0}{r_o}\left[\left(1-\frac{5\lambda}{3t_0}\right)^{1/5}-1\right]\csc\theta\right\}.
\end{align}
By using eq. (\ref{z_corr}) and eq. (\ref{ddl}), we can then infer the value of the redshift and luminosity distance.

\section{Multipoles of the analytical model cosmographic parameters} \label{app_alX_eqns}

We compute, at leading order of approximation, the explicit form of the multipoles of the covariant cosmographic parameters for the analytical model presented in Sec. \ref{sec_model}. We do this by evaluating 
 (\ref{Heff10}), (\ref{Qeff10}), (\ref{Reff10}) and (\ref{Jeff10}), using the analytical-model prescriptions
(\ref{Kvec1}), (\ref{pec_vel}) and (\ref{phi_r}). In order to simplify the equation, we define the dimensionless quantities $\xi_o\equiv R_s/r_o$ and $\xi_H\equiv R_s/R_H=R_s H_0$. All the multipoles $\ell\geq5$ vanish at linear order in the perturbation parameter $\delta_c$.  The only nonzero multipoles are
\begingroup
\allowdisplaybreaks
\begin{align}
\mathbb{H}_0 &=
H_0\left[1-\frac{1}{3}\delta_c \: \xi_o\left(\frac{\xi_o^2}{(1+\xi_o^2)^{3/2}}-\frac{9}{2}\xi_H^2 \csch^{-1}(\xi_o)\right)\right],
\\
\mathbb{H}_2 &=
\frac{2\delta_c H_0 \xi_o^3}{3(1+\xi_o)^{3/2}}\left[-4-3\xi_o^2+3(1+\xi_o^2)^{3/2}\csch^{-1}(\xi_o)\right],
\\
\mathbb{Q}_0 &=q_0+\frac{5\delta_c \xi_o^3}{6 \left(1+\xi_o^2\right)^{3/2}},\label{appb3}
\\
\mathbb{Q}_1 &=-\frac{9\delta_c \xi_o^4}{5 \xi_H (1+\xi_o^2)^{5/2}},
\\
\mathbb{Q}_2 &=-\frac{5}{2}\frac{\mathbb{H}_2}{H_0} \,,
\\
\mathbb{Q}_3&=\frac{2\delta_c \xi_o^4}{5 \xi_H (1+\xi_o^2)^{5/2}}\left[15(1+\xi_o^2)^{5/2}\csch^{-1}(\xi_o)-23-35\xi_o^2-15\xi_o^4\right],
\\
\mathbb{R}_0 &=1-2\mathbb{Q}_0\,,
\\
\mathbb{R}_1 &=\mathbb{Q}_1 \,,
\\
\mathbb{R}_2 &=-\frac{\mathbb{Q}_2}{5},
\\
\mathbb{R}_3 &=\mathbb{Q}_3 \,,
\\
\mathbb{J}_0 &=j_0-\frac{\delta_c \xi_o^3}{15 \xi_H^{2} (1+\xi_o^2)^{7/2}}\left[9\xi_o^2(2-3\xi_o^2)-25 \xi_H^2 (1+\xi_o^2)^2\right],
\\
\mathbb{J}_1&=-\mathbb{Q}_1\,,
\\
\mathbb{J}_2 &=- \frac{10\delta_c \xi_o^3}{21 \xi_H^{2} (1+\xi_o^2)^{7/2}}\left[18\xi_o^2-7(1+\xi_o^2)^2 (4+3\xi_o^2) \xi_H^2+21 \xi_H^2 (1+\xi_o^2)^{7/2}\csch^{-1}(\xi_o)\right],
\\
\mathbb{J}_3 &=-\mathbb{Q}_3\,,
\\
\mathbb{J}_4 &=\frac{8\delta_c \xi_o^5}{35 \xi_H^{2} (1+\xi_o^2)^{7/2}}\left[105(1+\xi_o^2)^{7/2}\csch^{-1}(\xi_o)-7\xi_o^2(58+50\xi_o^2+15\xi_o^4)-176\right].
\end{align}
\endgroup

Note that, in addition to the monopole of the deceleration parameter $\mathbb{Q}$, which mainly comes from the background, the dominant multipoles of $\mathbb{Q}$ are the dipole and the octupole -- since they are proportional to $1/ \xi_H$. They compensate for the smallness of $z$ in the Taylor expansion of eqs. (\ref{dipe}) and (\ref{octe}) when the size of the perturbations is comparable with the redshift of volume under study. For the jerk parameter $\mathbb{J}$, the dominant multipoles are the even multipoles (monopole, quadrupole, hexadecapole), since they are proportional to $1/ \xi_H^2$, which compensates the smallness of $z^2$ in eqs. (\ref{monom}) and (\ref{quade}).

\section{Legendre coefficients estimation}\label{forleg}

We show here how we calculate the Legendre coefficients of an axisymmetric field $\eta$.
For each spherical shell $S$ at redshift $z$ and with thickness $\delta z$, we calculate the angle $\theta$ between the angular position of each galaxy $j$ contained in the shell $(l_j,b_j)$, and the axis of symmetry $(l_a,b_a)$:
\begin{equation}
\cos(\theta_j)=\sin(b_j) \sin(b_a)+\cos(b_j) \cos(b_a) \cos(l_j-l_a)\,.
\end{equation}
We then split the shell into $N_{\rm bins}$ equal bins in $\cos\theta$ and calculate $\nu\equiv\log(z/d_L)=\log(z)+5-{\mu}/{5}$ for each galaxy in the bin, and calculate the weighted average of $\nu$:
\begin{equation}
\bar\nu(\cos\theta_i)={\sum_{j=1}^{(N_{\rm gal})_i} \frac{\nu(j)}{\delta_j^2}} \,\left[{\sum_{j=1}^{(N_{\rm gal})_i}\frac{1}{\delta_j^2}}\right]^{-1}\,.
\end{equation}
Here $(N_{\rm gal})_i$ is the number of galaxies in the bin $(\Delta\cos\theta)_i$, where $i$ runs from 1 to $N_{\rm bins}$ ($\theta_i$ is the angle calculated at the middle of the bin), and $\delta_j$ is the error of $\nu(j)$, which is induced by the uncertainty of the distance modulus ($\sigma_\mu$) of each galaxy ($\delta_j=\sigma_{\mu_ j}/5$). The zeroth order estimator of the Legendre coefficients is
\begin{equation}
\nu_{\ell}^{(0)}=\frac{2\ell +1}{N_{\rm bins}} \sum_{i=1}^{N_{\rm bins}} \bar\nu(\cos\theta_i)\, P_\ell(\cos\theta_i)\,.
 \label{alPl1}
\end{equation}
The number $N_{\rm bins}$ must be larger than the number of multipoles one wants to compute (we stop at $\ell=4$), but small enough such that all the bins have galaxies inside, so we set it to be $15$. We compute the higher-order refinements iteratively, according to a formula similar to that used by 
HEALPix \cite{Gorski:2004by}, in order to estimate the spherical harmonic coefficients of a generic, non-axisymmetric field:
\begin{equation}
\nu_{\ell}^{(k+1)}=\nu_{\ell}^{(k)}+ \frac{2\ell +1}{N_{\rm bins}}\sum^{N_{\rm bins}}_{i=1}\Big[\bar\nu(\cos\theta_i)-\bar\nu^{(k)}(\cos\theta_i)\Big] \, P_\ell(\cos\theta_i)\,,
  \label{almkestimate}
\end{equation}
where
\begin{equation}
  \bar\nu^{(k)}(\cos\theta_i)=\sum^{\ell_{\rm max}}_{\ell=0} \nu_\ell^{(k)}P_\ell(\cos\theta_i)\,.
\end{equation}
We choose $\ell_{\rm max}=4$, which is the maximum dominant multipole of the Jerk,  the highest order covariant cosmographic parameter considered in our analysis. Note that for the case under study the iteration  becomes stable after about 5 loops, so we stop the process for $k_{max}=10$. Concerning the variance of $\eta_\ell$, we proceed as follows. We calculate the variance of the zeroth order estimator $\nu_\ell^{(0)}$, which is
\begin{equation}
V[\nu_{\ell}^{(0)}]=\left(\frac{2\ell +1}{N_{\rm bins}}\right)^2\sum_{i=1}^{N_{\rm bins}}\big[\sigma_{\nu(i)} P_\ell(\cos\theta_i)\big]^2\,,
 \label{nulerr0}
\end{equation}
where $\sigma_{\nu(i)}^2$ is the variance of $\nu(\cos\theta_i)$: 
\begin{equation}	\sigma_{\nu(i)}^2=\left[{\sum_{j=1}^{(N_{\rm gal})_i}\frac{1}{\delta_j^2}}\right]^{-1}\,.
 \label{sigmanu}
\end{equation}
Since $\mathcal{M}=\nu_0^{(k_{\rm max})}$ and $\eta_\ell=\nu_\ell^{(k_{\rm max})}$,
we set 
\begin{equation}
\sigma_{\mathcal{M}}^2=V[\nu_0^{(0)}]\,,
 \label{a0err}
\end{equation}
\begin{equation}
\sigma_{\eta_{\ell}}^2=V[\nu_\ell^{(0)}]\,,
 \label{alerr}
\end{equation}
which provide excellent approximations, as we confirmed with simulations. This process can be used equivalently for reconstructing $\eta$ in the matter frame or in the CMB frame. If the distribution of the points is isotropic over the sky and the uncertainty in the distance modulus is the same for all objects, then we can approximate the variance of $\nu_\ell$ by
\begin{equation}	V[\nu_{\ell}]=\frac{2\ell+1}{N_{\rm gal}^{\text{inside the shell}}}\,\frac{\sigma_\mu^2}{25}\,,
 \label{alerr2}
\end{equation}
which does not depend on the choice of the number of bins, $N_{\rm bins}$.

\section{Correlation between multipoles of the expansion rate fluctuations}\label{app_corr}

The $\chi^2$ minimization (see eq. (\ref{chi2})) was performed by ignoring the correlation between the multipoles of $\eta$, i.e., by assuming that the covariance matrix and its inverse (the precision matrix) are diagonal. Here, for completeness, we justify this choice. We present the correlation matrix for some specific 
redshift shells of the model $M2$. We generate 50,000 samples with distance moduli randomly simulated around the measured value, and calculate the correlation matrix as 
\begin{equation}
\text{Corr}_{\ell\ell'}=\frac{\big\langle (\boldsymbol{\nu}_\ell-\langle\boldsymbol{\nu}_\ell\rangle) (\boldsymbol{\nu}_{\ell'}-\langle\boldsymbol{\nu}_{\ell'}\rangle)\big\rangle}{\sigma_{\boldsymbol{\nu}_{\ell}}\sigma_{\boldsymbol{\nu}_{\ell'}}}\,.
\end{equation}
Here angled brackets denote the average and $\sigma_{\boldsymbol{\nu}_{\ell}}$ is the standard deviation of the multipole $\ell$ of $\nu$, where $\nu$ is defined in Appendix \ref{forleg} ($\mathcal{M}=\nu_0$, $\eta_\ell=\nu_\ell$ for $\ell>0$). The elements of the correlation matrix at three different depths are
\begin{align}
  \bold{Corr}\,(\tilde z =0.015)&=\left(
\begin{array}{cccc}
 1 & 0.03 & -0.06 & -0.04 \\
 0.03 & 1 & -0.01 & 0.06 \\
 -0.06 & -0.01 & 1 & 0.04 \\
 -0.04 & 0.06 & 0.04 & 1 \\
\end{array}
\right),\\
  \bold{Corr}\,(\tilde z =0.055)&=\left(
\begin{array}{cccc}
 1 & 0.024 & 0.001 & 0.002 \\
 0.024 & 1 & 0.02 & -0.001 \\
 0.001 & 0.02 & 1 & 0.05 \\
 0.002 & -0.001 & 0.05 & 1 \\
\end{array}
\right),\\
  \bold{Corr}\,(\tilde z =0.095)&=\left(
\begin{array}{cccc}
 1 & -0.041 & -0.006 & -0.027 \\
 -0.041 & 1 & -0.062 & -0.007 \\
 -0.006 & -0.062 & 1 & -0.001 \\
 -0.027 & -0.007 & -0.001 & 1 \\
\end{array}
\right).
\end{align}
The off-diagonal elements are indeed small and can be safely neglected.

\newpage
\bibliographystyle{JHEP}
\bibliography{biblio}

\providecommand{\href}[2]{#2}\begingroup\raggedright\begin{thebibliography}{10}

\bibitem{DiValentino:2021izs}
E.~Di~Valentino, O.~Mena, S.~Pan, L.~Visinelli, W.~Yang, A.~Melchiorri et~al.,
  \emph{{In the realm of the Hubble tension\textemdash{}a review of
  solutions}}, \href{https://doi.org/10.1088/1361-6382/ac086d}{\emph{Class.
  Quant. Grav.} {\bfseries 38} (2021) 153001}
  [\href{https://arxiv.org/abs/2103.01183}{{\ttfamily 2103.01183}}].

\bibitem{Abdalla:2022yfr}
E.~Abdalla et~al., \emph{{Cosmology intertwined: A review of the particle
  physics, astrophysics, and cosmology associated with the cosmological
  tensions and anomalies}},
  \href{https://doi.org/10.1016/j.jheap.2022.04.002}{\emph{JHEAp} {\bfseries
  34} (2022) 49} [\href{https://arxiv.org/abs/2203.06142}{{\ttfamily
  2203.06142}}].

\bibitem{Perivolaropoulos:2021jda}
L.~Perivolaropoulos and F.~Skara, \emph{{Challenges for
  \ensuremath{\Lambda}CDM: An update}},
  \href{https://doi.org/10.1016/j.newar.2022.101659}{\emph{New Astron. Rev.}
  {\bfseries 95} (2022) 101659}
  [\href{https://arxiv.org/abs/2105.05208}{{\ttfamily 2105.05208}}].

\bibitem{Schoneberg:2021qvd}
N.~Sch\"oneberg, G.~Franco~Abell\'an, A.~P\'erez~S\'anchez, S.J.~Witte,
  V.~Poulin and J.~Lesgourgues, \emph{{The H0 Olympics: A fair ranking of
  proposed models}},
  \href{https://doi.org/10.1016/j.physrep.2022.07.001}{\emph{Phys. Rept.}
  {\bfseries 984} (2022) 1} [\href{https://arxiv.org/abs/2107.10291}{{\ttfamily
  2107.10291}}].

\bibitem{Schwarz:2007wf}
D.J.~Schwarz and B.~Weinhorst, \emph{{(An)isotropy of the Hubble diagram:
  Comparing hemispheres}},
  \href{https://doi.org/10.1051/0004-6361:20077998}{\emph{Astron. Astrophys.}
  {\bfseries 474} (2007) 717}
  [\href{https://arxiv.org/abs/0706.0165}{{\ttfamily 0706.0165}}].

\bibitem{Kashlinsky:2008ut}
A.~Kashlinsky, F.~Atrio-Barandela, D.~Kocevski and H.~Ebeling, \emph{{A
  measurement of large-scale peculiar velocities of clusters of galaxies:
  results and cosmological implications}},
  \href{https://doi.org/10.1086/592947}{\emph{Astrophys. J. Lett.} {\bfseries
  686} (2009) L49} [\href{https://arxiv.org/abs/0809.3734}{{\ttfamily
  0809.3734}}].

\bibitem{Antoniou:2010gw}
I.~Antoniou and L.~Perivolaropoulos, \emph{{Searching for a Cosmological
  Preferred Axis: Union2 Data Analysis and Comparison with Other Probes}},
  \href{https://doi.org/10.1088/1475-7516/2010/12/012}{\emph{JCAP} {\bfseries
  12} (2010) 012} [\href{https://arxiv.org/abs/1007.4347}{{\ttfamily
  1007.4347}}].

\bibitem{Cai:2011xs}
R.-G.~Cai and Z.-L.~Tuo, \emph{{Direction Dependence of the Deceleration
  Parameter}}, \href{https://doi.org/10.1088/1475-7516/2012/02/004}{\emph{JCAP}
  {\bfseries 02} (2012) 004} [\href{https://arxiv.org/abs/1109.0941}{{\ttfamily
  1109.0941}}].

\bibitem{Kalus:2012zu}
B.~Kalus, D.J.~Schwarz, M.~Seikel and A.~Wiegand, \emph{{Constraints on
  anisotropic cosmic expansion from supernovae}},
  \href{https://doi.org/10.1051/0004-6361/201220928}{\emph{Astron. Astrophys.}
  {\bfseries 553} (2013) A56}
  [\href{https://arxiv.org/abs/1212.3691}{{\ttfamily 1212.3691}}].

\bibitem{Wang:2014vqa}
J.S.~Wang and F.Y.~Wang, \emph{{Probing the anisotropic expansion from
  supernovae and GRBs in a model-independent way}},
  \href{https://doi.org/10.1093/mnras/stu1279}{\emph{Mon. Not. Roy. Astron.
  Soc.} {\bfseries 443} (2014) 1680}
  [\href{https://arxiv.org/abs/1406.6448}{{\ttfamily 1406.6448}}].

\bibitem{Yoon:2014daa}
M.~Yoon, D.~Huterer, C.~Gibelyou, A.~Kov\'acs and I.~Szapudi, \emph{{Dipolar
  modulation in number counts of WISE-2MASS sources}},
  \href{https://doi.org/10.1093/mnrasl/slu133}{\emph{Mon. Not. Roy. Astron.
  Soc.} {\bfseries 445} (2014) L60}
  [\href{https://arxiv.org/abs/1406.1187}{{\ttfamily 1406.1187}}].

\bibitem{Tiwari:2015tba}
P.~Tiwari and A.~Nusser, \emph{{Revisiting the NVSS number count dipole}},
  \href{https://doi.org/10.1088/1475-7516/2016/03/062}{\emph{JCAP} {\bfseries
  03} (2016) 062} [\href{https://arxiv.org/abs/1509.02532}{{\ttfamily
  1509.02532}}].

\bibitem{Javanmardi2015}
B.~{Javanmardi}, C.~{Porciani}, P.~{Kroupa} and J.~{Pflamm-Altenburg},
  \emph{{Probing the Isotropy of Cosmic Acceleration Traced By Type Ia
  Supernovae}},
  \href{https://doi.org/10.1088/0004-637X/810/1/47}{\emph{Astrophys. J.}
  {\bfseries 810} (2015) 47}
  [\href{https://arxiv.org/abs/1507.07560}{{\ttfamily 1507.07560}}].

\bibitem{Bengaly:2015nwa}
C.A.P.~Bengaly, Jr., \emph{{Constraining the local variance of $H_0$ from
  directional analyses}},
  \href{https://doi.org/10.1088/1475-7516/2016/04/036}{\emph{JCAP} {\bfseries
  04} (2016) 036} [\href{https://arxiv.org/abs/1510.05545}{{\ttfamily
  1510.05545}}].

\bibitem{Colin:2017juj}
J.~Colin, R.~Mohayaee, M.~Rameez and S.~Sarkar, \emph{{High redshift radio
  galaxies and divergence from the CMB dipole}},
  \href{https://doi.org/10.1093/mnras/stx1631}{\emph{Mon. Not. Roy. Astron.
  Soc.} {\bfseries 471} (2017) 1045}
  [\href{https://arxiv.org/abs/1703.09376}{{\ttfamily 1703.09376}}].

\bibitem{Rameez:2017euv}
M.~Rameez, R.~Mohayaee, S.~Sarkar and J.~Colin, \emph{{The dipole anisotropy of
  AllWISE galaxies}}, \href{https://doi.org/10.1093/mnras/sty619}{\emph{Mon.
  Not. Roy. Astron. Soc.} {\bfseries 477} (2018) 1772}
  [\href{https://arxiv.org/abs/1712.03444}{{\ttfamily 1712.03444}}].

\bibitem{Migkas:2020fza}
K.~Migkas, G.~Schellenberger, T.H.~Reiprich, F.~Pacaud, M.E.~Ramos-Ceja and
  L.~Lovisari, \emph{{Probing cosmic isotropy with a new X-ray galaxy cluster
  sample through the $L_{\text{X}}-T$ scaling relation}},
  \href{https://doi.org/10.1051/0004-6361/201936602}{\emph{Astron. Astrophys.}
  {\bfseries 636} (2020) A15}
  [\href{https://arxiv.org/abs/2004.03305}{{\ttfamily 2004.03305}}].

\bibitem{Migkas:2021zdo}
K.~Migkas, F.~Pacaud, G.~Schellenberger, J.~Erler, N.T.~Nguyen-Dang,
  T.H.~Reiprich et~al., \emph{{Cosmological implications of the anisotropy of
  ten galaxy cluster scaling relations}},
  \href{https://doi.org/10.1051/0004-6361/202140296}{\emph{Astron. Astrophys.}
  {\bfseries 649} (2021) A151}
  [\href{https://arxiv.org/abs/2103.13904}{{\ttfamily 2103.13904}}].

\bibitem{Secrest:2020has}
N.J.~Secrest, S.~von Hausegger, M.~Rameez, R.~Mohayaee, S.~Sarkar and J.~Colin,
  \emph{{A Test of the Cosmological Principle with Quasars}},
  \href{https://doi.org/10.3847/2041-8213/abdd40}{\emph{Astrophys. J. Lett.}
  {\bfseries 908} (2021) L51}
  [\href{https://arxiv.org/abs/2009.14826}{{\ttfamily 2009.14826}}].

\bibitem{Siewert:2020krp}
T.M.~Siewert, M.~Schmidt-Rubart and D.J.~Schwarz, \emph{{Cosmic radio dipole:
  Estimators and frequency dependence}},
  \href{https://doi.org/10.1051/0004-6361/202039840}{\emph{Astron. Astrophys.}
  {\bfseries 653} (2021) A9}
  [\href{https://arxiv.org/abs/2010.08366}{{\ttfamily 2010.08366}}].

\bibitem{Luongo:2021nqh}
O.~Luongo, M.~Muccino, E.O.~Colg\'ain, M.M.~Sheikh-Jabbari and L.~Yin,
  \emph{{Larger H0 values in the CMB dipole direction}},
  \href{https://doi.org/10.1103/PhysRevD.105.103510}{\emph{Phys. Rev. D}
  {\bfseries 105} (2022) 103510}
  [\href{https://arxiv.org/abs/2108.13228}{{\ttfamily 2108.13228}}].

\bibitem{Krishnan:2021jmh}
C.~Krishnan, R.~Mohayaee, E.O.~Colg\'ain, M.M.~Sheikh-Jabbari and L.~Yin,
  \emph{{Hints of FLRW breakdown from supernovae}},
  \href{https://doi.org/10.1103/PhysRevD.105.063514}{\emph{Phys. Rev. D}
  {\bfseries 105} (2022) 063514}
  [\href{https://arxiv.org/abs/2106.02532}{{\ttfamily 2106.02532}}].

\bibitem{Sorrenti:2022zat}
F.~Sorrenti, R.~Durrer and M.~Kunz, \emph{{The dipole of the
  Pantheon+SH0ES~data}},
  \href{https://doi.org/10.1088/1475-7516/2023/11/054}{\emph{JCAP} {\bfseries
  11} (2023) 054} [\href{https://arxiv.org/abs/2212.10328}{{\ttfamily
  2212.10328}}].

\bibitem{Aluri:2022hzs}
P.K.~Aluri et~al., \emph{{Is the observable Universe consistent with the
  cosmological principle?}},
  \href{https://doi.org/10.1088/1361-6382/acbefc}{\emph{Class. Quant. Grav.}
  {\bfseries 40} (2023) 094001}
  [\href{https://arxiv.org/abs/2207.05765}{{\ttfamily 2207.05765}}].

\bibitem{Cowell_Dhawan_Macpherson_2023}
J.A.~Cowell, S.~Dhawan and H.J.~Macpherson, \emph{Potential signature of a
  quadrupolar hubble expansion in pantheon+ supernovae},
  \href{https://doi.org/10.1093/mnras/stad2788}{\emph{Monthly Notices of the
  Royal Astronomical Society} {\bfseries 526} (2023) 1482–1494}.

\bibitem{Hu:2023eyf}
J.P.~Hu, Y.Y.~Wang, J.~Hu and F.Y.~Wang, \emph{{Testing the cosmological
  principle with the Pantheon+ sample and the region-fitting method}},
  \href{https://arxiv.org/abs/2310.11727}{{\ttfamily 2310.11727}}.

\bibitem{Dainotti:2021pqg}
M.G.~Dainotti, B.~De~Simone, T.~Schiavone, G.~Montani, E.~Rinaldi and
  G.~Lambiase, \emph{{On the Hubble constant tension in the SNe Ia Pantheon
  sample}}, \href{https://doi.org/10.3847/1538-4357/abeb73}{\emph{Astrophys.
  J.} {\bfseries 912} (2021) 150}
  [\href{https://arxiv.org/abs/2103.02117}{{\ttfamily 2103.02117}}].

\bibitem{Marinoni:2012ba}
C.~Marinoni, J.~Bel and A.~Buzzi, \emph{{The Scale of Cosmic Isotropy}},
  \href{https://doi.org/10.1088/1475-7516/2012/10/036}{\emph{JCAP} {\bfseries
  10} (2012) 036} [\href{https://arxiv.org/abs/1205.3309}{{\ttfamily
  1205.3309}}].

\bibitem{Hoffman:2017ako}
Y.~Hoffman, D.~Pomarede, R.~Brent~Tully and H.~Courtois, \emph{{The Dipole
  Repeller}},  \href{https://arxiv.org/abs/1702.02483}{{\ttfamily 1702.02483}}.

\bibitem{Strauss:1995fz}
M.A.~Strauss and J.A.~Willick, \emph{{The Density and peculiar velocity fields
  of nearby galaxies}},
  \href{https://doi.org/10.1016/0370-1573(95)00013-7}{\emph{Phys. Rept.}
  {\bfseries 261} (1995) 271}
  [\href{https://arxiv.org/abs/astro-ph/9502079}{{\ttfamily
  astro-ph/9502079}}].

\bibitem{kalbouneh_marinoni_bel_2023}
B.~Kalbouneh, C.~Marinoni and J.~Bel, \emph{{Multipole expansion of the local
  expansion rate}},
  \href{https://doi.org/10.1103/PhysRevD.107.023507}{\emph{Phys. Rev. D}
  {\bfseries 107} (2023) 023507}
  [\href{https://arxiv.org/abs/2210.11333}{{\ttfamily 2210.11333}}].

\bibitem{kristian_sachs_1966}
J.~Kristian and R.K.~Sachs, \emph{Observations in cosmology},
  \href{https://doi.org/10.1086/148522}{\emph{The Astrophysical Journal}
  {\bfseries 143} (1966) 379}.

\bibitem{MacCallum_Ellis_1970}
M.A.H.~MacCallum and G.F.R.~Ellis, \emph{A class of homogeneous cosmological
  models}, \href{https://doi.org/10.1007/bf01645496}{\emph{Communications in
  Mathematical Physics} {\bfseries 19} (1970) 31–64}.

\bibitem{ellis_2009}
G.F.R.~Ellis, \emph{Relativistic cosmology (republication of 1969 lectures)},
  \href{https://doi.org/10.1007/s10714-009-0760-7}{\emph{General Relativity and
  Gravitation} {\bfseries 41} (2009) 581–660}.

\bibitem{ellis_1983}
G.F.R.~{Ellis}, D.R.~{Matravers} and R.~{Treciokas}, \emph{{Anisotropic
  solutions of the Einstein-Boltzmann equations: I. General formalism}},
  \href{https://doi.org/10.1016/0003-4916(83)90023-4}{\emph{Annals of Physics}
  {\bfseries 150} (1983) 455}.

\bibitem{ellis85}
G.F.R.~{Ellis}, S.D.~{Nel}, R.~{Maartens}, W.R.~{Stoeger} and A.P.~{Whitman},
  \emph{{Ideal observational cosmology.}},
  \href{https://doi.org/10.1016/0370-1573(85)90030-4}{\emph{Physics Reports}
  {\bfseries 124} (1985) 315}.

\bibitem{Peebles1971}
P.J.~Peebles, \emph{Physical Cosmology}, Princeton University Press (1971).

\bibitem{Hasse:1999}
W.~Hasse and V.~Perlick, \emph{{On spacetime models with an isotropic Hubble
  flow}}, {\emph{Class. Quant. Grav.} {\bfseries 16} (1999) 2559}.

\bibitem{Clarkson_theses_2000}
C.A.~Clarkson, \emph{{On the observational characteristics of inhomogeneous
  cosmologies}},  \href{https://arxiv.org/abs/astro-ph/0008089}{{\ttfamily
  astro-ph/0008089}}.

\bibitem{clarkson_maartens_2010}
C.~Clarkson and R.~Maartens, \emph{Inhomogeneity and the foundations of
  concordance cosmology},
  \href{https://doi.org/10.1088/0264-9381/27/12/124008}{\emph{Classical and
  Quantum Gravity} {\bfseries 27} (2010) 124008}.

\bibitem{Umeh:2013}
O.~Umeh, \emph{{The influence of structure formation on the evolution of the
  Universe}}, PhD thesis, University of Cape Town (2013).

\bibitem{heinesen_2021}
A.~Heinesen, \emph{Multipole decomposition of the general luminosity distance
  hubble law — a new framework for observational cosmology},
  \href{https://doi.org/10.1088/1475-7516/2021/05/008}{\emph{Journal of
  Cosmology and Astroparticle Physics} {\bfseries 05} (2021) 008}.

\bibitem{Maartens:2023tib}
R.~Maartens, J.~Santiago, C.~Clarkson, B.~Kalbouneh and C.~Marinoni, \emph{{The
  observer-dependence of the Hubble parameter: a covariant perspective}},
  \href{https://arxiv.org/abs/2312.09875}{{\ttfamily 2312.09875}}.

\bibitem{Colin:2019ulu}
J.~Colin, R.~Mohayaee, M.~Rameez and S.~Sarkar, \emph{{A response to Rubin
  \textbackslash{}\& Heitlauf: ''Is the expansion of the universe accelerating?
  All signs still point to yes''}},
  \href{https://arxiv.org/abs/1912.04257}{{\ttfamily 1912.04257}}.

\bibitem{Rubin:2019ywt}
D.~Rubin and J.~Heitlauf, \emph{{Is the expansion of the universe accelerating?
  All signs still point to yes a local dipole anisotropy cannot explain dark
  energy}}, \href{https://doi.org/10.3847/1538-4357/ab7a16}{\emph{Astrophys.
  J.} {\bfseries 894} (2020) 68}
  [\href{https://arxiv.org/abs/1912.02191}{{\ttfamily 1912.02191}}].

\bibitem{Dhawan_Borderies_Macpherson_Heinesen_2022}
S.~Dhawan, A.~Borderies, H.J.~Macpherson and A.~Heinesen, \emph{The quadrupole
  in the local hubble parameter: First constraints using type ia supernova data
  and forecasts for future surveys},
  \href{https://doi.org/10.1093/mnras/stac3812}{\emph{Monthly Notices of the
  Royal Astronomical Society} {\bfseries 519} (2022) 4841–4855}.

\bibitem{Macpherson:2024zwu}
H.J.~Macpherson, \emph{{The impact of anisotropic sky-sampling on the Hubble
  constant in numerical relativity}},
  \href{https://arxiv.org/abs/2402.09659}{{\ttfamily 2402.09659}}.

\bibitem{amenouche:tel-04165406}
M.~Amenouche, \emph{{Probing local anisotropies with Type Ia Supernovae from
  the Zwicky Transient Facility}}, Ph.D. thesis, {Universit{\'e} Clermont
  Auvergne}, 2022.

\bibitem{Planck:2018nkj}
{\scshape Planck} collaboration, \emph{{Planck 2018 results. I. Overview and
  the cosmological legacy of Planck}},
  \href{https://doi.org/10.1051/0004-6361/201833880}{\emph{Astron. Astrophys.}
  {\bfseries 641} (2020) A1}
  [\href{https://arxiv.org/abs/1807.06205}{{\ttfamily 1807.06205}}].

\bibitem{Visser_2004}
M.~Visser, \emph{Jerk, snap and the cosmological equation of state},
  \href{https://doi.org/10.1088/0264-9381/21/11/006}{\emph{Classical and
  Quantum Gravity} {\bfseries 21} (2004) 2603–2615}.

\bibitem{hui_greene_2006}
L.~Hui and P.B.~Greene, \emph{Correlated fluctuations in luminosity distance
  and the importance of peculiar motion in supernova surveys},
  \href{https://doi.org/10.1103/physrevd.73.123526}{\emph{Physical Review D}
  {\bfseries 73} (2006) }.

\bibitem{Bonvin_Durrer_Gasparini_2006}
C.~Bonvin, R.~Durrer and M.A.~Gasparini, \emph{Fluctuations of the luminosity
  distance}, \href{https://doi.org/10.1103/physrevd.73.023523}{\emph{Physical
  Review D} {\bfseries 73} (2006) }.

\bibitem{Tully_2016}
R.B.~Tully, H.M.~Courtois and J.G.~Sorce, \emph{{COSMICFLOWS}-3},
  \href{https://doi.org/10.3847/0004-6256/152/2/50}{\emph{The Astronomical
  Journal} {\bfseries 152} (2016) 50}.

\bibitem{Scolnic2018}
D.M.~{Scolnic}, D.O.~{Jones}, A.~{Rest}, Y.C.~{Pan}, R.~{Chornock},
  R.J.~{Foley} et~al., \emph{{The Complete Light-curve Sample of
  Spectroscopically Confirmed SNe Ia from Pan-STARRS1 and Cosmological
  Constraints from the Combined Pantheon Sample}},
  \href{https://doi.org/10.3847/1538-4357/aab9bb}{\emph{Astrophys. J.}
  {\bfseries 859} (2018) 101}
  [\href{https://arxiv.org/abs/1710.00845}{{\ttfamily 1710.00845}}].

\bibitem{Horstmann:2021jjg}
N.~Horstmann, Y.~Pietschke and D.J.~Schwarz, \emph{{Inference of the cosmic
  rest-frame from supernovae Ia}},
  \href{https://doi.org/10.1051/0004-6361/202142640}{\emph{Astron. Astrophys.}
  {\bfseries 668} (2022) A34}
  [\href{https://arxiv.org/abs/2111.03055}{{\ttfamily 2111.03055}}].

\bibitem{marinoni_monaco_giuricin_costantini_1998}
C.~Marinoni, P.~Monaco, G.~Giuricin and B.~Costantini, \emph{Galaxy distances
  in the nearby universe: Corrections for peculiar motions},
  \href{https://doi.org/10.1086/306183}{\emph{The Astrophysical Journal}
  {\bfseries 505} (1998) 484–505}.

\bibitem{peebles_1980}
P.J.~Peebles, \emph{The large-scale structure of the universe}, Princeton
  University Press (1980).

\bibitem{Fleury:2015hgz}
P.~Fleury, \emph{{Light propagation in inhomogeneous and anisotropic
  cosmologies}}, Ph.D. thesis, Paris U., VI, IAP, 2015.
\newblock \href{https://arxiv.org/abs/1511.03702}{{\ttfamily 1511.03702}}.

\bibitem{durrer_2021}
R.~Durrer, \emph{The cosmic microwave background}, Cambridge University Press
  (2021).

\bibitem{Tully_Pomarède_Graziani_Courtois_Hoffman_Shaya_2019}
R.B.~Tully, D.~Pomarède, R.~Graziani, H.M.~Courtois, Y.~Hoffman and
  E.J.~Shaya, \emph{Cosmicflows-3: Cosmography of the local void},
  \href{https://doi.org/10.3847/1538-4357/ab2597}{\emph{The Astrophysical
  Journal} {\bfseries 880} (2019) 24}.

\bibitem{Dhawan_2022}
S.~Dhawan, A.~Goobar, J.~Johansson, I.S.~Jang, M.~Rigault, L.~Harvey et~al.,
  \emph{A uniform type ia supernova distance ladder with the zwicky transient
  facility: Absolute calibration based on the tip of the red giant branch
  method}, \href{https://doi.org/10.3847/1538-4357/ac7ceb}{\emph{The
  Astrophysical Journal} {\bfseries 934} (2022) 185}.

\bibitem{Kalbouneh_p3}
B.~Kalbouneh et~al., \emph{In preparation},  2024.

\bibitem{Heinesen_Macpherson_2022}
A.~Heinesen and H.J.~Macpherson, \emph{A prediction for anisotropies in the
  nearby hubble flow},
  \href{https://doi.org/10.1088/1475-7516/2022/03/057}{\emph{Journal of
  Cosmology and Astroparticle Physics} {\bfseries 2022} (2022) 057}.

\bibitem{Adi_Nusser_2015}
A.~Nusser, \emph{On methods of estimating cosmological bulk flows},
  \href{https://doi.org/10.1093/mnras/stv2099}{\emph{Monthly Notices of the
  Royal Astronomical Society} {\bfseries 455} (2015) 178–184}.

\bibitem{Tully:2022rbj_CF4}
R.B.~Tully et~al., \emph{{Cosmicflows-4}},
  \href{https://doi.org/10.3847/1538-4357/ac94d8}{\emph{Astrophys. J.}
  {\bfseries 944} (2023) 94}
  [\href{https://arxiv.org/abs/2209.11238}{{\ttfamily 2209.11238}}].

\bibitem{Brout:2022vxf_Pantheon+}
D.~Brout et~al., \emph{{The Pantheon+ Analysis: Cosmological Constraints}},
  \href{https://doi.org/10.3847/1538-4357/ac8e04}{\emph{Astrophys. J.}
  {\bfseries 938} (2022) 110}
  [\href{https://arxiv.org/abs/2202.04077}{{\ttfamily 2202.04077}}].

\bibitem{Gorski:2004by}
K.M.~G\'orski, E.~Hivon, A.J.~Banday, B.D.~Wandelt, F.K.~Hansen, M.~Reinecke
  et~al., \emph{{HEALPix - A Framework for high resolution discretization, and
  fast analysis of data distributed on the sphere}},
  \href{https://doi.org/10.1086/427976}{\emph{Astrophys. J.} {\bfseries 622}
  (2005) 759} [\href{https://arxiv.org/abs/astro-ph/0409513}{{\ttfamily
  astro-ph/0409513}}].

\end{thebibliography}\endgroup

\end{document}